\documentclass[fleqn,usenatbib]{mnras}

\usepackage{newtxtext,newtxmath}

\usepackage[T1]{fontenc}
\usepackage{ae,aecompl}

\usepackage{graphicx}	
\usepackage{amsmath}	
\usepackage[dvipsnames]{xcolor}
\usepackage{multirow}
\usepackage{delarray}
\usepackage{color}
\usepackage{here}
\usepackage{float}
\usepackage{tensor}
\usepackage{xspace}
\usepackage{orcidlink}
\usepackage[multiple]{footmisc}

\newcommand{\mach}{\mathcal{M}}

\newcommand{\Mdot}{\dot{M}}
\newcommand{\SFE}{\mathrm{SFE}}
\newcommand{\solarmass}{\mathrm{M}_{\rm \sun}}
\newcommand{\msun}{\solarmass}
\newcommand{\Msink}{M_\mathrm{sink}}
\newcommand{\Nsink}{N_\mathrm{sink}}
\newcommand{\Mmean}{M_{\rm mean}}
\newcommand{\Mmedian}{M_{\rm med}}
\newcommand{\Mmassmedian}{M_{\rm 50}}
\newcommand{\Msonic}{M_{\rm sonic}}
\newcommand{\MBE}{M_{\rm BE}^{\rm turb}}

\newcommand{\Lsonic}{L_{\rm sonic}}
\newcommand{\MJeans}{M_{\rm Jeans}}
\newcommand{\alphath}{\alpha_{\mathrm{th}}}
\newcommand{\alphaturb}{\alpha_{\mathrm{turb}}}
\newcommand{\alphaB}{\alpha_{\mathrm{B}}}
\newcommand{\cs}{c_{\rm s}}
\newcommand{\csmin}{c_{\rm s, min}}

\newcommand{\pc}{\mathrm{pc}}
\newcommand{\kelvin}{\mathrm{K}}

\newcommand{\dderiv}{\mathrm{d}}
\newcommand{\vvector}{\mathbf{v}}

\newcommand{\appropto}{\mathrel{\vcenter{
  \offinterlineskip\halign{\hfil$##$\cr
    \propto\cr\noalign{\kern2pt}\sim\cr\noalign{\kern-2pt}}}}}

\newcommand{\myquote}[1]{``#1''}

\usepackage[normalem]{ulem}

\defcitealias{Guszejnov_isoT_MHD}{Paper 0}
\defcitealias{grudic_starforge_methods}{Paper I}

\title[STARFORGE: STARFORGE: Protostellar outflows and the IMF]{STARFORGE: The effects of protostellar outflows on the IMF}

\author[]{
D\'avid Guszejnov\orcidlink{0000-0001-5541-3150}$^{1}$\thanks{guszejnov@utexas.edu},
Michael Y. Grudi\'{c}\orcidlink{0000-0002-1655-5604}$^{2}$\thanks{mike.grudic@northwestern.edu},
Philip F. Hopkins\orcidlink{0000-0003-3729-1684}$^{3}$,
\newauthor
Stella S. R. Offner\orcidlink{0000-0003-1252-9916}$^{1}$,
Claude-Andr{\'e} Faucher-Gigu{\`e}re\orcidlink{0000-0002-4900-6628}$^{2}$
\\
$^{1}$Department of Astronomy, University of Texas at Austin, TX 78712, USA \\
$^{2}${CIERA and Department of Physics and Astronomy, Northwestern University, 2145 Sheridan Road, Evanston, IL 60208, USA}\\
$^{3}$TAPIR, Mailcode 350-17, California Institute of Technology, Pasadena, CA 91125, USA \\
}

\date{\today \vspace{-0.6cm}}

\pubyear{2020}

\begin{document}
\label{firstpage}
\pagerange{\pageref{firstpage}--\pageref{lastpage}}
\maketitle

\begin{abstract}
The initial mass function (IMF) of stars is a key quantity affecting almost every field of astrophysics, yet it remains unclear what physical mechanisms determine it. We present the first runs of the STARFORGE project, using a new numerical framework to follow the formation of individual stars in giant molecular clouds (GMCs) using the {\small GIZMO} code. Our suite include runs with increasingly complex physics, starting with isothermal ideal magnetohydrodynamic (MHD) and then adding non-isothermal thermodynamics and protostellar outflows. We show that without protostellar outflows the resulting stellar masses are an order of magnitude too high, similar to the result in the base isothermal MHD run. Outflows disrupt the accretion flow around the protostar, allowing gas to fragment and additional stars to form, thereby lowering the mean stellar mass to a value similar to that observed. The effect of jets upon global cloud evolution is most pronounced for lower-mass GMCs and dense clumps, so while jets can disrupt low-mass clouds, they are unable to regulate star formation in massive GMCs, as they would turn an order unity fraction of the mass into stars before unbinding the cloud. Jets are also unable to stop the runaway accretion of massive stars, which could ultimately lead to the formation of stars with masses $>500\,\msun$. Although we find that the mass scale set by jets is insensitive to most cloud parameters (i.e., surface density, virial parameter), it is strongly dependent on the momentum loading of the jets (which is poorly constrained by observations) as well the the temperature of the parent cloud, which predicts slightly larger IMF variations than observed. We conclude that protostellar jets play a vital role in setting the mass scale of stars, but additional physics are necessary to reproduce the observed IMF.
\end{abstract}



 \section{Introduction}\label{sec:intro}

 Star formation involves a large set of interconnected complex physical processes, including gravity, turbulence, magnetic fields, chemistry and radiation \citep{Girichidis_2020_sf_processes_review}. While each of these processes is necessary for a full picture of star formation, it is important to understand what role each of them plays and how they interact with each other.
 
 Due to the complexity of the physics involved, star formation models often consider only a subset of the relevant physical processes to make the problem analytically (and even numerically) tractable. The simplest such model considers only the equations of isothermal hydrodynamics coupled to gravity, which models the dense, $\sim 10\rm K$ interstellar medium (ISM) found in molecular clouds in our Galaxy \citep[e.g.,][]{padoan_nordlund_2002_imf,hc08,core_imf}. Recent numerical works have shown that the mass spectrum of collapsed fragments in such systems does not converge with numerical resolution \citep[see e.g.][]{Martel_numerical_sim_convergence,Kratter10a,guszejnov_feedback_necessity,Federrath_2017_IMF_converge_proceedings, guszejnov_isothermal_collapse, Hennebelle_Lee_isoT_sim}, so additional physics must play a role.
 
 Observations suggest that molecular clouds have significant support from magnetic fields \citep{crutcher_2009_mc_magnetic_fields}. In theoretical and numerical works the addition of magnetic fields to isothermal star formation models have been shown to impose a resolution independent scale on the stellar mass spectrum \citep[see e.g.][]{padoan_2007, padoan_nordlund_2011_imf, Haugbolle_Padoan_isot_IMF}. While some of these studies claimed to reproduce the observed IMF, our recent study \citep{Guszejnov_isoT_MHD} showed that, for clouds similar to GMCs in the Milky Way, the mean stellar masses predicted by these magnetized, gravo-turbulent models are an order of magnitude higher than observed (i.e., the mean stellar mass is $\sim 4\,\msun$ in the simulations while $\sim 0.4\,\msun$ is observed). This study also found that stellar masses in isothermal MHD also increase with time and are sensitive to initial conditions (see analysis in \S4.2 of \citealt{Guszejnov_isoT_MHD}), leading to order of magnitude variations in the predicted characteristic scale of the IMF. Observations, however, have found the IMF to be near-universal within the Milky Way (MW), with variations in the IMF peak mass within a factor of <3 (see reviews of \citealt{imf_review} and \citealt{imf_universality}, as well as analysis of \citealt{Dib_2014_IMF_variations_bayesian}). 
 
Of course the ISM is not isothermal, one of the key assumptions of the above models is the gas can cool more rapidly than other relevant timescales, making it effectively isothermal for this problem. This behavior, however, is only a crude approximation of the real thermochemistry and radiative cooling, detailed calculations (e.g., \citealt{Glover_Clark_n_T}) have shown significant temperature differences between low density regions ($\sim 10^2\,\mathrm{cm}^{-3}$, $T\sim 30\,\kelvin$) and high density regions where collapse occurs ($\sim 10^5\,\mathrm{cm}^{-3}$, $T\sim 10\,\kelvin$). Even at high densities, the isothermal assumption inevitably breaks down completely at high densities, when the cloud becomes opaque to its own cooling radiation, leading to an increase in temperature and thus a suppression of fragmentation (for the original idea see \citealt{lowlyndenbell1976, rees1976}, for modern interpretations see \citealt{Lee_Hennebelle_2018_EOS,Colman_Teyssier_2019_tidal_screening}). 

Another key feature of the simple models above is that they neglect feedback from the forming protostar and the stars that previously formed. These processes can dramatically effect the star formation process, as accreting protostars heat their surroundings \citep{Offner_2009_radiative_sim, krumholz_stellar_mass_origin, bate12a, Myers_2013_ORION_radiation_IMF,guszejnov_gmc_imf,guszejnov_feedback_necessity}. Previously formed massive stars can also heat 
a large portion of their progenitor cloud and shut down star formation altogether \citep{grudic_2016, kim_2018_gmc_raytrace, Li_Vogelsberger_2019_GMC_disrupt}. The mass loss of accreting protostars is dominated by high velocity bipolar outflows that can significantly affect their environment \citep[see reviews of][]{Frank_2014_jets_preview, Bally_2016_outflows_review}. These outflows are thought to be driven by highly collimated bipolar jets that entrain the ambient gas \citep{Rosen_krumholz_2020_outflows_massive_stars}. These jets in turn are launched by MHD interactions between the protostar and the accretion disk \citep{Shu_1988_X_winds,Pelletier_1992_D_wind}, with radiation pressure also contributing to their driving \citep{Kuiper_2010_massive_star_radiation_outflow,Vaidya_2011_massive_jet_sim}. These jets not only reduce the accretion rates of stars but also disrupt local accretion flows and drive turbulence on small scales (\citealt{Nakamura_2007_outflow_turbulence_driving,Matzner_2007_otflow_tub_obs,Wang_2010_outflow_regulated_SF,Cunningham_2011_outflow_sim,Offner_Arce_2014,Federrath_2014_jets,Offner_Chaban_2017_jets_sfe, murray_2018_jets}).

Past work has shown that protostellar jets significantly reduce the global  star formation rate in a cloud \citep{hansen_lowmass_sf_feedback, Federrath_2014_jets}. Protostellar jets have been shown to play a role in setting the mass scale of stars, preventing \myquote{over-accretion} from stars heating up their surroundings, thus preventing the gas from fragmenting and forming new stars \citep{krumholz_2012_orion_sims,li_2018_sf_mhd_jets,Cunningham_2018_feedback}.

Simulations that take into account the above processes are necessary to understand the effects of each physical process, but so far these have generally been limited to simple physics or a very narrow range of cloud initial conditions (ICs). In this paper we introduce the first results from the STAR FORmation in Gaseous Environments (STARFORGE) project\footnote{\url{http://www.starforge.space}}. These MHD simulations achieve a dynamic range in mass resolution that is an order of magnitude higher than any previous star cluster simulation, allowing us to simulate the detailed evolution of Giant Molecular Clouds (GMCs) while following the formation of individual low-mass stars (see companion methods paper of \citealt{grudic_starforge_methods}, henceforth referred to as \citetalias{grudic_starforge_methods}). In this study we perform and analyze a set of simulations with different initial conditions and levels of physics to identify the effects of non-isothermality and protostellar jets on the IMF. 

We present our results in \S\ref{sec:results} with a focus on how the characteristic masses of sink particles (stars) change with the inclusion of additional physics and variations in the initial conditions (e.g., cloud temperature, surface density, level of turbulence). In \S\ref{sec:jet_peak_model} we introduce a simple toy model to explain the effects of protostellar outflows. The implications of these result as well as the potential role of further physics are discussed \S\ref{sec:discussion}. We summarize our conclusions in \S\ref{sec:conclusions} and leave the details on how exactly our results vary with initial conditions to Appendix \ref{sec:scaling_dependencies}


 \section{Methods}\label{sec:methods}
 
  \subsection{Physics}\label{sec:physics}
   A full description and presentation of our methods including a variety of tests and algorithm details are given in a companion methods paper \citepalias{grudic_starforge_methods}, therefore we only briefly summarize them here. 
   
  \subsubsection{Core Physics}\label{sec:shared_physics}
   Similar to our previous studies of isothermal collapse with and without magnetic fields (\citealt{guszejnov_isothermal_collapse} and \citealt{Guszejnov_isoT_MHD}, to the latter of which we will henceforth refer to as \citetalias{Guszejnov_isoT_MHD}), we simulate star-forming clouds with the {\small GIZMO} code\footnote{\url{http://www.tapir.caltech.edu/~phopkins/Site/GIZMO.html}} (\citealt{hopkins_mfm_2015}), using the Lagrangian meshless finite-mass (MFM) method for magnetohydrodynamics \citep{hopkins_gizmo_mhd}, assuming ideal MHD (with the constrained gradient scheme of \citealt{Hopkins_2016_divb_cleaning} to ensure that $\nabla\cdot \mathbf{B}=0$).
  

  Gravity is solved with an improved version of the Barnes-Hut tree method from \citet{Springel_2005_gadget} with high-order integration of sink particle trajectories to accurately follow multiple sink systems (see \citetalias{grudic_starforge_methods}). Force softening is fully adaptive for gas cells \citep{price_monaghan_softening, hopkins2015_gizmo}. Sink particles (representing stars) have a fixed Plummer-equivalent softening radius of $7.56\,\rm AU$. We adopt the sink formation and accretion algorithm from \citet{Bate_1995_accretion}, while accurately accounting for thermal, magnetic, kinetic and gravitational energies and angular momentum, again described in \citetalias{grudic_starforge_methods}. As such we are able to follow the formation and evolution of binaries and multiples with separations larger than $\sim 10\,\rm AU$.  
  
Once sinks form, they follow the protostellar evolution model from \citet{Offner_2009_radiative_sim},  which is also used in the {\small ORION} code. In this model the protostar is treated as a collapsing polytrope: the collapse is divided into distinct phases during which the qualitative behavior changes. These phases are \myquote{pre-collapse}, \myquote{no burning}, \myquote{core deuterium burning at fixed temperature}, \myquote{core deuterium burning at variable temperature}, \myquote{shell deuterium burning} and \myquote{main sequence}. This module dynamically evolves stellar properties (e.g., radius, accretion and internal luminosities) throughout the simulation. For details see Appendix B of \citet{Offner_2009_radiative_sim} and \citetalias{grudic_starforge_methods}.
  

\subsubsection{Thermodynamics}\label{sec:cooling}
We compare simulations with two different thermodynamics modules. Our \myquote{isothermal} simulations enforce an isothermal equation of state (EOS) with $c_s=0.2\,\rm km/s$ (effective gas temperature $T\sim10\,\kelvin$). Our \myquote{non-isothermal} or \myquote{cooling} simulation runs utilize the radiative cooling and themrochemistry module presented in \citet{hopkins2017_fire2} that contains detailed metallicity-dependent cooling and heating physics from $T=10-10^{10}\,$K, including recombination, thermal bremsstrahlung, metal lines (following \citealt{Wiersma2009_cooling}), molecular lines, fine structure (following \citealt{CLOUDY}) and dust collisional processes. The cooling module self-consistently solves for the internal energy and ionization state of the gas (see Appendix B of \citealt{hopkins2017_fire2}). The gas adiabatic index is calculated from a fit to density based on the results of \citet{Vaidya_2015_EOS}. Note that a constant dust temperature of $T_\mathrm{dust}=10\,\mathrm{K}$ and a temperaure floor of $T_\mathrm{floor}=10\,\mathrm{K}$ are assumed here. As detailed in \citetalias{grudic_starforge_methods}, this module does not explicitly evolve radiation-hydrodynamics (RHD), but it does attempt to approximately capture the transition between optically thick and optically thin cooling regimes. It does so following \citet{rafikov2007} and modeling each gas cell as a plane-parallel atmosphere with with optical depth to escape integrated using the {\small TreeCol} algorithm \citep{treecol}.





\subsubsection{Protostellar jets}\label{sec:jets}

Protostars eject a significant portion of the accreting material in bipolar jets. To represent this process we adopt the following jet model: each accreting protostar launches an $f_w$ fraction of the accreting mass in bipolar jets along its rotational axis with a velocity of 
\begin{equation}
v_\mathrm{jet} = f_K \sqrt{G M_\ast/R_*},
\label{eq:jet_velocity}
\end{equation}
which is just $f_K$ times the Keplerian velocity at the surface of the star, where the $R_*$ stellar radius is evolved using the protostellar evolution model of \citet{Offner_2009_radiative_sim}. Observations estimate the $f_w$ mass loading parameter to be in the range of 0.1-0.4 \citep[see review by][]{Frank_2014_jets_preview}, while simulations found values 0.1-0.6 \citep[e.g.][]{Seifried_2012_magnetized_outflow_sim}. The $f_K$ velocity scaling parameter is not observed directly, however $f_K f_w$ can be derived from the observed momentum injection rate by assuming a constant protostellar radius (see \S2.4 \citealt{Cunningham_2011_outflow_sim}), which yields the constraint $f_K f_w \sim 0.05-0.4$. In our runs we adopt $f_w=0.3$ and $f_K=0.3$, similar to the values used by \cite{Cunningham_2011_outflow_sim} and many other works, which puts $f_{\rm w} f_{\rm K}$ in the middle of the observed range. It is useful to introduce the $\Gamma$ \emph{momentum loading parameter}
\begin{equation}
\Gamma = \frac{2}{3} \frac{f_w f_K}{(1-f_w)},
\label{eq:gamma_def}
\end{equation}
which describes how the momentum output of the jets per unit accreted mass scales with these parameters (see \S\ref{sec:jet_peak_model} for a derivation).

The numerical implementation of jets is described in \citetalias{grudic_starforge_methods}, briefly we spawn new gas cells around the sink particle and launch them along the sink particle's angular momentum axis using the same angular distribution model as \cite{Matzner_McKee_1999_jet_model} and \cite{Cunningham_2011_outflow_sim}, which corresponds to a vanishingly small opening angle. We find that the exact value of the opening angle has little effect on the results, provided that it is $<<1$ (see \citetalias{grudic_starforge_methods} for details). These gas cells are spawned in pairs (to conserve momentum and centre of mass exactly) and in mass quanta of $\Delta m_\mathrm{jet} = 0.1 \Delta m$, where $\Delta m$ is the mass resolution element of our simulation, for which our fiducial value is $\Delta m=10^{-3}\,\msun$, sufficient to predict the shape of the IMF in the stellar ($\gtrsim 0.1\msun$) mass range (see \citetalias{grudic_starforge_methods} for resolution study).

In \citetalias{grudic_starforge_methods} we find that for $\Delta m_\mathrm{jet}/\Delta m\leq 1$ the sink mass spectrum is insensitive to our choice of $\Delta m_\mathrm{jet}$, so we adopt $\Delta m_\mathrm{jet}/\Delta m=0.1$ in our simulations. We will show that the effects from the jet module are primarily determined by the $\Gamma$ momentum loading parameter, see \S\ref{sec:jet_mom_loading} for a details.

 
\subsection{Initial Conditions \& Parameters of Clouds }\label{sec:physics_ladder}

\subsubsection{Initial conditions}\label{sec:IC}
The main aim of the STARFORGE project is to identify the roles different physical processes play in star formation from the protostellar to the GMC scale (AU to $100\,\pc$). This investigation requires simulations of GMC scale clouds with individual star formation and progressively more complicated physics: starting with magnetized, isothermal gas (see \citetalias{Guszejnov_isoT_MHD}), then enabling gas thermodynamics without stellar feedback and finally adding protostellar outflows, see Table \ref{tab:physics_ladder} for the different \myquote{rungs} of this \myquote{physics ladder}. To explore the dependency of our results on initial conditions and simulation parameters, we also carry out a detailed parameter study. Note that the STARFORGE numerical framework can incorporate many other important feedback physics (e.g., radiative heating, winds, and supernovae: for methods see \citetalias{grudic_starforge_methods}), which will be explored in future papers.

\begin{table}
\setlength\tabcolsep{3.0pt} 
\centering
	\centering
	\begin{tabular}{ | c | c | c | c | c | c |c |c | }
	\hline
	\textbf{Physics label} & MHD & Thermodynamics & Protostellar Jets \\
	\hline
	\textbf{I\_M} & Ideal (M) & Isothermal (I) & Not included  \\
	\hline
	\textbf{C\_M} & Ideal (M) & ApproxRad (C) & Not included   \\
	\hline
	\textbf{C\_M\_J} & Ideal (M) & ApproxRad (C) & Included (J) \\
	\hline
	\end{tabular}
	\caption{Labels used by through this paper to identify simulations with different physics. See \S\ref{sec:physics} and \citetalias{grudic_starforge_methods} for details on the individual physics modules.
	}
\label{tab:physics_ladder}
\end{table}

We generate our ICs using {\small MakeCloud}\footnote{\url{https://github.com/mikegrudic/MakeCloud}}. Unless otherwise specified our runs utilize \emph{\myquote{Sphere} ICs}, meaning that we initialize a spherical cloud ($T=10\,\kelvin$, radius $R_\mathrm{cloud}$ and mass $M_\mathrm{0}$) with uniform density, surrounded by diffuse gas with a density contrast of 1000. The cloud is placed at the center of a periodic $10 R_\mathrm{cloud}$ box. The initial velocity field is a Gaussian random field with power spectrum $E_k\propto k^{-2}$ \citep{ostriker_2001_mhd}, scaled to the value prescribed by $\alpha_{\rm turb}$. The initial clouds have a uniform $B_z$ magnetic field whose strength is set by the parameter $\mu$. There is no external driving in these simulations.

We also run simulations using \emph{\myquote{Box} ICs}, similar to the driven boxes used in e.g., \citet{Federrath_2014_jets,Cunningham_2018_feedback}. These are initialized as constant density, zero velocity periodic cubic box with $T=10\,\kelvin$. This periodic box is then \myquote{stirred} using the driving algorithm from by \citealt{federrath_sim_compare_2010,bauerspringel2012}. This involves a spectrum of $E_k\propto k^{-2}$ of driving modes in Fourier space at wavenumbers 1/2 - 1 the box size, with an appropriate decay time for driving mode correlations ($t_{\mathrm{decay}}\sim t_{\mathrm{cross}}\sim L_{\rm box}/\sigma_{\rm 3D}$). This stirring is initially performed without gravity for five global freefall times $\left(t_{\mathrm{ff}}\equiv\sqrt{\frac{3 \pi}{32 G \rho_0}}\right)$, to achieve saturated MHD turbulence. The normalization of the driving spectrum is set so that in equilibrium the gas in the box has a turbulent velocity dispersion ($\sigma_{\rm 3D}$) that gives the desired $\mach$ and $\alpha_{\rm turb}$. We use purely solenoidal driving, which remains active throughout the simulation after gravity is switched on. We take the box side length $L_\mathrm{box}$ to give a box of equal volume to the associated {\it Sphere} cloud model, and thus define $\alphaturb$ using the volume-equivalent $R_\mathrm{cloud}$ in Equation \ref{eq:alphaturb_sphere}. An important difference between the \textit{Sphere} and \textit{Box} runs is that in the case of driven boxes the magnetic field is enhanced by a turbulent dynamo \citep{federrath_2014_dynamo} and saturates at about $\alphaB\sim 0.1$ (see \citetalias{Guszejnov_isoT_MHD}), so for Box runs the \myquote{pre-stirring} magnetic field strength (defined by $\mu$) does not directly specify the actual initial magnetic field strength when gravity is turned on (however the \myquote{pre-stirring} flux in the box will still affect the large-scale geometry of the magnetic field).

\subsubsection{Parameters Surveyed}\label{sec:cloud_params}

To describe our initial conditions we introduce several parameters, such as the \emph{3D sonic Mach number}
\begin{equation}
\mach^2\equiv\langle ||\vvector_\mathrm{turb}||^2/\cs^2\rangle,
\label{eq:mach}
\end{equation}
where $\cs$ is the gas sound speed and $\vvector_\mathrm{turb}$ is the turbulent velocity field, while $\langle...\rangle$ denotes mass-weighted averaging. It is also useful to introduce the \emph{turbulent virial parameter} $\alphaturb$, which measures the relative importance of turbulence to gravity, following the convention in the literature (e.g., \citealt{Bertoldi_McKee_1992, federrath_sim_2012}),
\begin{equation}
\alphaturb \equiv \frac{5 ||\vvector_\mathrm{turb}||^2 R_\mathrm{cloud}}{3 G M_\mathrm{0}} = \frac{5 \mach^2 \cs^2 R_\mathrm{cloud}}{3 G M_\mathrm{0}},
\label{eq:alphaturb_sphere}
\end{equation} 
where $R_\mathrm{cloud}$ and $M_\mathrm{0}$ are the cloud (spherical-equivalent) radius and total mass. The relative importance of the magnetic field is commonly described by the \emph{normalized magnetic flux} (or mass-to-flux ratio), which for a uniform magnetic field can expressed as:
\begin{equation}
\mu = c_1\sqrt{\frac{-E_{\rm grav}}{E_{\rm mag}}},
\label{eq:mu}
\end{equation}
where $E_{\rm grav}$ and $E_{\rm mag}$ are the gravitational and magnetic energy (assuming a uniform initial field), respectively, while the normalization constant $c_1\approx 0.4$. With this normalization $\mu=1$ corresponds to the critical point in the stability of a homogeneous sphere in a uniform magnetic field \citep{Mouschovias_Spitzer_1976_magnetic_collapse}.

Clouds have several characteristics mass scales defined by initial conditions. Such a scale is the \emph{Jeans mass}, representing the scale below which thermal pressure can prevent the gravitational collapse of a fluid element:
\begin{equation}
\MJeans \equiv \frac{4 \pi\cs^3}{3\sqrt{G^3\rho_0}},
\label{eq:MJeans}
\end{equation}
where $\rho_0$ is the density of the gas in the cloud. The initial turbulence also has a characteristic length scale: the sonic length, $\Lsonic$, on which the turbulent dispersion becomes supersonic. The corresponding mass scale is the \emph{sonic mass}:
\begin{equation}
\Msonic \equiv \frac{\cs^2 \Lsonic}{G}=\frac{\cs^2 R_\mathrm{cloud}}{\mach^2 G},
\label{eq:Msonic}
\end{equation}
where we used the supersonic linewidth-size relation ($\sigma^2(L)\propto L$). Another mass scale of an isothermal turbulent flow is the \textit{turbulent Bonnor-Ebert mass}, the maximum gas mass that can support itself against its own self-gravity plus external pressure in post-shock compressed gas with $\rho/\rho_0 \sim 1+\frac{1}{3}\mach^2$  \citep{padoan_nordlund_1997_imf}, which scales as
\begin{equation}
\MBE \sim 2 \MJeans\left(1+\frac{1}{3}\mach^2\right)^{-1/2}=\frac{8 \pi\cs^3}{3\sqrt{G^3\rho_0\left(1+\frac{1}{3}\mach^2\right)}}.
\label{eq:MBE}
\end{equation}
Note that many other parameters are also used in the literature that can be expressed in terms of the ones introduced in this subsection (see \S2 in \citetalias{Guszejnov_isoT_MHD} for how they relate to each other).

Table \ref{tab:IC} shows the target parameters for the runs we present in this paper. The input parameters are the cloud mass $M_0$, size $R_0$, turbulent virial parameter $\alphaturb$, normalized magnetic flux $\mu$ and initial temperature. Since our primary goal is to study the IMF in similar environments to the Milky Way, we set up our fiducial runs as clouds between $2000$ - $2\times 10^5\,\msun$ that lie along a mass-size relation similar to observed GMCs in the Milky Way (e.g. \citealt{larson_law}, specifically assuming $\Sigma\equiv M_\mathrm{0}/ \uppi R_\mathrm{cloud}^2 = 63 \msun\,\mathrm{pc}^{-2}$). These clouds are marginally bound ($\alphaturb=2$) and start out at $T=10\,\kelvin$, the temperature of the cold ISM. For the initial magnetization we assumed $-E_\mathrm{mag}/E_\mathrm{grav}=0.01$, which translates to $\mu=0.4$ (note that this choice has little effect on the results, see \S\ref{sec:sensitivity}). For the treatment of protostellar jets we use our fiducial parameters of $f_w=0.3$ and $f_K=0.3$ (see \S\ref{sec:jets}). Since observed clouds can deviate from the observed linewidth-size relation \citep{heyer_2009_larson}, we also simulate several clouds with different surface densities, turbulence and magnetic support. Note that for these studies we use clouds with a $2\times 10^4\,\msun$ initial mass (\textbf{M2e4}), due to the high computational cost of larger runs. Also, since most MW GMCs achieve a star formation efficiency ($\SFE=M_{\star}/M_{0}$) of 1\%-10\% over their lifetime (see \citealt{sf_big_problems} for a discussion, and note that some clouds have <1\%, see \citealt{Federrath_density_distrib}), we restrict our analysis to times when SFE is below 10\%.

\begin{table*}
    \setlength\tabcolsep{2.0pt} 
	\centering
		\begin{tabular}{|ccccccc|ccccccccc|c c|}
		 \multicolumn{1}{c}{}&
		 \multicolumn{7}{c}{\bf Input Parameters} &
		 \multicolumn{8}{c}{\bf Derived Parameters}&
		 \multicolumn{2}{c}{\textbf{Highest Resolution Run}} \\
		\hline
		\bf Cloud label & $M_0$ [$\msun$] & $R_{\mathrm{cloud}}$ [pc] & $L_{\mathrm{box}}$ [pc] & $\alphaturb$ & $\mu$ & $T_0$ [K] & $\mach$  &  $\alphath$ & $\alpha$ & $\mach_{\rm A} $ & $\beta$ & $\alphaB$ & $\frac{\MJeans}{M_0}$ & $\frac{\Msonic}{M_0}$ & $\frac{M_{\Phi}}{M_0}$ &  $M_0/\Delta m$ &  $\Delta x_\mathrm{J}$ [AU] \\
		\hline
		\multicolumn{18}{|c|}{\bf MW cloud analogues} \\
		\hline
		\bf M2e2 & $2\times 10^2$ & 1 &  & 2 & 4.2 & 10 
		& 5 & 0.02 & 2.04 & 10 &  7.8 & 0.02 & 
		$6\times 10^{-2}$ & $7 \times 10^{-3}$ & 0.1 &  $2\times10^{5}$ & 36 
		\\
		\hline
		\bf M2e3 & $2\times 10^3$ & 3 & 4.8 & 2 & 4.2 & 10  
		& 9.3 & 0.02 & 2.04 & 10 &  2.3 & 0.02 & 
		$1\times 10^{-2}$ & $6 \times 10^{-4}$ & 0.1 &  $2\times10^{7}$ & 3.6 \\
		\hline
		\bf M2e4 & $2\times 10^4$ & 10 & 16 & 2 & 4.2 & 10 
		& 16 & 0.008 & 2.03 & 10 &  0.78 & 0.02 & 
		$3\times 10^{-3}$ & $7 \times 10^{-5}$ & 0.1 &  $2\times10^{7}$ & 36 \\
		\hline
		\bf M2e5 & $2\times 10^5$ & 30 &  & 2 & 4.2 & 10 
		& 29 & 0.002 & 2.02 & 10 &  0.23 & 0.02 & 
		$5\times 10^{-4}$ & $7 \times 10^{-6}$ & 0.1 &  $2\times10^{8}$  & 36 \\
		\hline
		\multicolumn{18}{|c|}{\bf Parameter variation tests} \\
		\hline
		\bf M2e4\_a4 & $2\times 10^4$ & 10 &  & 4 & 4.2 & 10 
		& 22.6 & 0.008 & 4.03 & 10 &  0.78 & 0.02 & 
		$3\times 10^{-3}$ & $4 \times 10^{-5}$ & 0.1 &  $2\times10^{7}$  & 36 \\
		\hline
		\bf M2e4\_a1 & $2\times 10^4$ & 10 &  & 1 & 4.2 & 10 
		& 11.3 & 0.008 & 1.01 & 10 &  0.78 & 0.02 & 
		$3\times 10^{-3}$ & $1 \times 10^{-4}$ & 0.1 &  $2\times10^{7}$  & 36 \\
		\hline
		\bf M2e4\_a05 & $2\times 10^4$ & 10 &  & 0.5 & 4.2 & 10 
		& 8 & 0.008 & 0.51 & 10 &  0.78 & 0.02 & 
		$3\times 10^{-3}$ & $5 \times 10^{-5}$ & 0.1 &  $2\times10^{7}$  & 36 \\
		\hline
		\bf M2e4\_R3 & $2\times 10^4$ & 3 &  & 2 & 4.2 & 10 
		& 29 & 0.002 & 2.02 & 10 &  0.23 & 0.02 & 
		$5\times 10^{-4}$ & $7 \times 10^{-6}$ & 0.1 &  $2\times10^{7}$  & 36 \\
		\hline
		\bf M2e4\_R30 & $2\times 10^4$ & 30 &  & 2 & 4.2 & 10 
		& 9.3 & 0.02 & 2.04 & 10 &  2.3 & 0.02 & 
		$1\times 10^{-2}$ & $6 \times 10^{-4}$ & 0.1 &  $2\times10^{7}$  & 36 \\
		\hline
		\bf M2e4\_mu13 & $2\times 10^4$ & 10 &  & 2 & 13 & 10 
		& 16 & 0.008 & 2.01 & 31 &  7.8 & 0.002 & 
		$3\times 10^{-3}$ & $7 \times 10^{-5}$ & 0.04 &  $2\times10^{7}$  & 36 \\
		\hline
		\bf M2e4\_mu1.3 & $2\times 10^4$ & 10 &  & 2 & 1.3 & 10 
		& 16 & 0.008 & 2.21 & 3.1 &  0.078 & 0.2 & 
		$3\times 10^{-3}$ & $7 \times 10^{-5}$ & 0.4 &  $2\times10^{7}$  & 36 \\
		\hline
		\bf M2e4\_T30 & $2\times 10^4$ & 10 &  & 2 & 4.2 & 30 
		& 9.3 & 0.024 & 2.04 & 10 &  2.3 & 0.02 & 
		$1\times 10^{-2}$ & $6 \times 10^{-4}$ & 0.1 &  $2\times10^{7}$ & 36  \\
		\hline
		\bf M2e4\_T60 & $2\times 10^4$ & 10 &  & 2 & 4.2 & 60 
		& 6.6 & 0.048 & 2.07 & 10 &  4.6 & 0.02 & 
		$5\times 10^{-2}$ & $2 \times 10^{-3}$ & 0.1 &  $2\times10^{7}$ & 36 \\
		\hline
		\end{tabular}
        \vspace{-0.1cm}
 \caption{Initial conditions of clouds used in our runs, with $M_0$, $R_{\mathrm{cloud}}$, $\alphaturb$, $\mu$ and $T_0$ being the initial cloud mass, size, virial parameter, mass to magnetic flux ratio and temperature respectively (note that in all our runs $T_\mathrm{floor}=T_0$, see \S\ref{sec:cooling}). We also report the initial 3D sonic Mach number $\mach$, thermal virial parameter $\alphath$, total virial parameter $\alpha$, Alfv\'{e}n Mach number $\mach_{\rm A}$, plasma $\beta$, magnetic virial parameter $\alphaB$, as well as the relative Jeans, sonic and magnetic mass scales (see \S2 in \citetalias{Guszejnov_isoT_MHD} for definitions). Note that the parameters in this table apply to both \textit{Box} and \textit{Sphere} runs as they are set up to have identical initial global parameters, with  $L_{\mathrm{box}}$ being the box size for \textit{Box} runs and $R_{\mathrm{cloud}}$ being the cloud radius for \textit{Sphere} runs. Note that \textit{Box} runs have slightly different initial parameters (e.g., Mach number, virial parameter) due to the non-exact scaling of the driving, so the values shown here are the target values. Many of the above clouds have been run at different mass resolutions as part of the resolution study in \citetalias{grudic_starforge_methods}, in the table we note for each the highest resolution that was run ($\Delta m$ mass resolution and $\Delta x_\mathrm{J}$ minimum resolved Jeans length, see \S 2 in \citetalias{grudic_starforge_methods} for details).}
 \label{tab:IC}\vspace{-0.5cm}
\end{table*}

  \section{Results}\label{sec:results}
We carried out a suite of simulations using the initial conditions from Table \ref{tab:IC} and the different physics combinations from Table \ref{tab:physics_ladder}. 
  
\begin{figure*}
\begin {center}
\includegraphics[width=\linewidth]{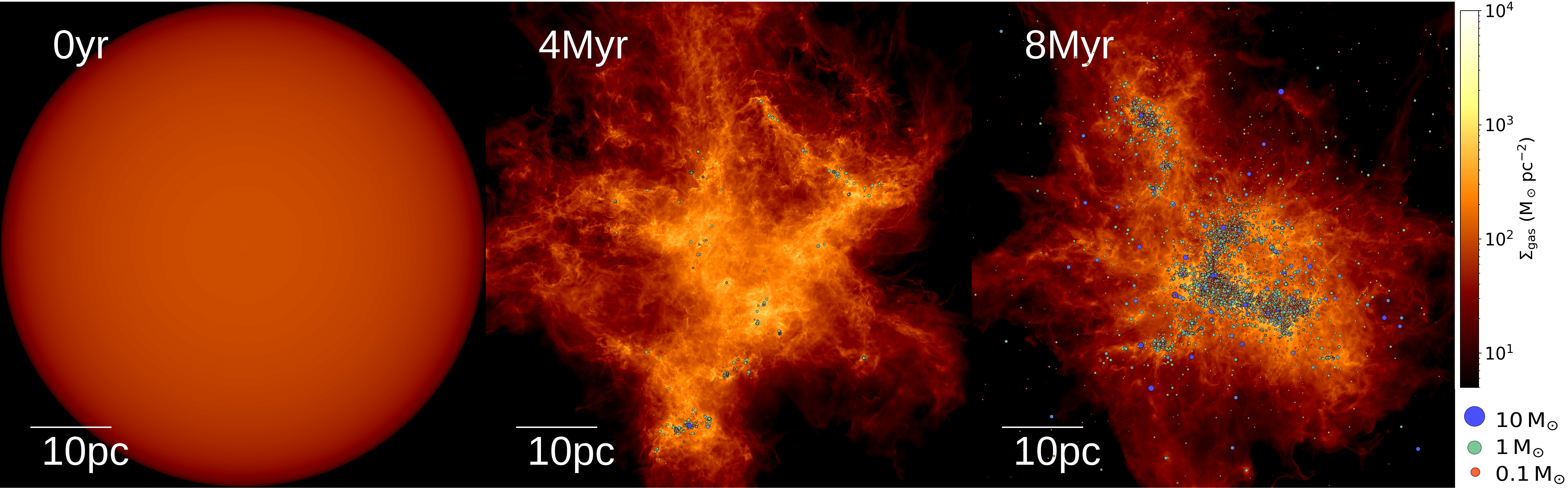}
\caption{Surface density maps for \textbf{M2e5\_C\_M\_J} with $M_0/\Delta m=2\times 10^8$ initial gas cells (see Tables \ref{tab:physics_ladder} and \ref{tab:IC}) at different times. The color scale is logarithmic and the circles represent sink particles (stars) that form in high-density regions where fragmentation can no longer be resolved, their size increasing with mass as well as their color changing from red ($M\sim0.1\,\msun$) to blue ($M\sim10\,\msun$). This simulation resolves a dynamic range from $\sim\!\mathrm{50\,pc}$ down $\sim\!\mathrm{30\,AU}$.}
\label{fig:M2e5_C_M_J_series}
\vspace{-0.5cm}
\end {center}
\end{figure*}

\begin{figure*}
\begin {center}
\includegraphics[width=0.95\linewidth]{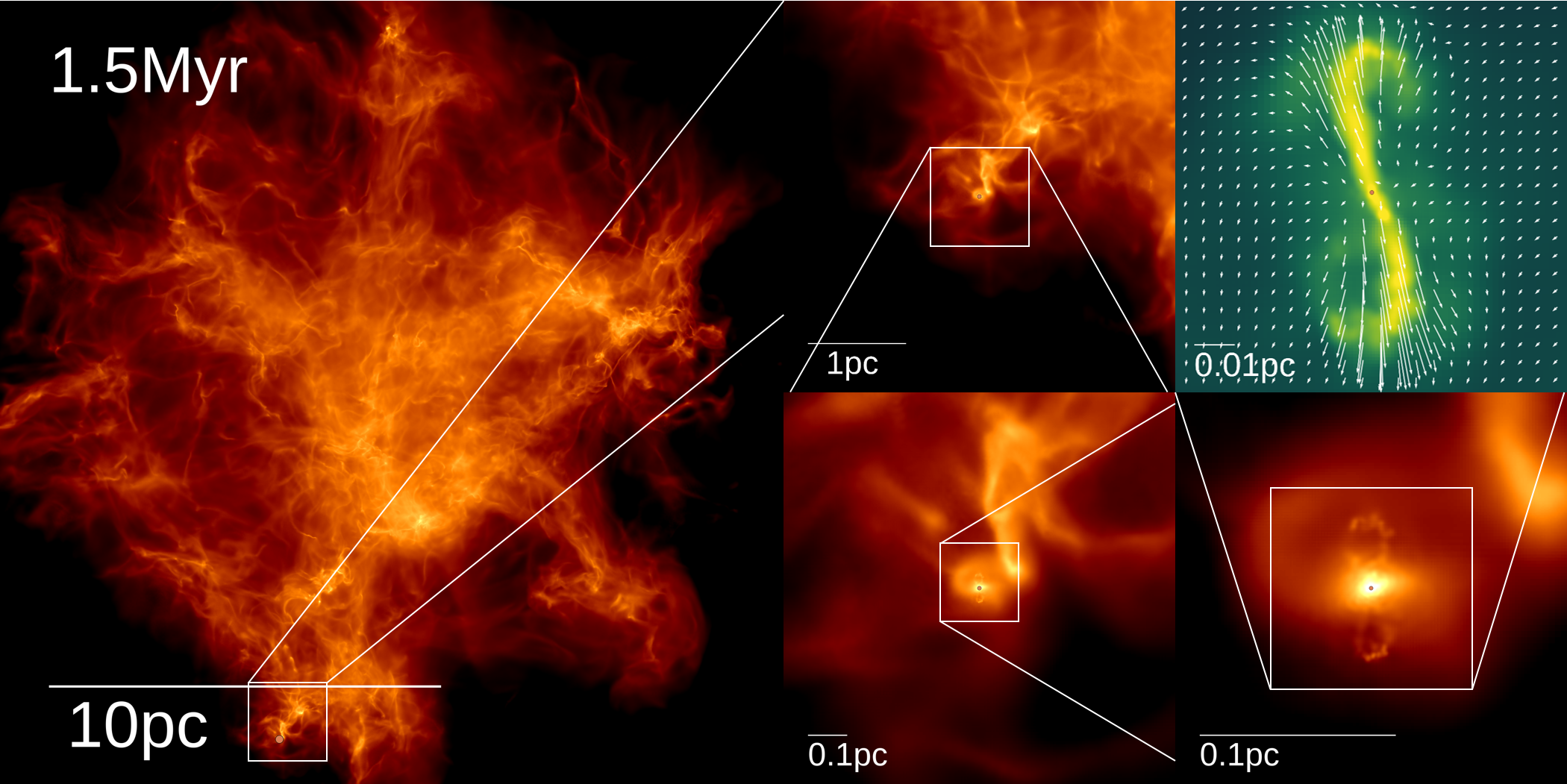}
\caption{Zoomed-in surface density maps for a medium-sized cloud with protostellar jets enabled (\textbf{M2e4\_C\_M\_J}). Symbols and color maps are similar to Figure \ref{fig:M2e5_C_M_J_series}. The final image (top, right) shows the kinetic energy weighted surface density ($\mathrm{weight}=m(1+[v/v_0]^2)$, where $v_0=1\,\mathrm{km/s}$), as well as the local velocity field (white arrows), whose length scales with velocity, to highlight the the jet (which has high velocity but low density, making it challenging to see in surface density maps).}
\label{fig:M2e4_C_M_J_zoom}
\vspace{-0.5cm}
\end {center}
\end{figure*}

All simulations develop filaments, clumps, and cores, and begin global collapse (see Figure \ref{fig:M2e5_C_M_J_series} for the case with protostellar jets). In the runs with protostellar jets, once star formation begins jets disrupt the flow around newly formed stars (see Figure \ref{fig:M2e4_C_M_J_zoom}), reducing their accretion rates and allowing new stars to form. In the following subsections we investigate different aspects of star formation with different physics enabled.

  \subsection{Star formation history}
  
\begin{figure*}
\begin {center}
\includegraphics[width=0.33\linewidth]{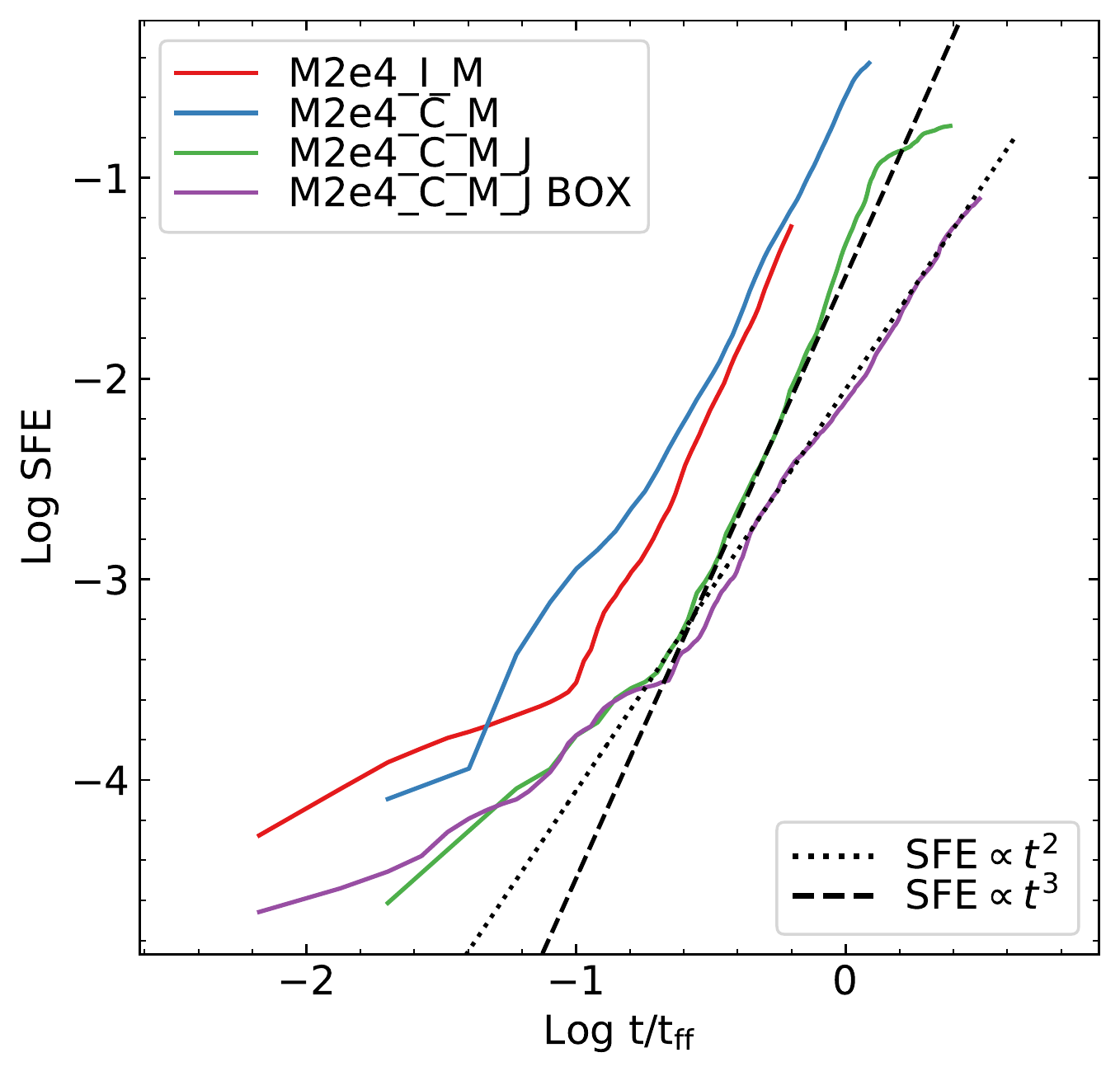}
\includegraphics[width=0.33\linewidth]{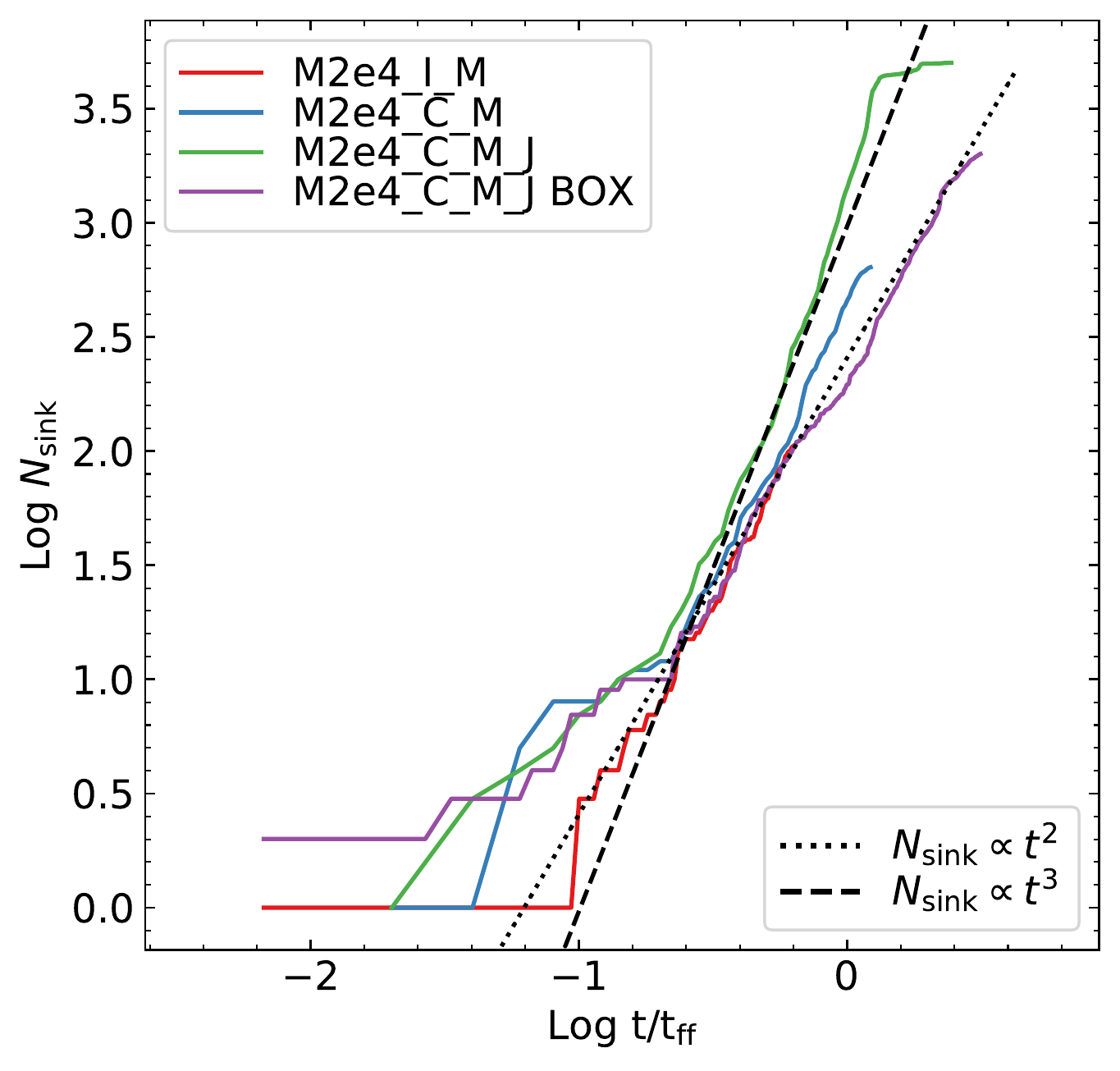}
\includegraphics[width=0.33\linewidth]{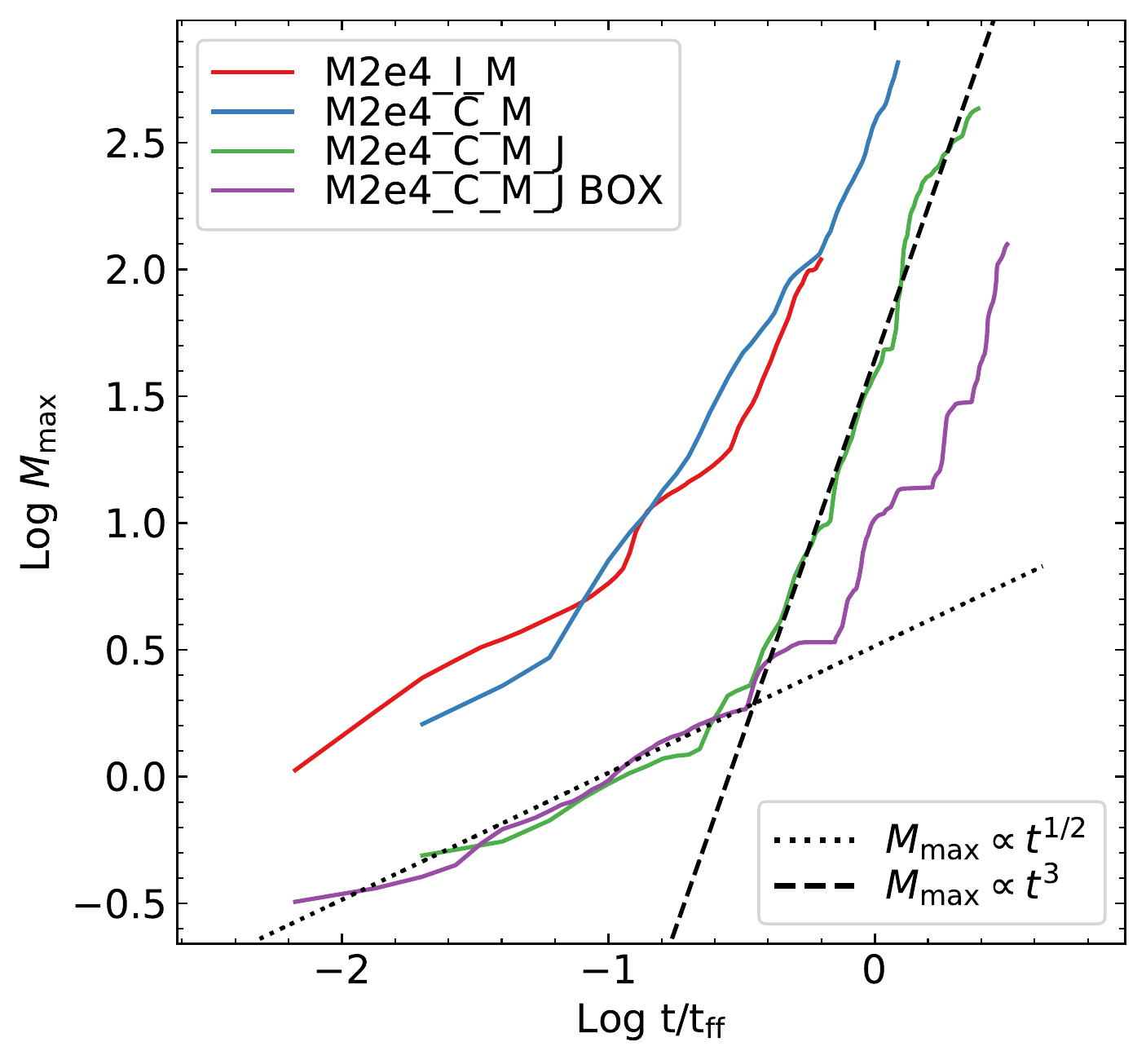}
\vspace{-0.4cm}
\caption{\emph{Left:} Evolution of the star formation efficiency ($\SFE(t)=\sum{\Msink(t)}/M_{0}$) as function of time for a subset of runs with \textbf{M2e4} runs but different physics (see Tables \ref{tab:physics_ladder}-\ref{tab:IC}). A run with \textit{Box} initial conditions is also included for comparison. Note that $t=0$ is set to the start of star formation and that the right panel ($\alphaturb$) uses a linear $x$-axis for time $t$, as opposed to a log axis, with negative values $t < 0$ representing time before the first sink forms. The SFE rises as a broken power-law of time and reaches about 10\% in 1-2 freefall times ($t_{\mathrm{ff}}=\sqrt{\frac{3 \pi}{32 G \rho_0}}$, which is about 3.5 Myr in these runs). Note that the overall shape of the SFE is unchanged by enabling feedback physics, however, jets shift the curve, effectively delaying the star formation process by a factor of 2 in $t/t_\mathrm{ff}$. Meanwhile, the slope of the curve is sensitive to the type of initial condition used, we find $\SFE\propto t^3$ for \textit{Sphere} ICs and $\SFE\propto t^2$ for \textit{Box} ICs. \emph{Middle:} Number of sink particles in the simulations as a function of time. Non-isothermal thermodynamics suppresses the formation of low-mass sink particles while the additional turbulence on small scales enhances it. Switching to Box ICs leads to a shallower exponent, similar to the SFE case above. \emph{Right:} Maximum sink mass in the same simulations as a function of time. In all cases the maximum mass asymptotes to $\propto t^3$ once massive stars have formed.}
\label{fig:sf_history}
\vspace{-0.5cm}
\end {center}
\end{figure*}

 Figure \ref{fig:sf_history} shows the star formation history of several clouds with identical initial conditions (\textbf{M2e4}) but with different physics modules and turbulent driving (see \textit{Sphere} vs \textit{Box} ICs in \S\ref{sec:IC}). For the \textit{Sphere} runs we find that the star formation efficiency (SFE) in all cases follows a similar broken power-law, which starts linearly (note that this \myquote{early time} slope is potentially sensitive to the definition of the time zero-point) and transitions to $\SFE(t)\propto t^3$ at later times, similar to the findings of \citetalias{Guszejnov_isoT_MHD} for the isothermal case and other simulations without turbulent driving from the literature \citep[e.g.][]{myers2014_sim}. Note that while protostellar jets do reduce the star formation rate, their net effect is only a shift in the curve, delaying the onset of the cubic regime from roughly 10\% of the freefall time to about 20\%. The results for the Box runs are qualitatively similar, but their star formation rates are slower: they scale as $\SFE \propto t^2$, similar to previous results with driven turbulent boxes \citep[e.g.,][]{federrath_sim_2012,murray_2015_turb_sim,murray_2018_jets}. 
 
 Figure \ref{fig:sf_history} also shows the number of sink particles, $\Nsink$, over time. For most runs $\Nsink$ follows a similar trend to the SFE, which produces a roughly time-invariant mean sink mass (see Figure \ref{fig:mass_scale_evol}). Note that even though switching to driven turbulence (\textit{Box} IC) reduces the star formation rate, the mean sink mass remains roughly similar ($\SFE\propto \Nsink$). This implies that the sink mass distribution (IMF) in the simulation is determined by local physics (e.g., jets) instead of large scale boundary conditions (i.e., turbulent driving spectrum).
 
 We find that the maximum sink mass increases over time, starting as a $M_\mathrm{max}\propto \sqrt{t}$ power-law, which steepens to $M_\mathrm{max}\propto t^3$ once massive sinks (stars) form, as they undergo runaway accretion. This plays out qualitatively similarly in all runs here, regardless of physics or turbulent driving. The main effect of protostellar jets is that they reduce the maximum sink mass by about an order of magnitude at fixed total sink mass in the simulation.
  
\begin{figure*}
\begin {center}
\includegraphics[width=0.33\linewidth]{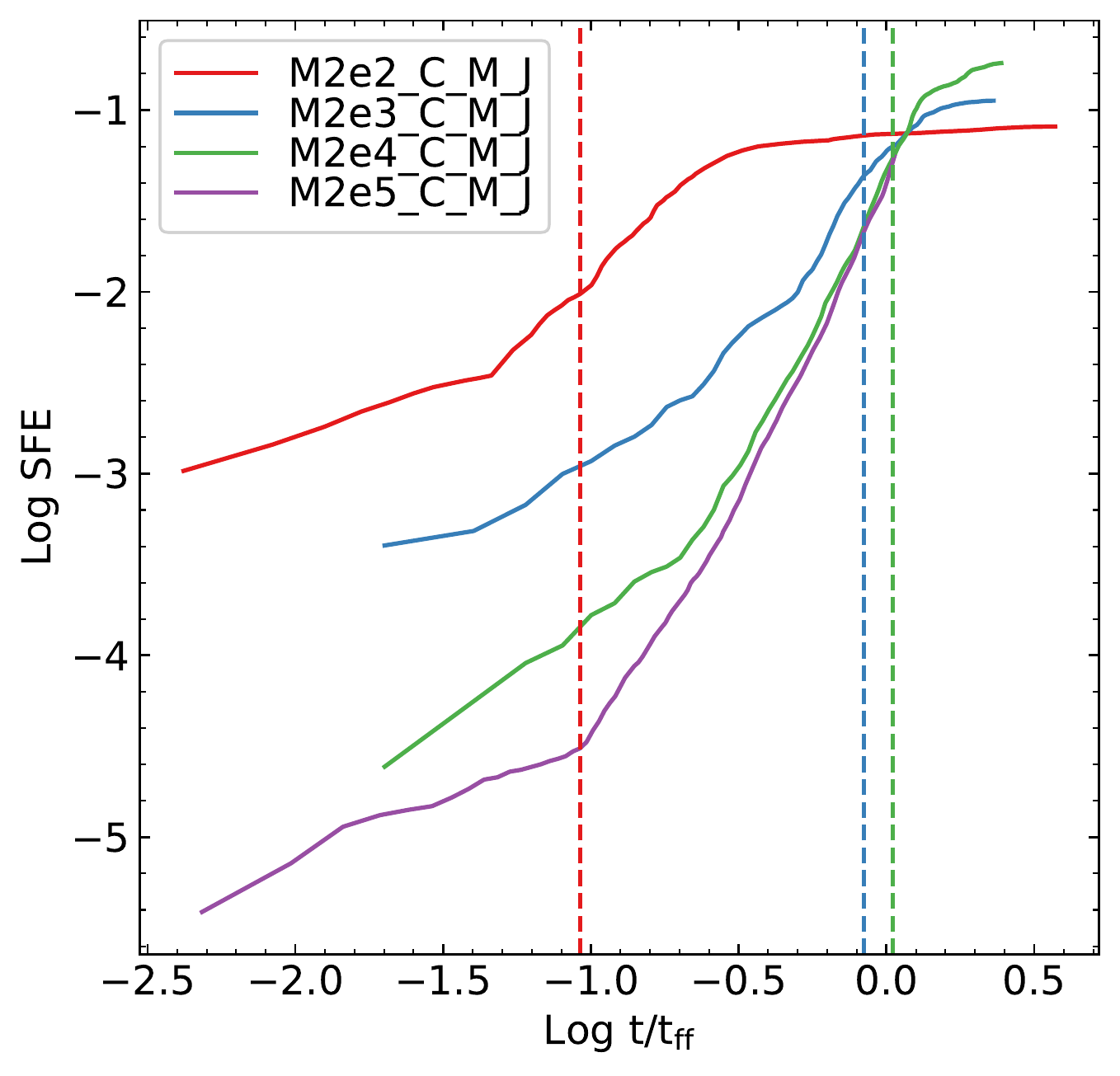}
\includegraphics[width=0.33\linewidth]{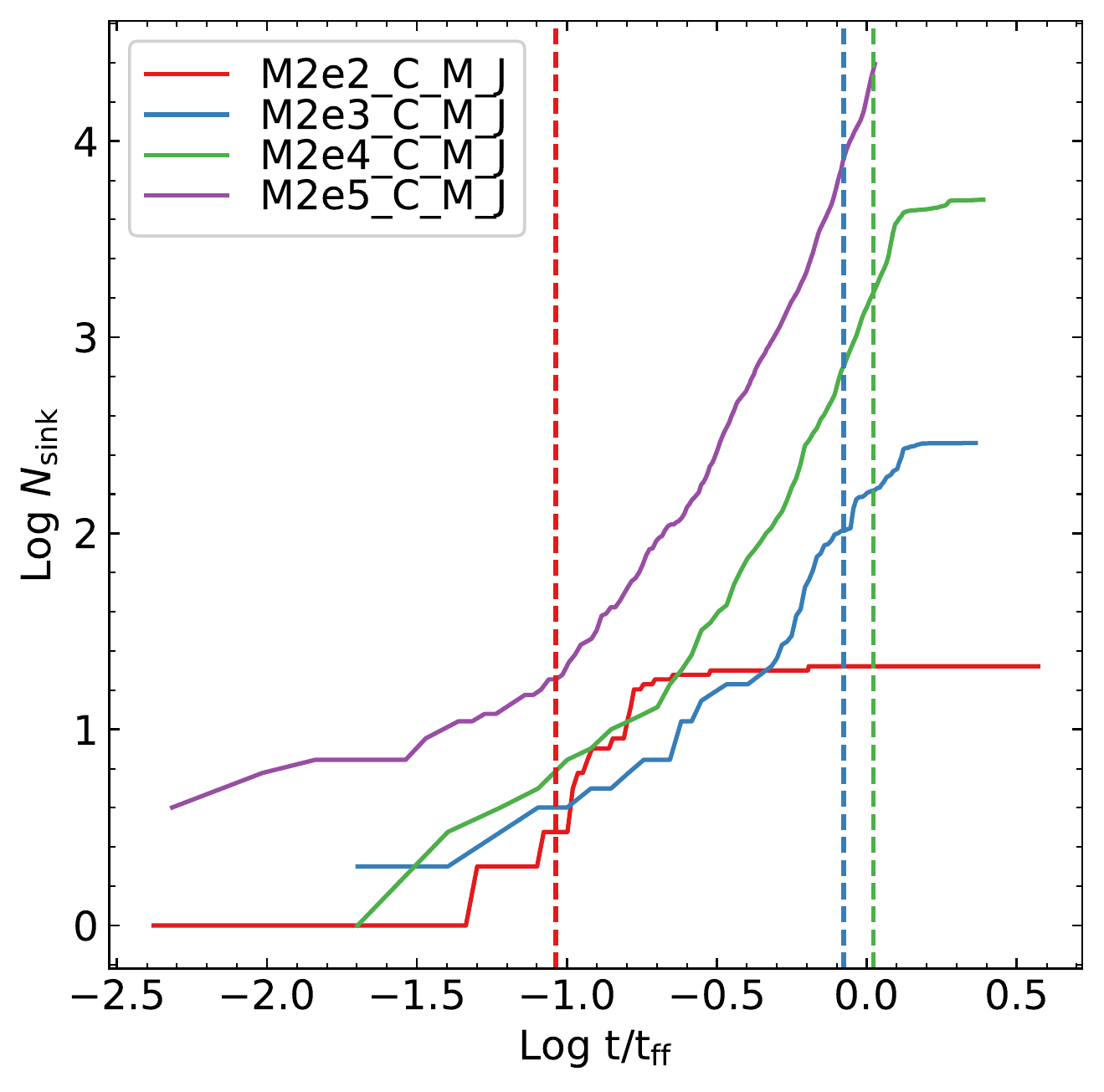}
\includegraphics[width=0.33\linewidth]{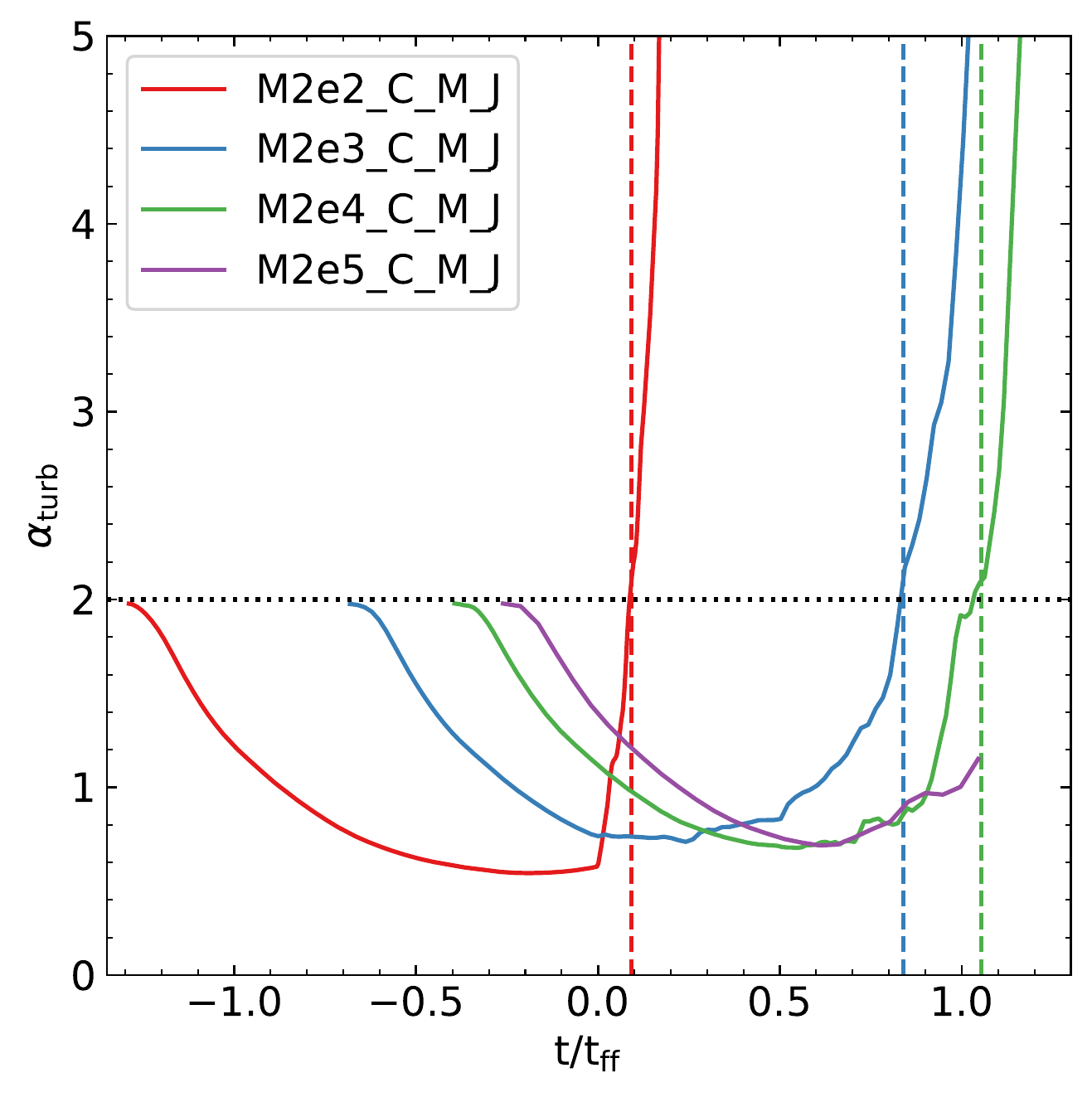}
\vspace{-0.4cm}
\caption{The evolution of the star formation efficiency (left), number of sink particles $\Nsink$ (middle) and turbulent virial parameter $\alphaturb$ ($\alphaturb=2$ being equivalent marginal gravitational boundedness) as function of time for a subset of runs with protostellar jets enabled (\textbf{C\_M\_J}) that have clouds with increasing initial masses and a constant surface density ($\Sigma\approx 60\,\msun/\pc^2$) similar to MW GMCs (\textbf{M2e2}-\textbf{M2e5}, see Table \ref{tab:IC}). Note that here $t=0$ is set to the start of star formation. We find that after reaching a sufficiently high star formation efficiency, protostellar jets are able to unbind low-mass clouds the time of which is marked with a vertical dashed line. After this jets are able to quench star formation, almost completely stopping the formation of new sink particles.}
\label{fig:sf_regulation_jets}
\vspace{-0.5cm}
\end {center}
\end{figure*}

Figure \ref{fig:sf_regulation_jets} shows that the inclusion of protostellar jets (\textbf{C\_M\_J}) can lead to the disruption of the parent cloud and subsequently preventing the formation of new stars. In more massive clouds ($>10^4\msun$, similar to MW GMCs), protostellar jets show no sign of arresting star formation before the SFE exceeds $\sim 10\%$. Note that SFE is challenging to measure observationally, but observed clouds in the range of sizes and masses we have simulated are generally believed to have a typical SFE of only a few \% \citep{Lee_eve_2016_GMC_SFE,Vutisalchavakul_2016_MW_GMC_SFR,grudic_2018_mwg_gmc,kruijssen_2019_gmcs, chevance:2020.gmcs}. 

\subsection{Sink mass distribution (IMF)}
  Sink particles represent stars (or systems with separations below the resolution limit) in our simulations, so we use their mass spectrum as an analogue of the IMF. Since it is possible for the sink mass spectrum (IMF) not to converge numerically at the lowest masses, while still converging on shape at higher masses or providing characteristic mass scales, we investigate the effects of different physics on both the various characteristic mass scales and the shape of the sink mass spectrum.

  \subsubsection{Characteristic mass of stars}
\begin{figure*}
\begin {center}
\includegraphics[width=0.33\linewidth]{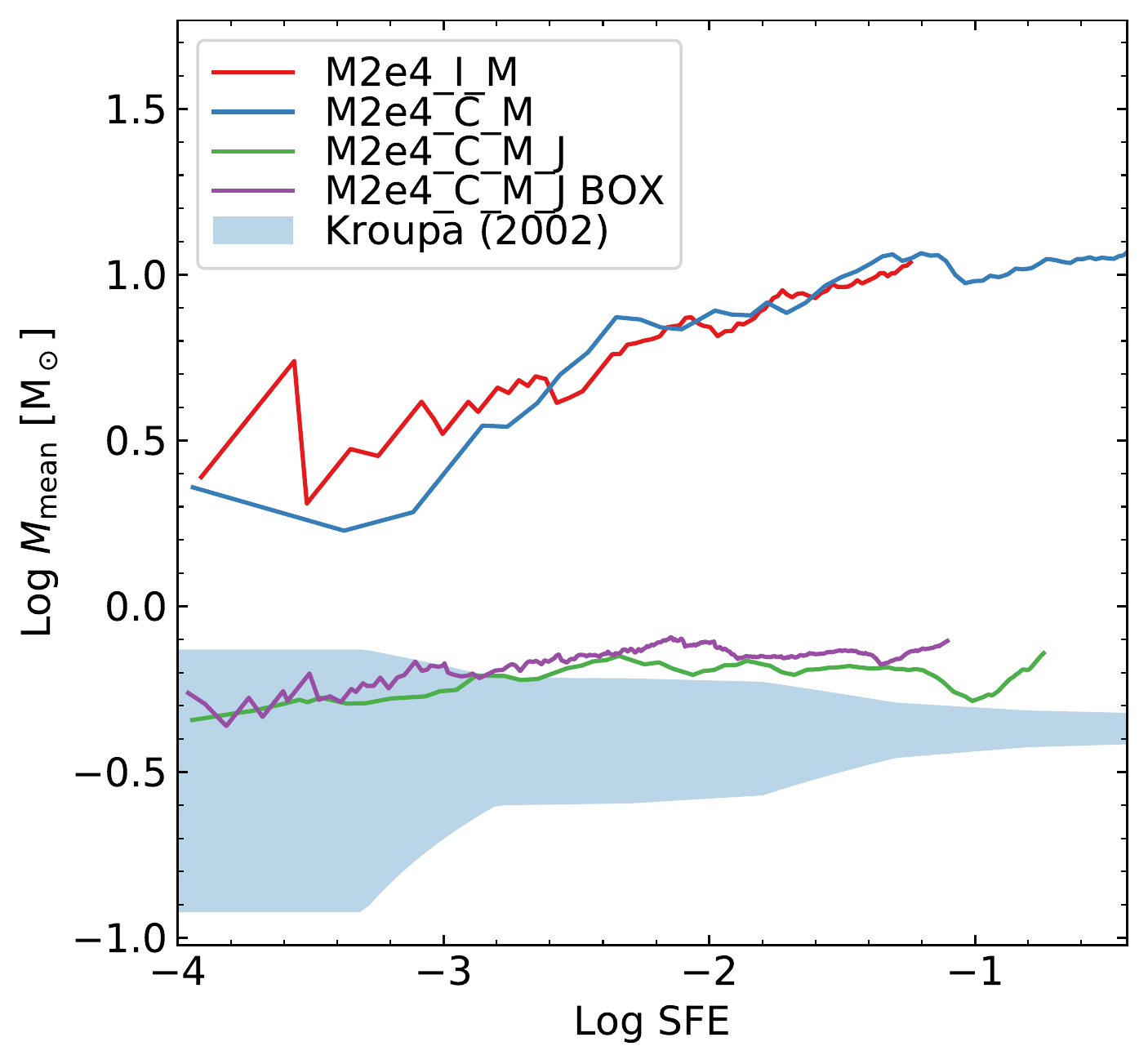}
\includegraphics[width=0.33\linewidth]{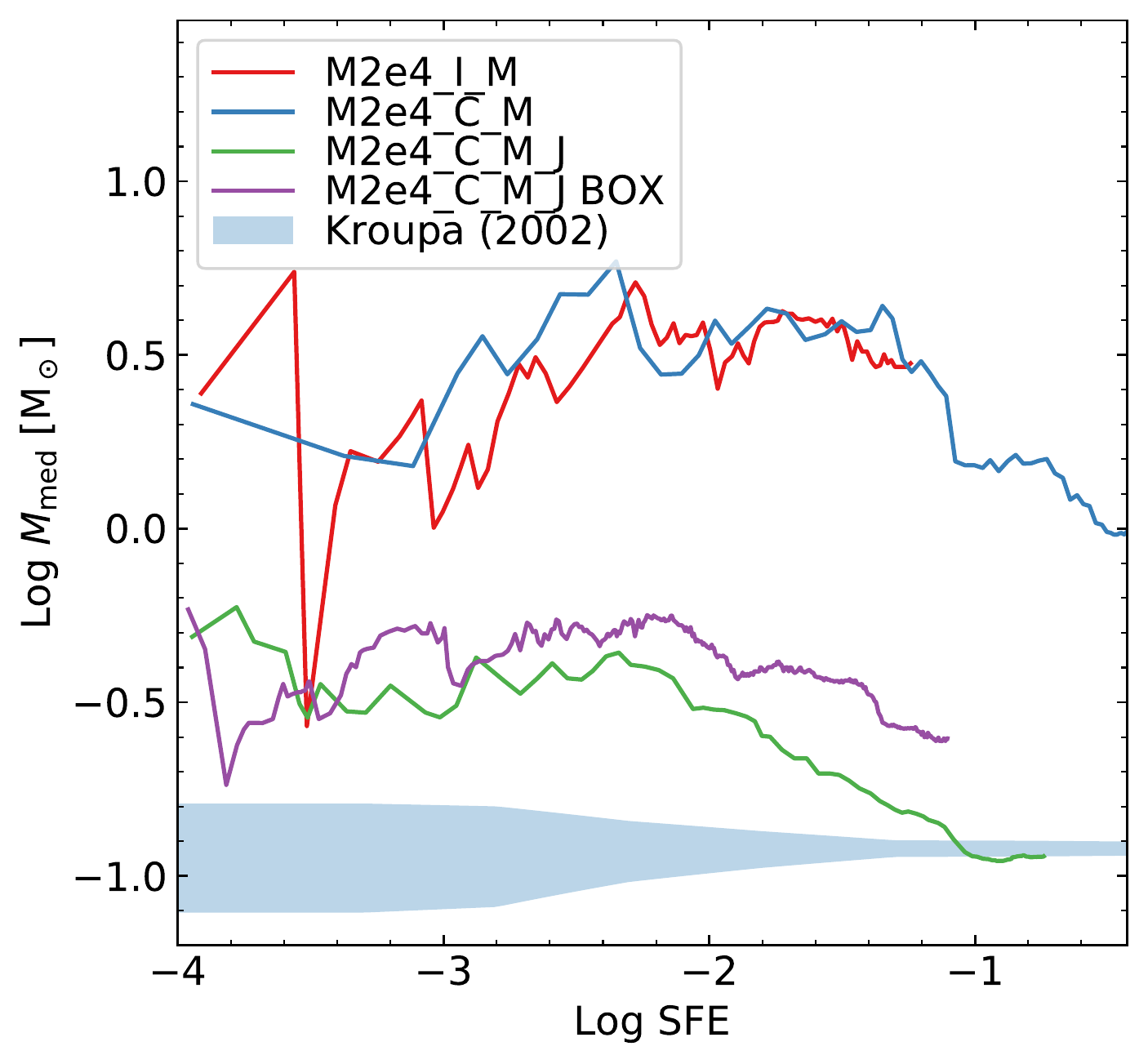}
\includegraphics[width=0.33\linewidth]{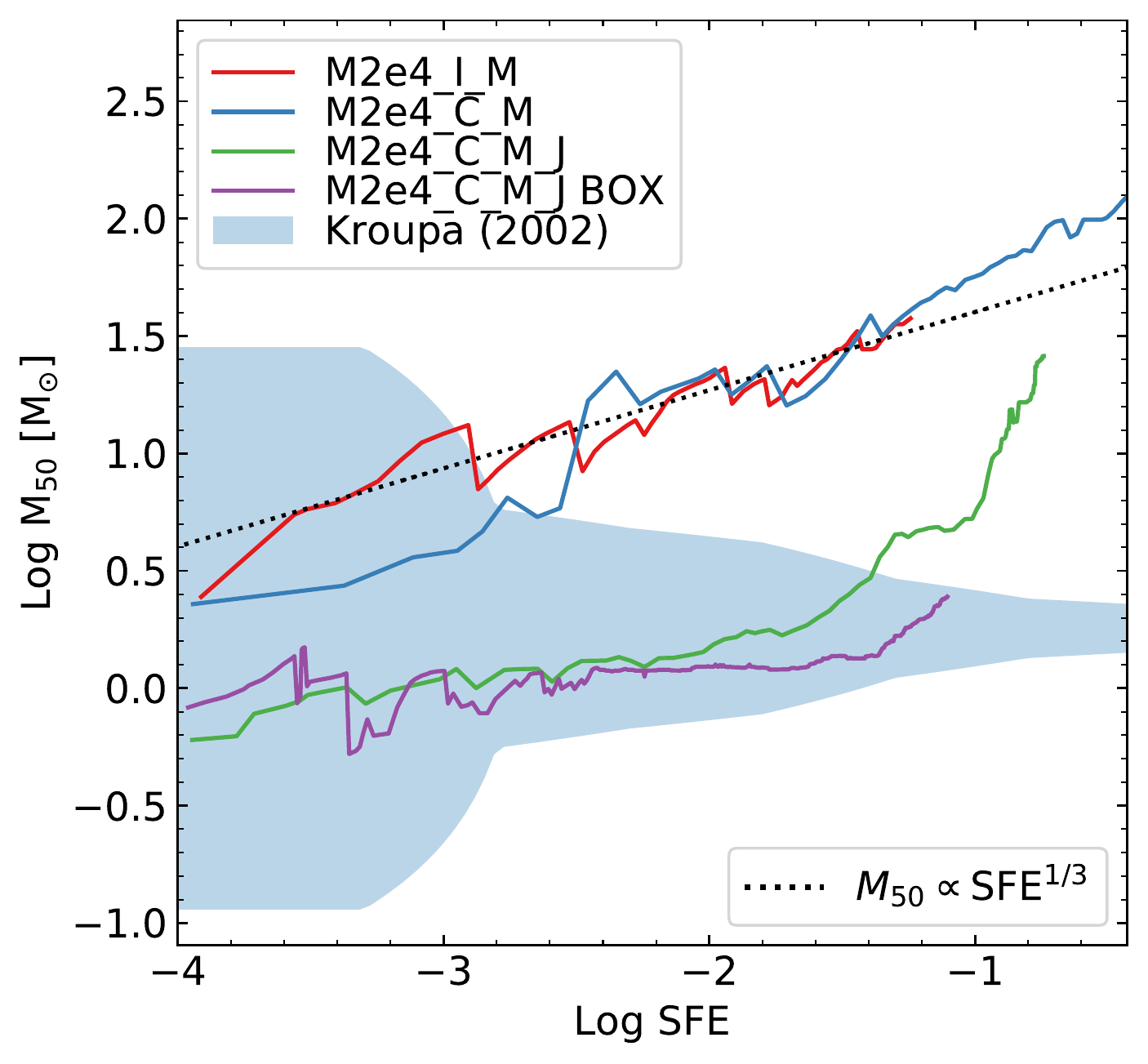}
\vspace{-0.4cm}
\caption{The evolution of the number-weighted mean ($\Mmean = \sum{\Msink}/\Nsink$, left), number-weighted median (defined such that $\Nsink(M>\Mmedian)=\Nsink/2$, center) and mass-weighted median ($\Mmassmedian$, the mass scale above which half the total sink mass resides, right) sink mass as a function of star formation efficiency for the runs shown in Figure \ref{fig:sf_history}. We also show with a shaded region the 95\% confidence interval for these values if one sampled the \citet{kroupa_imf} IMF at the current SFE value in the cloud. Protostellar jets reduce sink masses and bring all three mass scales closer to those of the observed IMF, however, $\Mmassmedian$ increases with time, diverging from observations at higher SFE values.}
\label{fig:mass_scale_evol}
\vspace{-0.5cm}
\end {center}
\end{figure*}

  A common issue in numerical simulations is that the low-mass end of the sink mass spectrum is sensitive to numerical resolution and simulations often have a large number of very low-mass objects near their resolution. While in most cases these objects represent a vanishingly small fraction of the total sink mass (see \citetalias{Guszejnov_isoT_MHD} for an example and \citealt{guszejnov_isothermal_collapse} for a counterexample), their large number skews the mean and median sink masses. Adopting the mass-weighted median mass of sinks $\Mmassmedian$ as the characteristic mass scale mitigates this effect (see \citealt{krumholz_2012_orion_sims} and \citetalias{Guszejnov_isoT_MHD}), but this choice makes the mass scale overly sensitive to the most massive sinks that can undergo runaway accretion (see Figure \ref{fig:mass_scale_evol}).
  
  Figure \ref{fig:mass_scale_evol} shows the evolution of the mean and median sink masses along with that of $\Mmassmedian$ as a function of SFE. For runs without jets we find that the mean sink mass and $\Mmassmedian$ both increase with time due to the runaway accretion of the massive sinks. Note the introduction of non-isothermal physics has little effect on the three mass scales and without jets they are all significantly larger than those observed in the MW. The introduction of jets allows low-mass stars to form again, such that the mean mass is roughly time invariant while all three mass scales are near their observed values. But as star formation progresses (SFE>1\%), we find that all simulations show an increasing trend in $\Mmassmedian$ due to the runaway accretion of massive sinks, similar to the $\Mmassmedian\propto \SFE^{1/3}$ scaling found in the isothermal case in \citetalias{Guszejnov_isoT_MHD}. Switching to \textit{Box} ICs has little effect on the evolution of $\Mmassmedian$ or the mean sink mass, except for a delay in the runaway accretion of massive stars. For the median mass, however, turbulent driving appears to suppress the formation of very low mass stars.

At our fiducial resolution both the mean sink mass and $\Mmassmedian$ are insensitive to numerical resolution (see \citetalias{grudic_starforge_methods}). We also find that the mean sink mass exhibits a nearly time invariant trend between 1\%-10\% SFE in most simulations (see Figure \ref{fig:mass_scale_evol} and Figures \ref{fig:imf_sensitivity_1}-\ref{fig:imf_sensitivity_2}), while $\Mmassmedian$ increases with time in nearly all cases, so we adopt it as a proxy for the characteristic scale of the IMF for the remainder of the paper.
 
\subsubsection{The IMF}
\begin{figure}
\begin {center}
\includegraphics[width=0.95\linewidth]{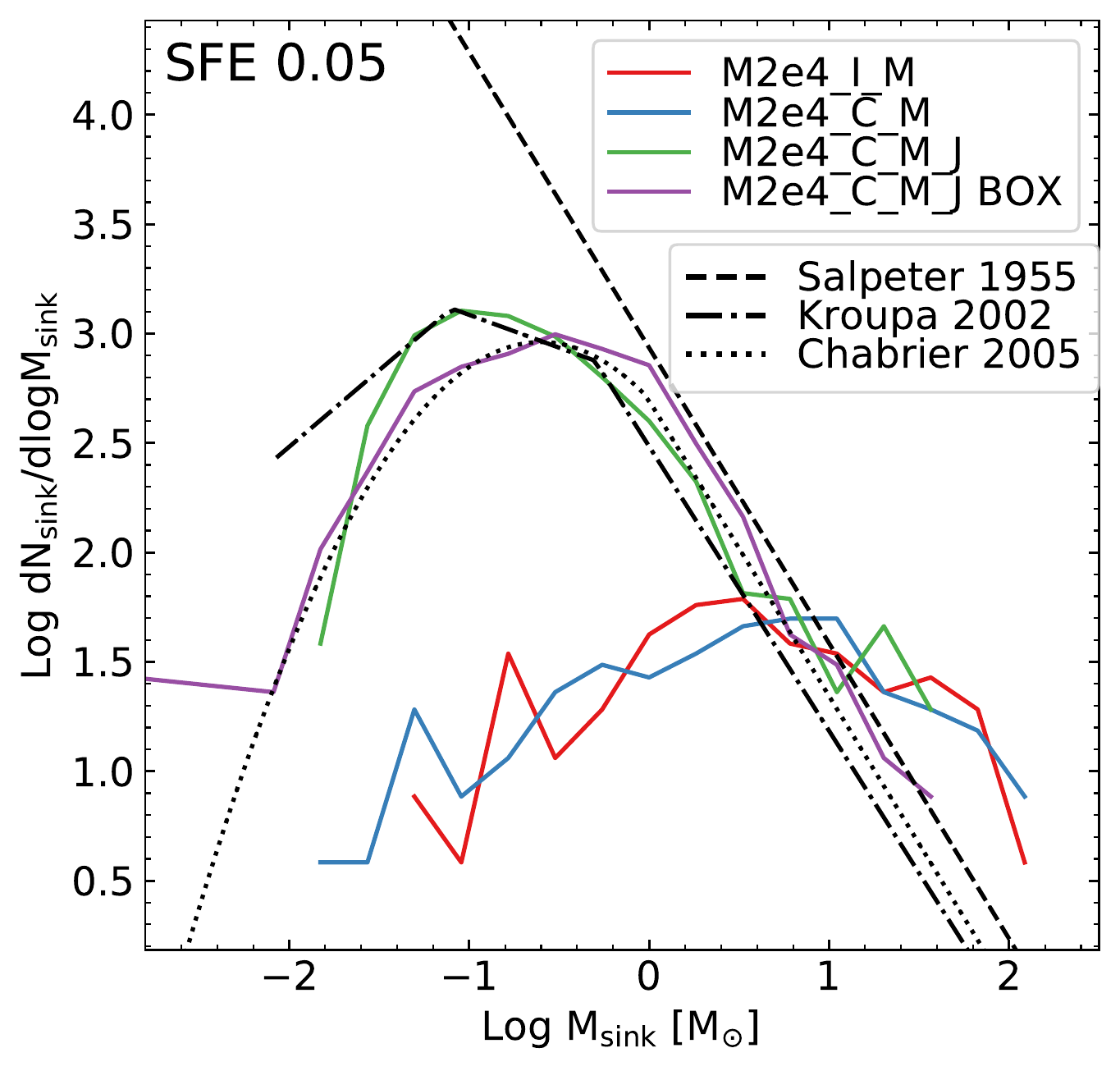}
\vspace{-0.4cm}
\caption{Distribution of sink particle masses measured in each simulation at 5\% star formation efficiency ($\SFE=\sum M_{\mathrm{sink}}/M_{0}$) for the runs shown in Figure \ref{fig:sf_history}. We also show the \citet{salpeter_slope}, \citet{kroupa_imf} and \citet{chabrier_imf} fitting functions for the IMF. Again, jets greatly improve the agreement with the IMF for our chosen jet parameters (see Appendix \ref{sec:scaling_dependencies} for results with different values).}
\label{fig:imf_compare}
\vspace{-0.5cm}
\end {center}
\end{figure}

While the various characteristic masses provide some information on the sink mass distribution, a holistic view of the IMF is necessary to understand the effects of each physical process. Figure \ref{fig:imf_compare} shows the mass distribution of sink particles at 5\% star formation efficiency (SFE), which we will use as a proxy for the IMF. We find that the addition of non-isothermal physics alone has little effect on the IMF \footnote{Note that since \citetalias{Guszejnov_isoT_MHD} improvements on the sink formation and accretion algorithms (see \citetalias{grudic_starforge_methods}) have reduced the population of very low-mass sinks in isothermal MHD runs compared to \citetalias{Guszejnov_isoT_MHD}, suggesting that a sub-population of these was unphysical in origin (strengthening our conclusions about the necessity of additional physics to prevent an overly top-heavy IMF).} leaving the IMF top-heavy (see \citetalias{Guszejnov_isoT_MHD}). We find that the inclusion of protostellar jets dramatically changes the distribution, shifting the turnover to mass scales comparable to that observed in the MW. Switching to Box ICs does not qualitatively change the IMF apart from slightly suppressing the formation of very low mass objects. Driven turbulence also delays the runaway accretion of massive stars; that is why the IMF is not yet top-heavy at 5\% SFE for the Box run in Figure \ref{fig:imf_compare}.
  
\subsubsection{Role of jet momentum loading}\label{sec:jet_mom_loading}

\begin{figure*}
\begin {center}
\includegraphics[width=0.95\linewidth]{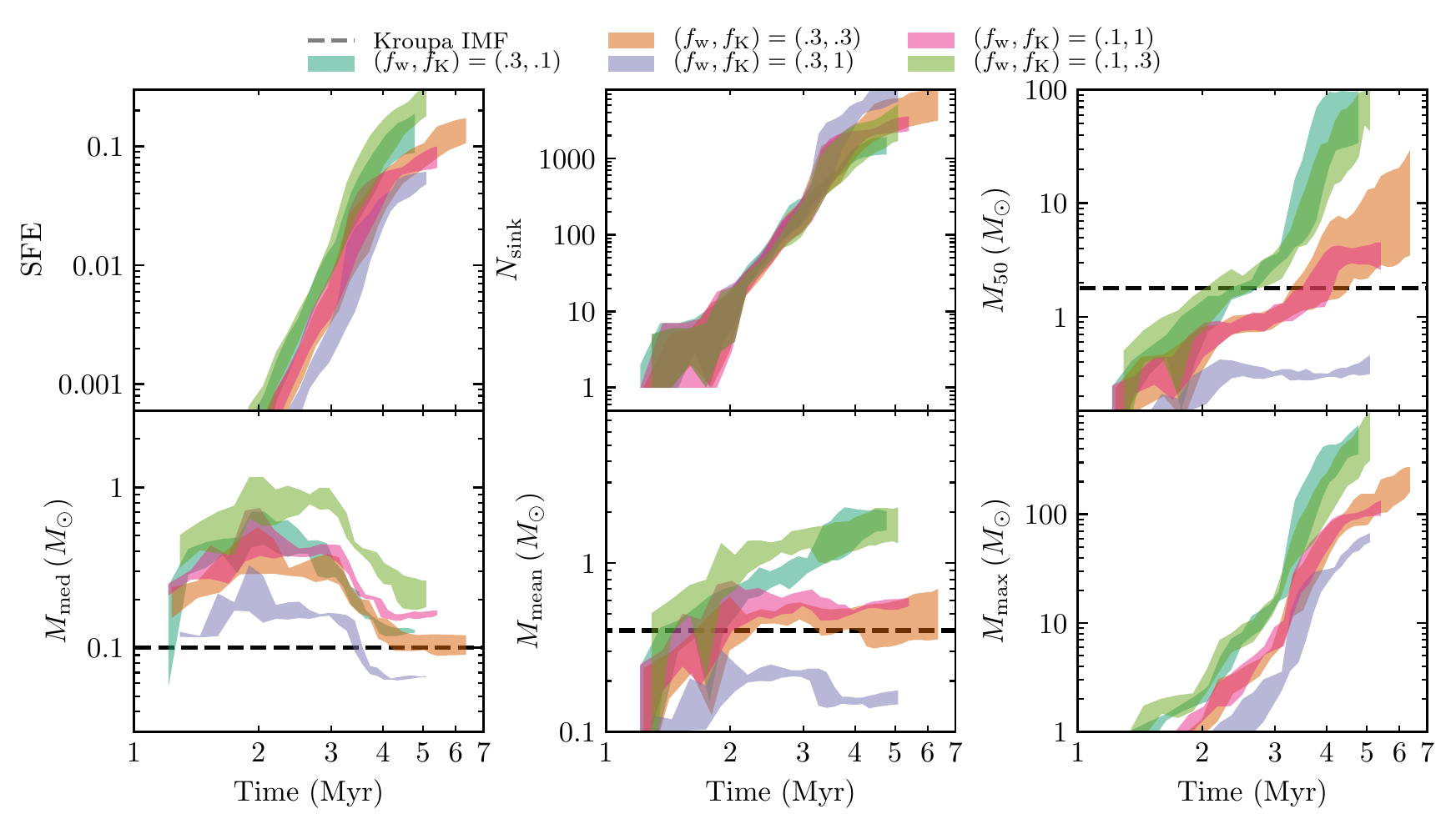}
\vspace{-0.4cm}
\caption{Star formation efficiency SFE (top left), number of sink particles $\Nsink$ (top center), mass-weighted median sink mass $\Mmassmedian$ (top right), median sink mass $\Mmedian$ (bottom left), mean sink mass $\Mmean$ (bottom center) and the maximum sink mass $M_\mathrm{max}$ (bottom right) in \textbf{M2e4\_C\_M\_J} runs with different jet mass loading ($f_w=\Mdot_\mathrm{jet}/\Mdot_\mathrm{acc}$) and velocity scaling ($f_K=v_\mathrm{jet}/\sqrt{G M_\ast/R_\ast}$) parameters for protostellar jets (see \S\ref{sec:jets}). The shaded region represents the minimum and maximum values at a given time over 3 different initial turbulence realizations. Dashed lines indicate statistics of the \citet{kroupa_imf} IMF. The IMF shape is sensitive to jet model parameters, particularly to $\Gamma\propto f_w f_K$, which determines the momentum loading of jets.}
\label{fig:jet_param_test}
\vspace{-0.5cm}
\end {center}
\end{figure*}

Since jets have a dramatic effect on the IMF (see Figure \ref{fig:imf_compare}), we examine how our results depend on the $f_w$ and $f_K$ jet parameters (see \S\ref{sec:jets}). Figure \ref{fig:jet_param_test} shows the results of varying these parameters for an \textbf{M2e4\_C\_M\_J} run. We find that the evolution of the cloud and the sink mass spectra depend primarily on $\Gamma=f_w f_K/(1-f_w)$, which determines the momentum loading of the jets (e.g. the results obtained for $f_{\rm w}=0.1$ and $f_{\rm K}=1$ are very similar to the results for our fiducial $f_{\rm w}=0.3$ and $f_{\rm K}=0.3$). 
Furthermore, we find that the number of sink particles appears to be insensitive to the values of the jet parameters, but there is a factor 2-3 difference between jet and non-jet runs (see Figure \ref{fig:sf_history}).
 
\subsubsection{Sensitivity to initial conditions}\label{sec:sensitivity}
We investigate the sensitivity of the predicted $\Mmean$ in our \textbf{C\_M\_J} runs (as these produce the most realistic IMF) to initial conditions by systematically varying cloud parameters around our \textbf{M2e4} reference cloud, as shown in Table \ref{tab:IC}. We also vary the momentum loading of protostellar jets. Using a least-squares fit for $\Mmean$ as a function of each varied parameter (at fixed 4\% SFE), we obtain 
\begin{equation}
\Mmean \propto  \Gamma^{-0.65\pm 0.15}\, \csmin^{2.5\pm 0.5}\, \Sigma^{-0.3\pm 0.1}\, \alphaturb^{0.15\pm 0.19}\, M_0^{-0.12\pm 0.07},
\label{eq:1Dfitting_func_sigma}
\end{equation}
which can also be expressed as
\begin{equation}
\Mmean \propto  \Gamma^{-0.65\pm 0.05}\, \csmin^{2.5\pm 0.5}\, \rho^{-0.2\pm 0.07}\, \alphaturb^{0.15\pm 0.19}\, M_0^{-0.22\pm 0.10},
\label{eq:1Dfitting_func_rho}
\end{equation}
where $\Gamma$ is the momentum loading of jets (see Eq. \ref{eq:gamma_def} and \S\ref{sec:jet_peak_model}), $\csmin$ is the adiabatic sound speed at the $T_\mathrm{floor}$ temperature floor, while $\rho$, $\alphaturb$ and $M_0$ are the initial density, virial parameter and mass of the parent cloud, see Appendix \ref{sec:scaling_dependencies} for a detailed presentation of the results and the derivation of the exponents and their errors. Assuming a mass-size relation similar to that in the MW (corresponding to $\Sigma\sim 60\,\msun/\pc^2$, see \citealt{larson_law}), we can simplify Eq. \ref{eq:1Dfitting_func_sigma} as 
\begin{equation}
\Mmean \propto  \Gamma^{-0.65\pm 0.15}\, \csmin^{2.5\pm 0.5}\, \alphaturb^{0.15\pm 0.19}\, M_0^{-0.12\pm 0.07}.
\label{eq:1Dfitting_func_rho_MW}
\end{equation}
Equations \ref{eq:1Dfitting_func_sigma}-\ref{eq:1Dfitting_func_rho_MW} imply that the number-weighted mean sink mass for clouds is only weakly dependent on most cloud properties and is primarily set by the jet momentum loading factor $\Gamma$ and the $\csmin$ sound speed at the cloud temperature floor.

\subsection{Effects of jets on the accretion flow}\label{sec:jets_effect_flow}

\begin{figure}
\begin {center}
\includegraphics[width=0.95\linewidth]{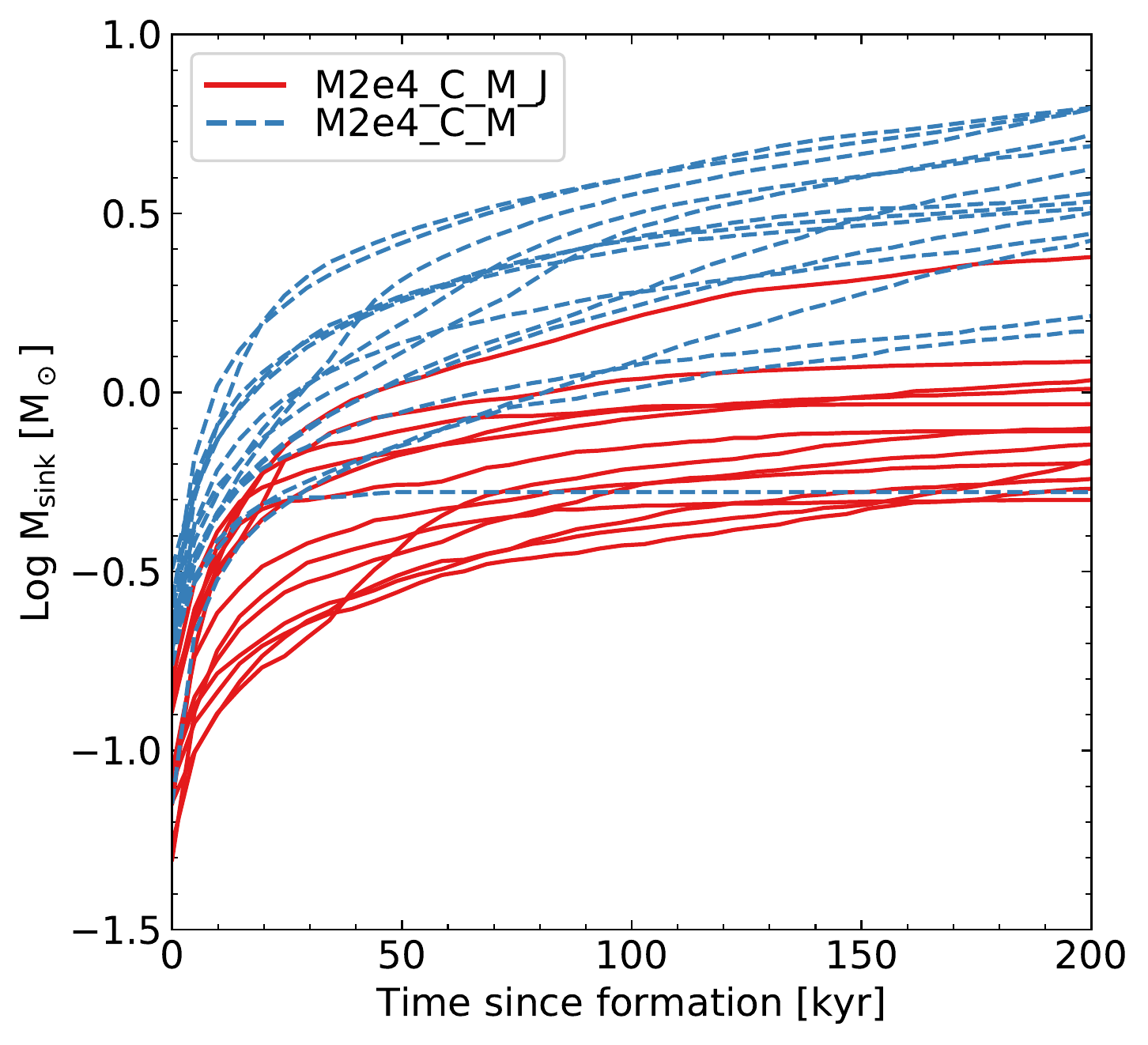}
\includegraphics[width=0.95\linewidth]{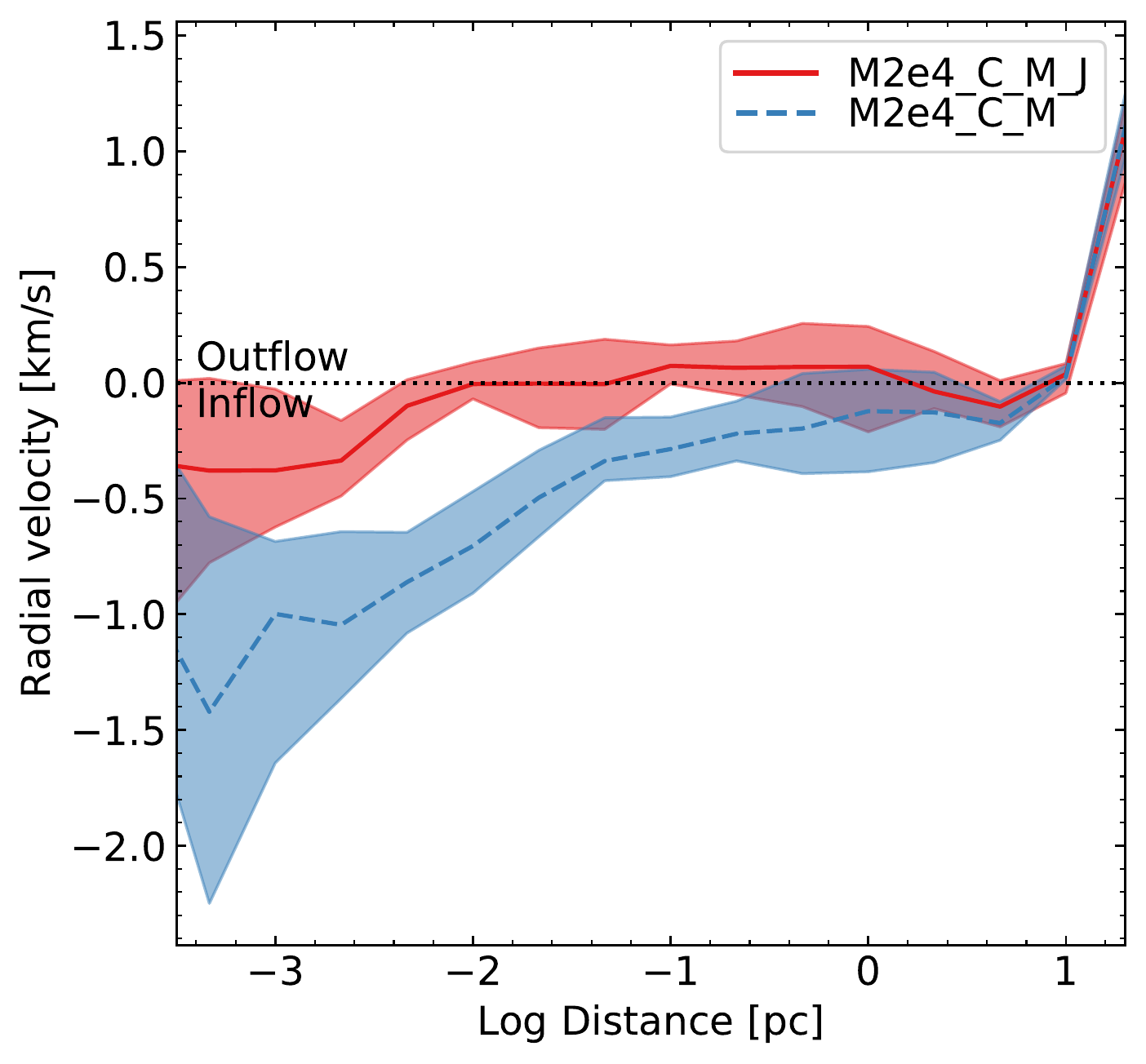}
\vspace{-0.4cm}
\caption{\textit{Top:} Evolution of the mass of a sink particles that form within the first Myr of star formation in a run with and without jets (\textbf{M2e4\_C\_M\_J} and \textbf{M2e4\_C\_M}). Note that we exclude sink particles that do not reach $0.5\,\msun$ within the first Myr of their lifetime to avoid including \myquote{failed sinks} that form around a massive sink particle that prevents them from growing. The addition of protostellar jets greatly reduces the growth rate of the sink, in some cases shutting down accretion. \textit{Bottom:} The average radial velocity in various radial shells around the same sink particles as above, 100 kyr after their formation. The solid line shows the value averaged over sinks, while the shaded area shows the interquartile range of values. On pc scales and above the two runs are essentially identical as the mass flux is set by the global collapse of the cloud. On $<0.1\,\pc$ scales jets disrupt the local accretion flow, dramatically reducing the magnitude of the mean radial velocity, which in turn leads to a dramatically reduced mass flux.}
\label{fig:mdot}
\vspace{-0.5cm}
\end {center}
\end{figure}

\begin{figure*}
\begin {center}
\begin{tabular}{c|c}
{\Large\textbf{M2e2}} & {\Large\textbf{M2e5}}\\
\includegraphics[width=0.24\linewidth]{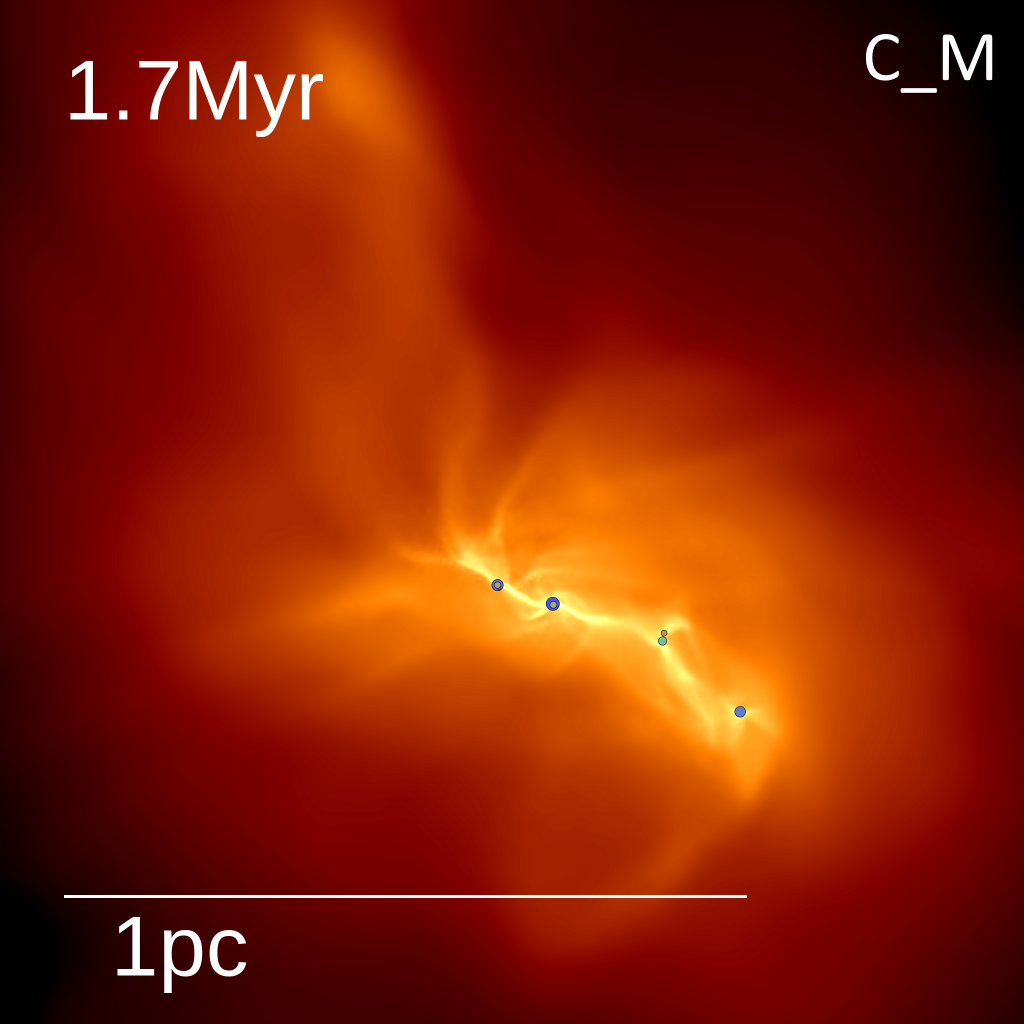}
\includegraphics[width=0.24\linewidth]{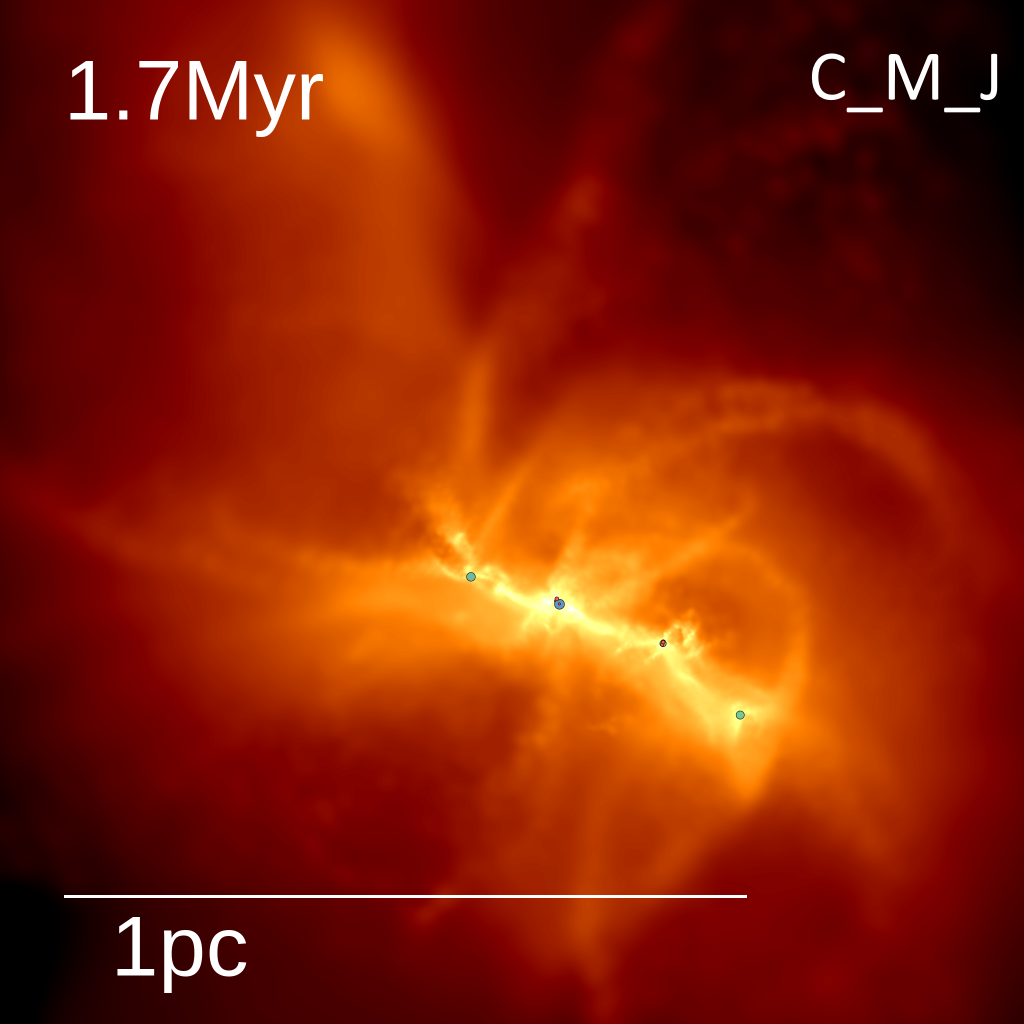} &
\includegraphics[width=0.24\linewidth]{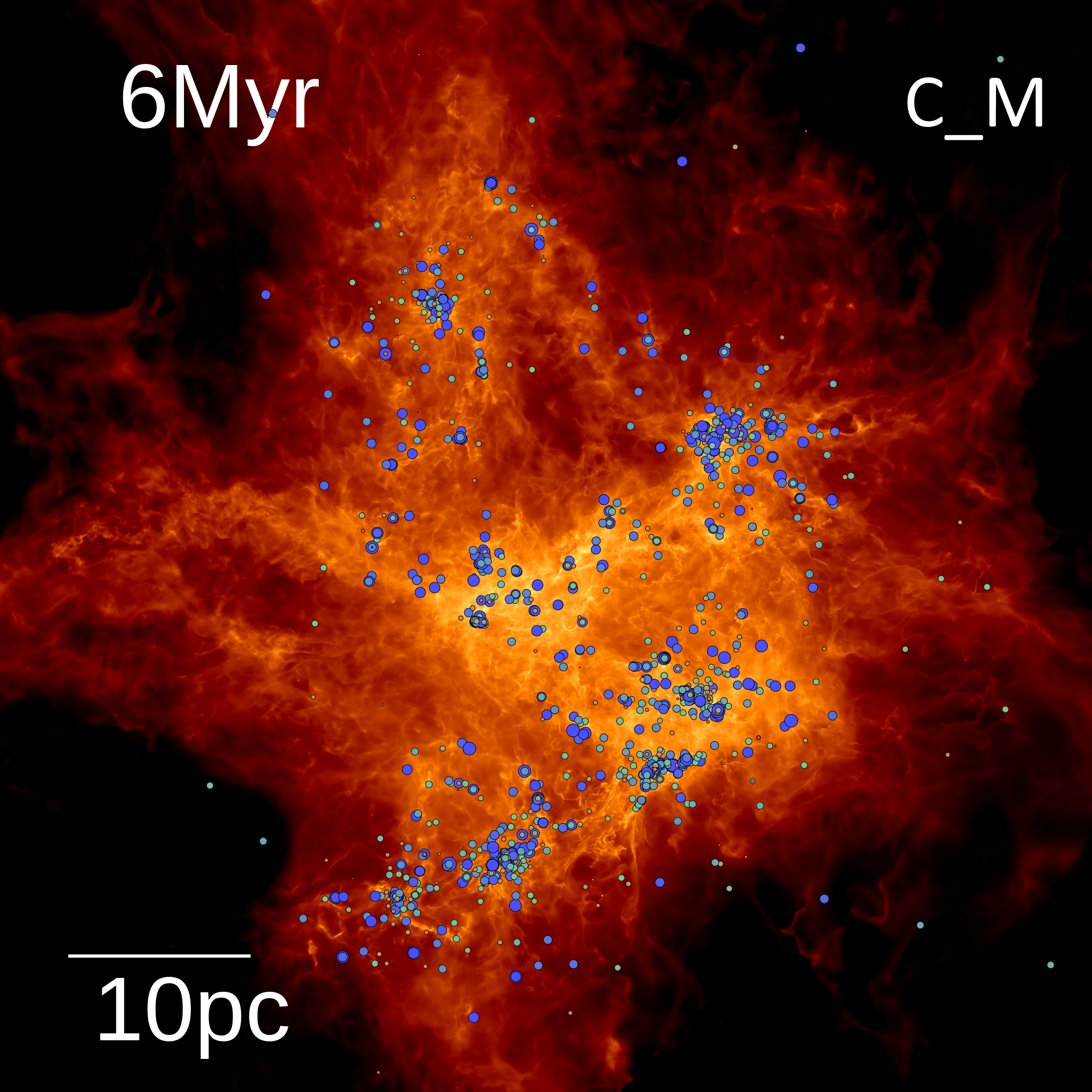}
\includegraphics[width=0.24\linewidth]{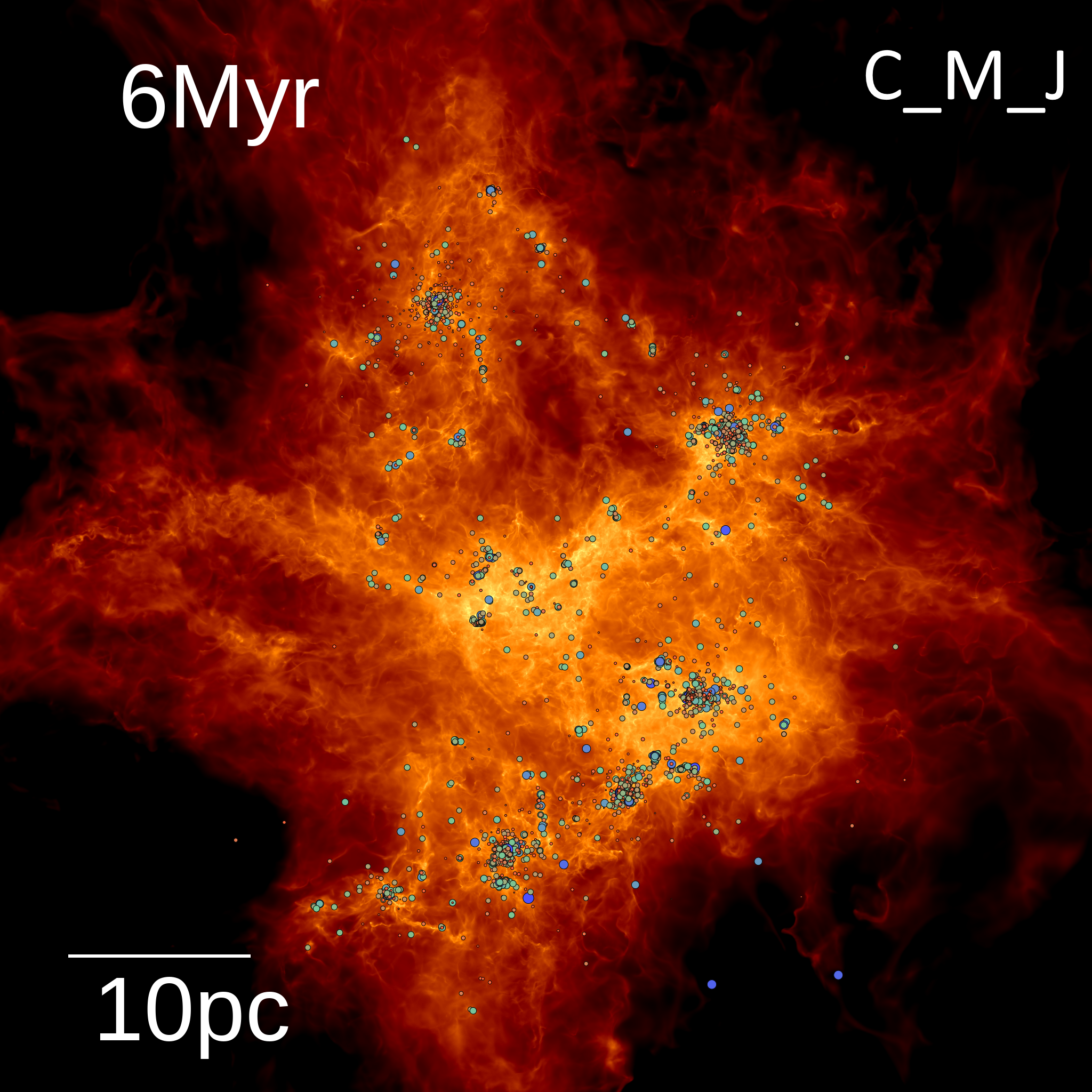}\\
\includegraphics[width=0.24\linewidth]{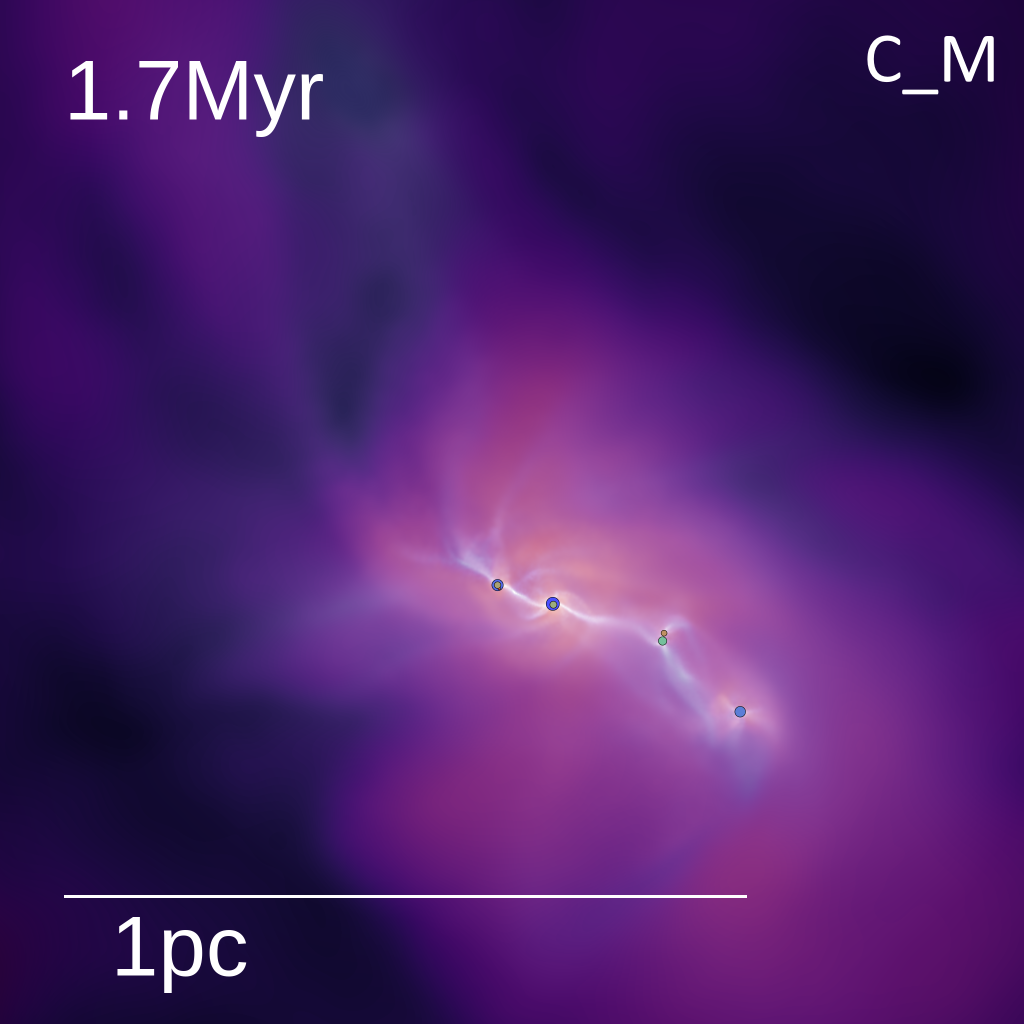}
\includegraphics[width=0.24\linewidth]{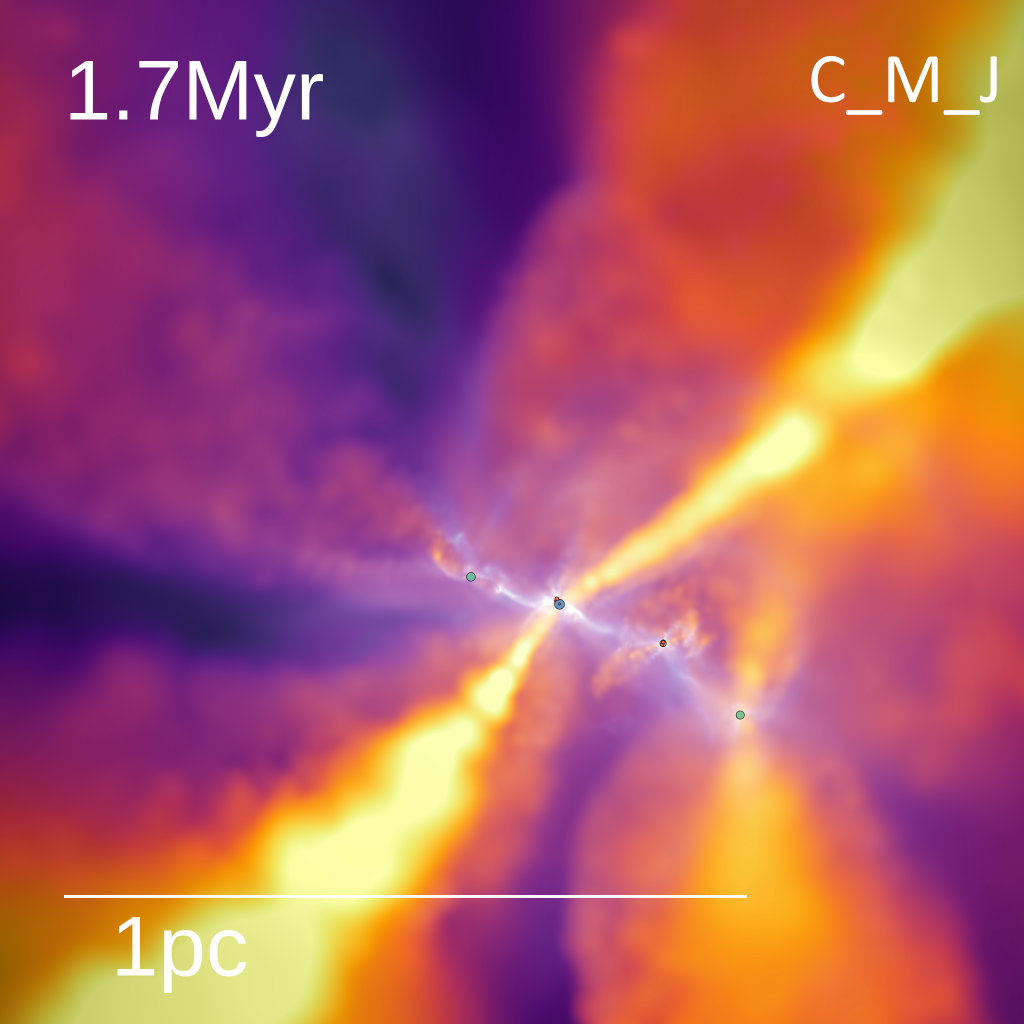}&
\includegraphics[width=0.24\linewidth]{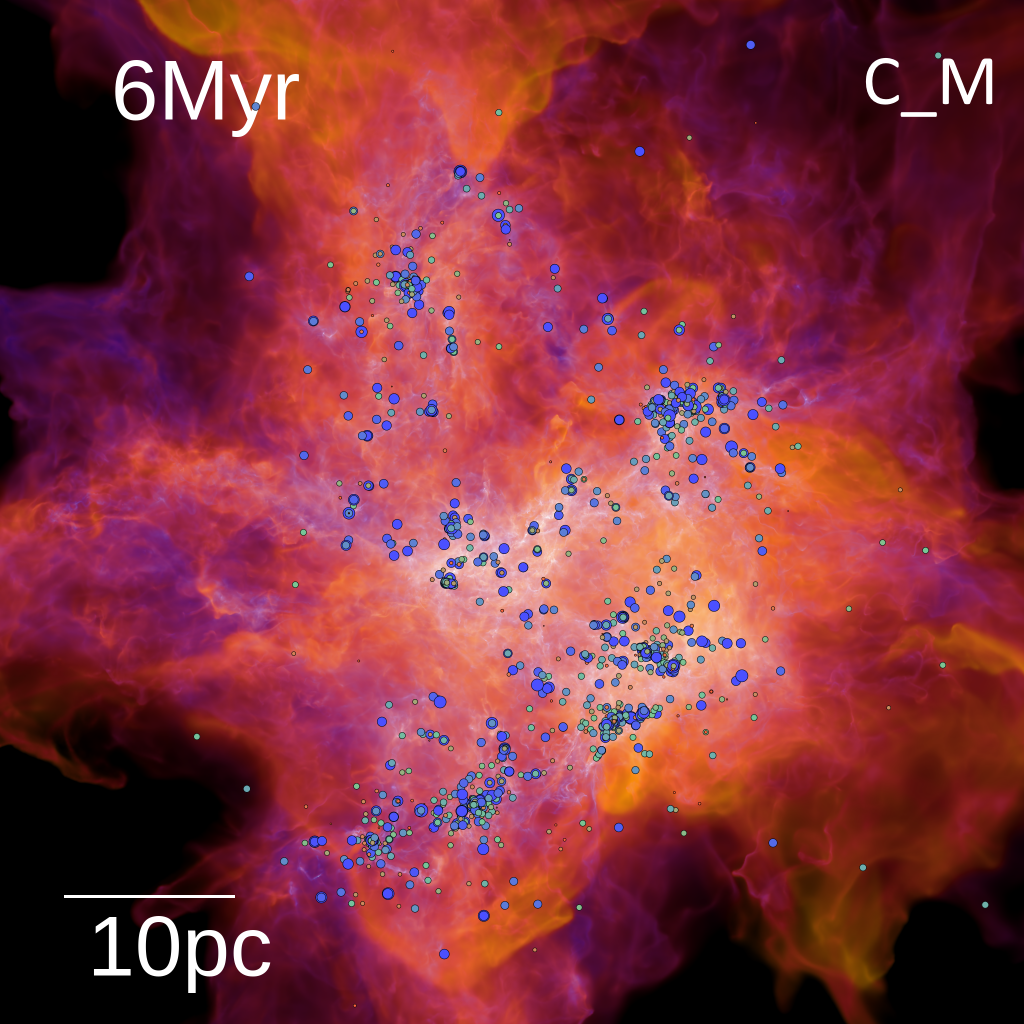}
\includegraphics[width=0.24\linewidth]{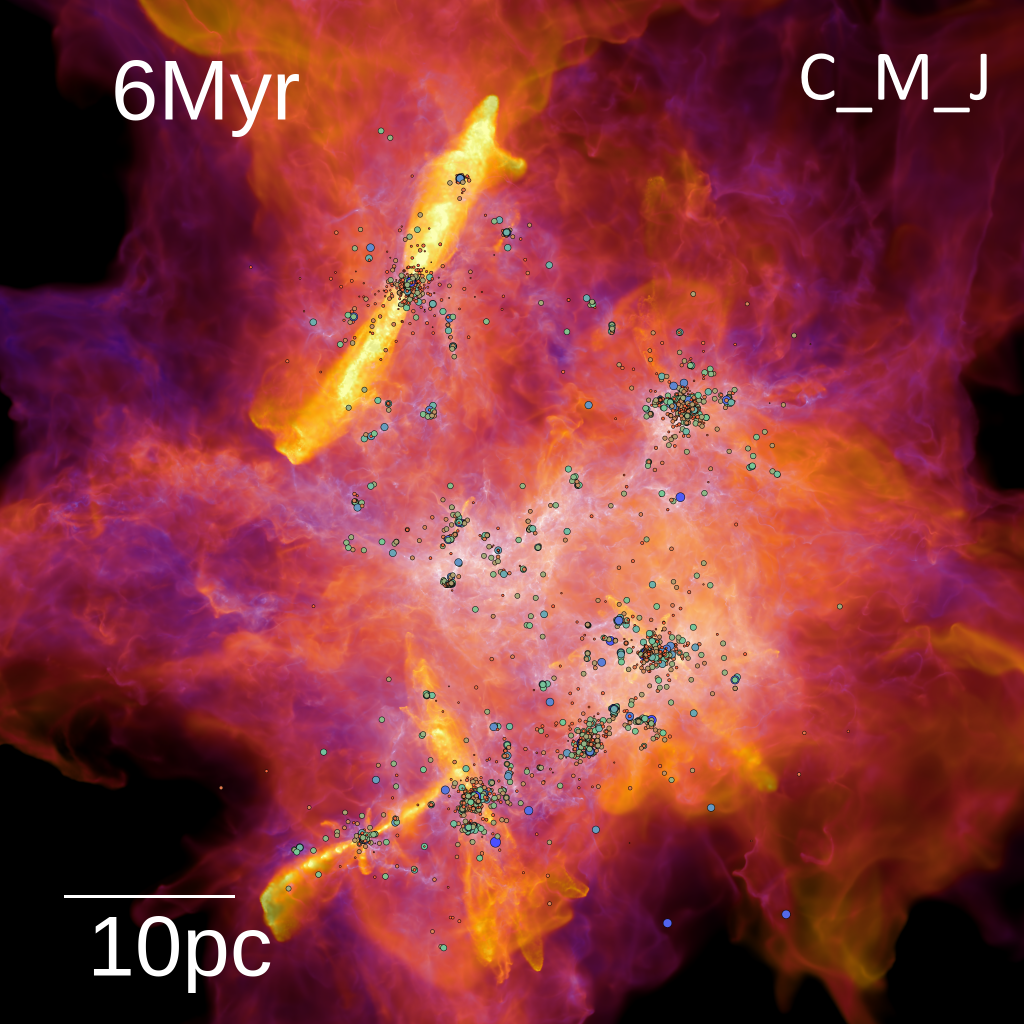}
\end{tabular}
\caption{\textit{(Top row)} Surface density maps for a very small (\textbf{M2e2}, \textit{left}) and a large cloud (\textbf{M2e5}, \textit{right}) with and without jets (\textbf{C\_M} and \textbf{C\_M\_J}). Colors (encoding projected gas surface density) and symbols (representing sink particles) are similar to Figure \ref{fig:M2e5_C_M_J_series}. While the gas and star distribution on larger scales is almost identical between the runs, the masses of sink particles are different as illustrated by the relative sizes and colors of the circles. \textit{(Bottom row)} Same as above, but now shown with a color map that encodes the 1D line-of-sight velocity dispersion (increasing from purple ($0.1\rm km\,s^{-1}$) to orange ($10\rm km\,s^{-1}$) and encodes surface density information in lightness (lighter is denser). While the protostellar jets are almost invisible in surface density maps (and hence dust and CO maps as well), due to their low density, they stand out in these kinematic maps.}
\label{fig:surfdens_compare}
\vspace{-0.5cm}
\end {center}
\end{figure*}

Figure \ref{fig:mdot} shows that protostellar jets dramatically change the accretion history of sink particles. Their effects are more than just removing some fraction of the accreted gas (i.e., multiplying the accretion rates by a constant factor), as the ejected jets entrain local gas and thus disrupt the accretion flow. This dramatically reduces the mass flux towards the sink particles on $<0.1\,\pc$ scales, slowing their growth (but not preventing the runaway accretion of massive stars, see Figure \ref{fig:Mmean_mass_scale_compare}). The nature of jet feedback is also showcased by Figure \ref{fig:surfdens_compare}. Looking at the surface density map, we find that the large-scale ($>0.1\pc$) gas structure is almost identical between runs with and without jets (\textbf{C\_M} and \textbf{C\_M\_J}), but the sink mass spectrum is dramatically different (see Figure \ref{fig:imf_compare}). This is due to the dramatic effect jets have on gas kinematics, disrupting accretion flows around stars and creating outflows that extend up to $\sim10\,\pc$ in scale (bottom row of Figure \ref{fig:surfdens_compare}).

\section{A simple model for the characteristic mass scale set by jets}\label{sec:jet_peak_model}

In this section we present a simple, plausible (but not necessarily unique) model that may explain the scaling of the mean sink (stellar) mass in our simulations (see Eq.\ref{eq:1Dfitting_func_rho} and Appendix \ref{sec:scaling_derivation}). The jet model in our simulation launches an $f_w$ fraction of the accreted mass at $f_K$ times the Keplerian velocity (see Eq. \ref{eq:jet_velocity} and \S\ref{sec:jets}). The total momentum output by the jet per unit time is therefore
\begin{equation}
\dot{P_J} = v_\mathrm{jet} \Mdot_\mathrm{jet} = f_K f_w \Mdot_\mathrm{acc} \sqrt{G M_\ast/R_*}, 
\end{equation}
where $\Mdot_\mathrm{acc}$ is the mass accretion rate. Let us further assume that $\Mdot_\mathrm{acc}=\mathrm{const.}$ and $R_*=\mathrm{const.}$\footnote{Note that our results in Figure \ref{fig:imf_sensitivity_2} show that the results are insensitive to whether we have an evolving or a constant $R_*$.} so that $M_\ast=\Mdot_\mathrm{acc} t$. This will simplify the above equation to
\begin{equation}
\dot{P_J} =  f_K f_w \left(1-f_w\right)^{1/2} G^{1/2}\Mdot_\mathrm{acc}^{3/2} R_*^{-1/2}t^{1/2},
\end{equation}
which we can integrate to get the total amount of momentum injected by jets over time $t$. Replacing $t=M_\ast/\Mdot_\mathrm{acc}$, we obtain:
\begin{equation}
P_J(M_\ast) = \frac{2}{3}\frac{f_w f_K}{1-f_w} G^{1/2}M_\ast^{3/2} R_*^{-1/2} = \Gamma G^{1/2}M_\ast^{3/2} R_*^{-1/2},
\end{equation}
where we have also used the $\Gamma$ momentum loading parameter from Eq. \ref{eq:gamma_def}.

Let us assume that this protostar forms in a cloud of uniform density $\rho$ that is much larger than the jet (i.e. GMC) and that there is a spherical gas reservoir of mass $M_g$ around the protostar that would eventually be accreted onto it without feedback. Let us also assume that protostellar jets are the only feedback process and that all the momentum injected by jets is deposited uniformly in the mass reservoir. The reservoir will become unbound if enough momentum is injected for its gas to reach escape velocity $v_\mathrm{esc}\sim \sqrt{G (M_\ast+M_g)/R}$, where $R=\left(M_g/(4\pi/3 \rho)\right)^{1/3}$ is the radius of the reservoir. This means:
\begin{equation}
\left(\frac{P_J(M_\ast)}{M_g}\right)^2 = \frac{G(M_g+M_\ast)}{R},
\end{equation}
so
\begin{equation}
\Gamma^2 G M_\ast^{3} R_*^{-1} M_g^{-2} = G(M_g+M_\ast)M_g^{-1/3}\left(\frac{4\pi\rho}{3}\right)^{1/3}.
\end{equation}
Assuming $M_g \gg M_\ast$ and a fixed $\rho$, we can solve for the star formation efficiency of the gas reservoir before it becomes unbound:
\begin{equation}
\frac{M_\ast}{M_g} = \Gamma^{-2/3} \left(\frac{4\pi}{3}\right)^{1/9} \rho^{1/9} M_g^{-1/9} R_*^{1/3}
\end{equation}
Substituting in typical values for GMCs this becomes:
\begin{equation}
\frac{M_\ast}{M_g} = 0.02 \left(\frac{\Gamma}{0.085}\right)^{-2/3} \left(\frac{n}{100\,\mathrm{cm^{-3}}}\right)^{1/9} \left(\frac{M_g}{\msun}\right)^{-1/9}\left(\frac{R_*}{2 \mathrm{R_\odot}}\right)^{1/3},
\label{eq:jet_peak_efficiency}
\end{equation}
where we used the $n$ number density instead of $\rho$ for convenience, as well as our fiducial parameters of $f_w=0.3$ and $f_K=0.3$ to normalize $\Gamma$.

\begin{figure*}
\begin {center}
\includegraphics[width=0.33\linewidth]{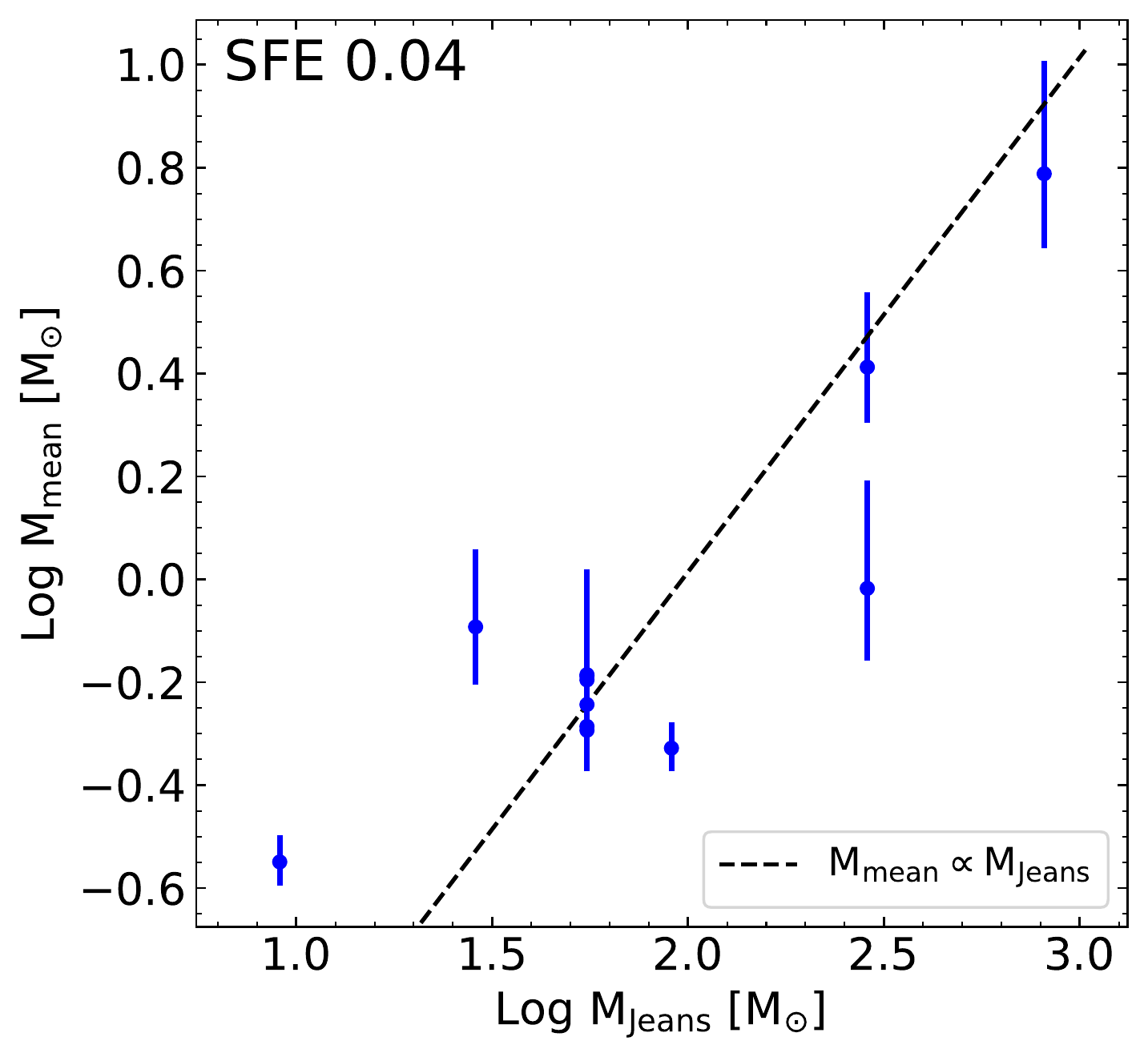}
\includegraphics[width=0.33\linewidth]{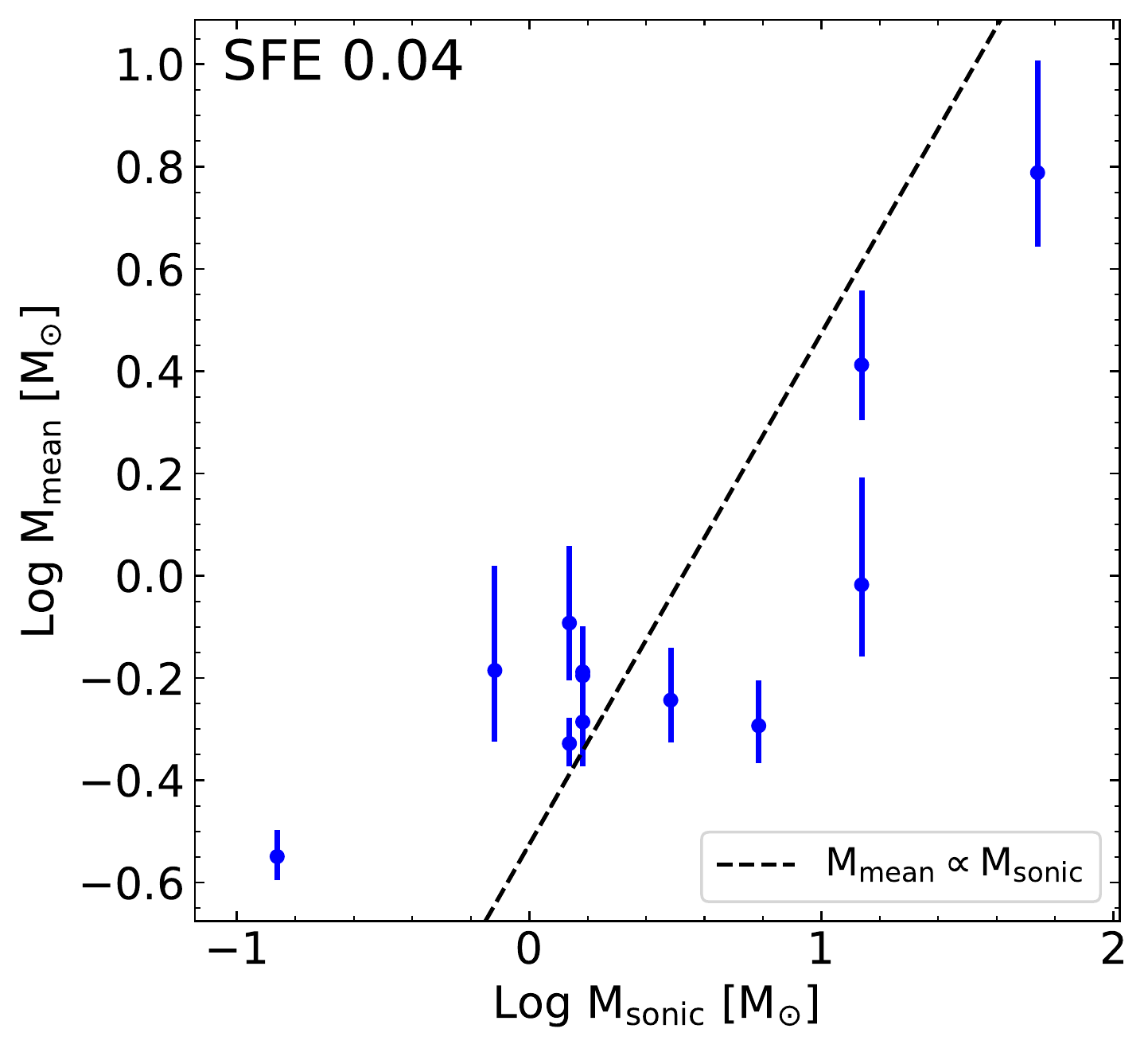}
\includegraphics[width=0.33\linewidth]{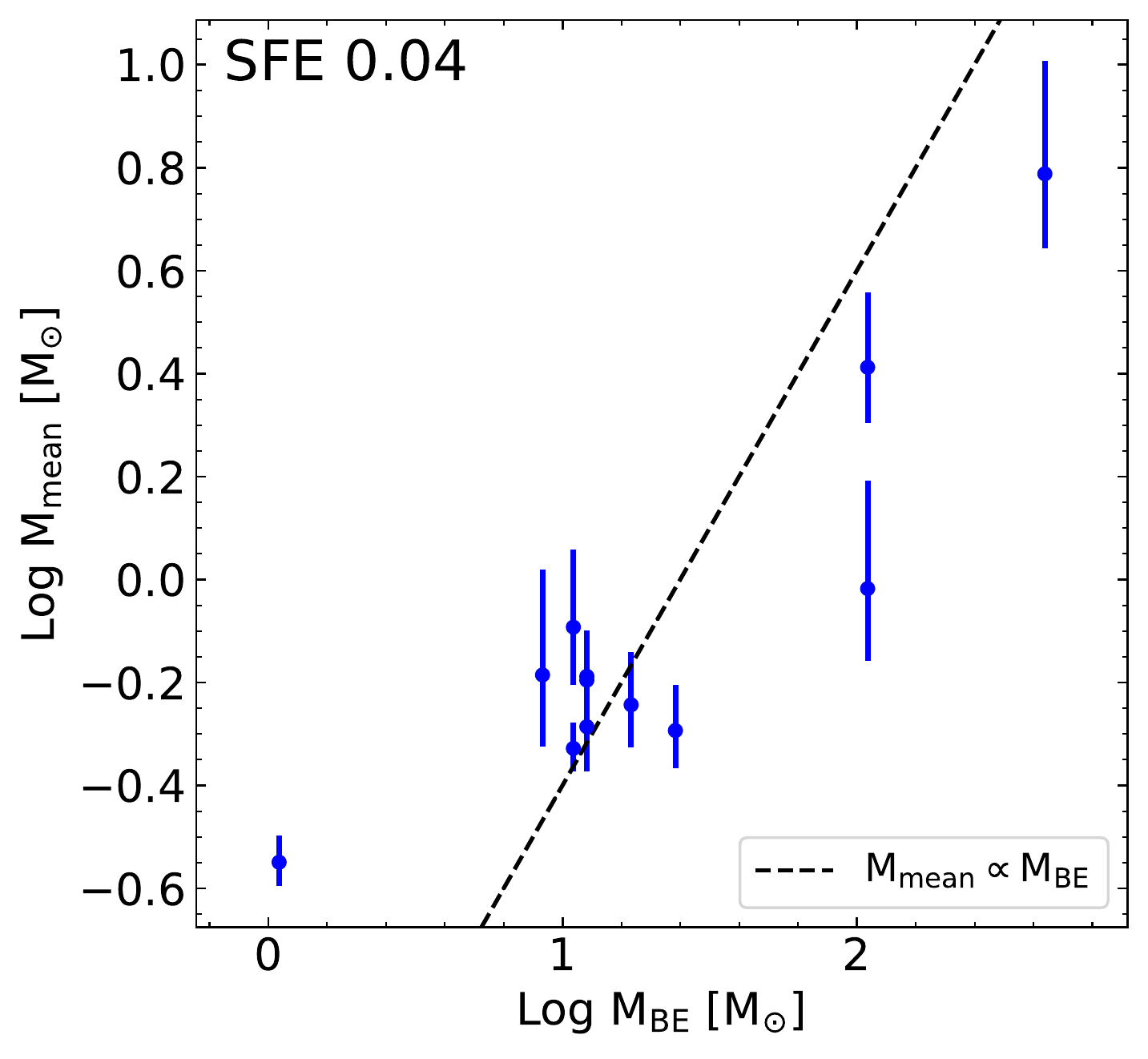}\\
\includegraphics[width=0.33\linewidth]{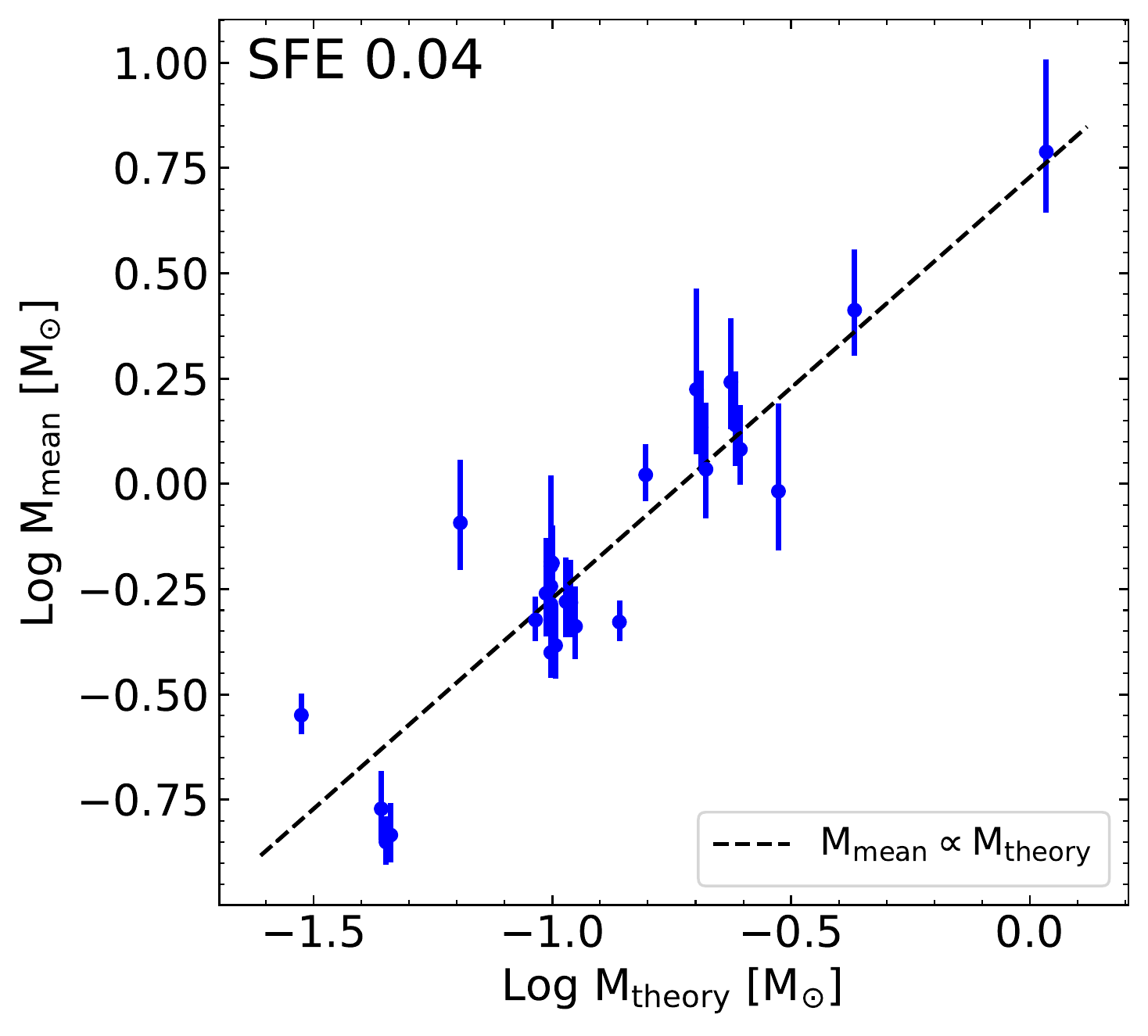}
\includegraphics[width=0.33\linewidth]{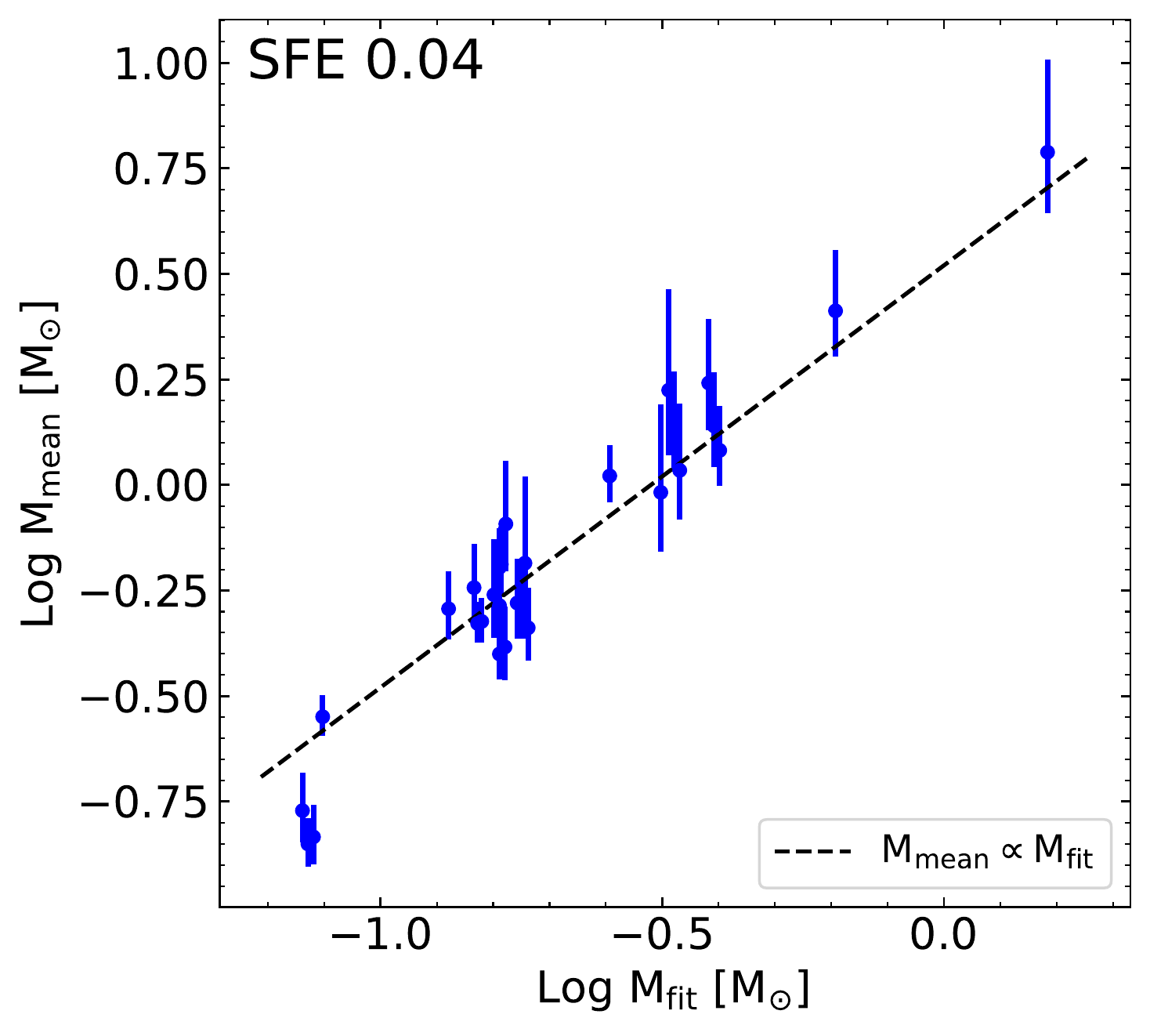}
\vspace{-0.4cm}
\caption{Comparison of the mean sink mass $\Mmean$ measured in different simulations using C\_M\_J physics with different initial cloud conditions (see Table \ref{tab:IC}) with the  initial Jeans mass $\MJeans$ (Eq. \ref{eq:MJeans}), sonic mass $\Msonic$ (Eq. \ref{eq:Msonic}), turbulent Bonnor-Ebert mass $\MBE$ (Eq. \ref{eq:MBE}) as well as the $M_\mathrm{theory}$ mass scale predicted by our toy model (Eq. \ref{eq:jet_peak_mass}) and the result $M_\mathrm{fit}$ obtained by arbitrary least-squares fit marginalized over these parameters (Eq. \ref{eq:1Dfitting_func_sigma}). A dashed line shows the best linear fit between these mass scales and the mean sink mass. Note that we chose 4\% as the reference SFE as some runs never reach higher values as jets disrupt the cloud. The errors are estimated by bootstrapping: we resample the sink mass distribution at fixed total sink mass and calculate the 95\% confidence interval of $\Mmassmedian$ over these realizations, which we denote with errorbars. Note that simulation runs with variable momentum loading are only shown in the bottom row and are slightly offset to make the plot easier to parse.}
\label{fig:Mmean_mass_scale_compare}
\vspace{-0.5cm}
\end {center}
\end{figure*}

To get the mass scale of the IMF we formulate an ansatz for the $M_g$ gas reservoir mass. Possible candidates are the Jeans, sonic and turbulent Bonnor-Ebert masses. Based on our scaling results from \S\ref{sec:sensitivity} and Appendix \ref{sec:scaling_dependencies} we know that the characteristic mass scales of the IMF ($\Mmassmedian$ and the $\Mmean$) both show weak dependence with the cloud virial parameter $\alphaturb$, consistent with an exponent between 0 and 1/3. Of these mass scales $\Msonic\propto \mach^{-2} \propto \alphaturb^{-1}$ and $\mathrm{M_{BE}}\propto \mach^{-1}\propto \alphaturb^{-1/2}$, while $\MJeans$ is independent, so we adopt $M_g=\MJeans$ in this model. Plugging it into Eq. \ref{eq:jet_peak_efficiency} we get
\begin{equation}
M_\ast = 0.12 \msun \left(\frac{\Gamma}{0.085}\right)^{-2/3}  \left(\frac{\csmin}{200\,\mathrm{m/s}}\right)^{24/9} \left(\frac{n}{100\,\mathrm{cm^{-3}}}\right)^{-1/3}\left(\frac{R_*}{2 \mathrm{R_\odot}}\right)^{1/3}.
\label{eq:jet_peak_mass}
\end{equation}
We find that the parameters of this model all fall within the uncertainty thresholds we found by fitting in Eq. \ref{eq:1Dfitting_func_rho}. Figure \ref{fig:Mmean_mass_scale_compare} shows how that the mass scales commonly used in the literature ($\MJeans$, $\Msonic$ and $\MBE$) are all correlated with the mean sink mass $\Mmean$ in our simulations with jets. Meanwhile, our toy model from Eq. \ref{eq:jet_peak_mass} provides a surprisingly good fit to the results with only a few outliers. Of course, it is only a toy model and makes several strong assumptions (e.g., constant $R_*$, $M_g\sim\MJeans$). Essentially, in this model $M_\ast$ is set by the characteristic reservoir mass (i.e., core mass) with a feedback efficiency factor that varies only weakly with gas properties and primarily depends on the jet momentum-loading as $M_\ast/M_g\propto \Gamma^{-2/3}$. 

We stress this particular model is not unique and should not be over-interpreted. For example, the time-integral above implies jets accelerate gas slowly on timescales long compared to core dynamical times. If this is not true, the criterion for unbinding gas becomes $\dot{P}_{J} > |{\bf F}_{\rm grav}|$ where ${\bf F}_{\rm grav}$ is the gravitational force. If we assume also (unlike our derivation above) that the protostar is sufficiently massive that its gravity is important in the envelope so $\dot{M}_{\rm acc}$ follows a Bondi-like scaling, then (following similar logic as before) the core would be unbound when ${M}_{\ast} \sim c_{s}^{2}\,\Gamma^{-2/3}\,G^{-1}\,R_{\ast}^{1/3}\,(M_{g}/\rho)^{2/9}$. This gives a comparably good fit to the scaling we empirically extract from the simulations, but without reference to the Jeans mass (in fact it depends quite weakly on whatever physics sets $M_{g}$). Instead, the $c_{s}$ dependence in this model comes from the fact that higher $c_{s}$ (all else equal) slows accretion and therefore reduces the instantaneous strength of feedback. What is robust is that in any momentum-feedback-regulated model, we expect $M_{\ast}$ to scale inversely with $\Gamma$. We also note that the above toy model is not unique to protostellar jet and can easily be adapted to derive the characteristic stellar mass for other feedback mechanisms.

\section{Discussion}\label{sec:discussion}
  
In isothermal MHD runs \citetalias{Guszejnov_isoT_MHD} found that magnetic fields impose a well-defined characteristic mass (related to the initial sonicmass) on the sink mass distribution that is insensitive to numerical resolution (unlike the non-magnetized isothermal hydrodynamics case, see \citealt{guszejnov_isothermal_collapse}), similar to the results of \citet{Haugbolle_Padoan_isot_IMF}. Above this mass scale the sink mass distribution roughly follows a $\dderiv N/\dderiv M \propto M^{-2}$ trend, similar to the observed IMF \citep{salpeter_slope}, ad likely arising as a general consequence of scale-free physics on this dynamic range \citep{guszejnov_scaling_laws}. \citetalias{Guszejnov_isoT_MHD} found that this characteristic mass of stars is an order of magnitude higher than what is observed, and is sensitive to initial conditions in a way that violates the apparent near-universality of the IMF in the MW \citep{imf_universality, guszejnov_imf_var}. 
 
\subsection{Role of non-isothermal thermodynamics}

Isothermality is often assumed in star formation theories and simulations due to the highly efficient cooling of molecular gas \citep{Girichidis_2020_sf_processes_review}, even though there is a significant scatter in the gas temperature with a clear density dependence (see \citealt{Glover_Clark_n_T}). At high densities the isothermality assumption must eventually break down, allowing for the formation of hydrostatic cores \citep{larson_1969} that are the progenitors of protostars. This transition from near-isothermal to adiabatic behavior was originally proposed to be responsible for setting the peak of the IMF (see \citealt{lowlyndenbell1976, rees1976}), but the corresponding mass scale ($\sim 0.008\msun$) was too low to explain observations. The idea has recently been revived by taking into account the tidal screening effect around the first Larson core \citep{Lee_Hennebelle_2018_EOS,Colman_Teyssier_2019_tidal_screening}, which increases the relevant mass scale to be comparable to the observed IMF peak. 

It is important to note that most of these simulations have been run on non-magnetized clouds, so the only unique mass scale in the sink mass spectrum arises from non-isothermal physics at high densities\footnote{Note that the runs in \citealt{Lee_Hennebelle_2019_T_B} did include magnetic fields and did not produce a top-heavy IMF. This is due to dense, highly turbulent initial conditions, which dramatically lowers the magnetic mass scale compared to what it would be in MW-like clouds (see \S4.3 in \citetalias{Guszejnov_isoT_MHD}), hence the opacity limit does dominate in this regime.}. Including magnetic fields, however, in MW-like cloud conditions shifts the turnover mass of the IMF to much larger $>20\,\msun$ scales (see Figure \ref{fig:imf_compare} and \citetalias{Guszejnov_isoT_MHD}). Thus, for MW-like clouds, gas thermodynamics (i.e., the opacity limit) do not set the \myquote{mean} characteristic or turnover mass scale of the IMF (which is of order $\sim \msun$), their effects are likely limited to the lowest mass scales of the IMF ($<0.1\,\msun$, see Figures \ref{fig:mass_scale_evol}-\ref{fig:imf_compare}).


\subsection{Role of protostellar jets}\label{sec:discussion_jets}

Previous work has shown that protostellar jets can expel a significant portion of accreting material, directly reducing stellar masses \citep[e.g.,][]{Federrath_2014_jets,Offner_Chaban_2017_jets_sfe} and potentially driving small scale turbulence \citep[e.g.,][]{Nakamura_2007_outflow_turbulence_driving,Wang_2010_outflow_regulated_SF,Offner_Arce_2014,Offner_Chaban_2017_jets_sfe, murray_2018_jets}. We do find that jets disrupt the local accretion flow, which greatly changes gas dynamics on $<0.1\,\pc$ scales, but this has little effect on the global evolution of a massive GMC. Previous work has shown that protostellar outflows reduce the star formation rate of the parent cloud \citep{Cunningham_2011_outflow_sim, hansen_lowmass_sf_feedback,Federrath_2014_jets,murray_2018_jets}, which we confirm. 

Previous non-MHD simulations \citep[e.g.,][]{bate_2009_rad_importance,krumholz_2012_orion_sims} argued that radiation (specifically radiative heating by local protostars) and jets are the key ingredients to the IMF, where radiation heats the gas surrounding the star, preventing it from fragmenting and forming new stars, thus creating a mass reservoir that the protostar can almost fully accrete. This, however, can lead to an \myquote{over-accretion} problem that is resolved by the addition of protostellar jets \citep{hansen_lowmass_sf_feedback, krumholz_2012_orion_sims}. Later works also included MHD processes and produced IMFs similar to that observed \citep{li_2018_sf_mhd_jets,Cunningham_2018_feedback}, but the combination of protostellar jets and MHD without radiation on cloud scales ($>\pc$) was not investigated. Our simulations suggest that radiation may not be necessary to reproduce the observed IMF, as magnetic fields naturally provide support against fragmentation near newly formed stars. This is true regardless of the initial magnetization of the cloud as the turbulent dynamo drives the system towards a common $B-\rho$ relation at high densities (see Figure 7 in \citetalias{Guszejnov_isoT_MHD} and Appendix \ref{sec:scaling_dependencies}). However, several caveats are in order. (1) We focus primarily on statistics insensitive to the lowest-mass stars (which may be most sensitive to radiation), so long as the IMF is shallower than Salpeter at low masses. We have not rigorously demonstrated that the low-mass IMF is numerically converged in our \textbf{C\_M\_J} simulations, even if $\Mmean$ is (see \citetalias{grudic_starforge_methods}). (2) Our cooling/non-isothermal simulations include simple approximations to account for the transition between optically thin and thick cooling, rather than explicit radiation-MHD; if these underestimate the cooling rates at high densities we might underestimate the need for radiative heating. (3) We enforce a constant dust and \myquote{floor} temperature $T_{\rm floor} = T_{\rm dust} = 10\,$K. In future work we will replace this with more realistic assumptions, but in Appendix \ref{sec:scaling_derivation} we show the IMF shape at $\lesssim 1\,M_{\odot}$ is quite sensitive to this value (this is essentially $c_{s,\,\rm min}$, in our Eq.\ref{eq:1Dfitting_func_rho}). So the IMF {\em is} sensitive to thermodynamics, and it remains to be seen whether more physical models for cooling and dust temperatures below $\sim 100\,$K can robustly reproduce the observed IMF without local radiative heating. (4) These simulations do not resolve protostellar disks (let alone disk fragmentation), whose stability may be critically impacted by radiative feedback. Also, our treatment neglects non-ideal MHD terms, so we see disks lose their angular momentum rapidly and simply accrete entirely onto the central sink owing to strong magnetic braking (\citealt{hennebelle_fromang_2008_corecollapse, wurster_2016_nonideal_braking}), artificially avoiding fragmentation (see \citealt{Wurster_2019_no_magnetic_break_catastrophe} for a counter-argument).

In addition to their effects upon sub-pc accretion flows and the IMF, we find that jets can have a significant global effect upon GMC kinematics and evolution in smaller clouds (Figure \ref{fig:sf_regulation_jets}). Specifically, protostellar jets {\it alone} appear sufficient to unbind initially-bound clouds at least as massive as $2\times 10^4\msun$, once a sufficiently-high SFR and momentum injection rate are achieved. However, in Figure \ref{fig:sf_regulation_jets} we see that for all but our least-massive clouds ($M=200\,\msun$), by the time jets begin to unbind the parent cloud (causing a sharp rise to $\alphaturb\gg2$) the integrated SFE has already reached $\gg10\%$ values, much larger than observed in MW clouds that motivate our ICs \citep{sf_big_problems}. Thus, for clouds with masses $>1000\,\msun$ some other process (e.g., radiation from massive stars) must dominate cloud disruption. Even if some other feedback mechanism is ultimately responsible for GMC disruption (\S\ref{sec:otherfeedback}), the contribution of jets alone to the cloud kinematics can be significant. Therefore it is likely that jet feedback has important nonlinear interactions with other feedback mechanisms, potentially making it easier for e.g. stellar radiation to disrupt the cloud by increasing the initial turbulence or reducing the initial density at the time that massive stars break out from their envelopes. For this reason, previous simulations of feedback and cluster formation on GMC scales that neglected jets (including previous works by the present authors, e.g. \citealt{grudic_2016,grudic_2018_mwg_gmc,grudic_2020_cluster_formation}) should be revisited.

\subsection{Apparent sensitivity to initial conditions}
We find that the mass scale set by protostellar jets exhibits significant sensitivity to variations in the initial conditions (see Eq. \ref{eq:1Dfitting_func_rho}).  Even if one could argue that parameters like the $\Gamma$ momentum loading factor are set by atomic and nuclear physics in a way that they vary little between star forming regions (despite differences in local metallicities), observed clouds, even in the Solar neighborhood, have a wide range of masses ($10^4-10^6\,\msun$), densities ($10-1000\,\mathrm{cm}^{-3}$), virial parameters ($0.1-10$) and temperatures ($10-30\,\mathrm{K}$), see \citet{kauffmann_pillai_2013, heyer_dame_2015, Miville_Deschenes_2017_MW_GMCs}. The properties of star-forming gas in more extreme environments (e.g., Galactic Center, ULIRGS) can vary much more wildly (e.g., densities $>10^5\,\mathrm{cm}^{-3}$, surface densities at $\sim 1000$ times higher values, molecular temperatures $\sim 70-100\,\kelvin$, see \citealt{Dame_2001_MW_clouds_CO,Gao_2004_ULIRG_gas_SF, Longmore_2012_the_Brick}). Meanwhile the IMF is observed to be near-universal, with variations, even in extragalactic sources, within a factor of 3 or less in both the IMF peak and mass-to-light ratio. In equation \ref{eq:1Dfitting_func_rho} the strong dependence on temperature is perhaps most concerning here, as that alone would predict $\sim4$ variations among local clouds and factor $\sim20-80$ variations  between the Solar neighborhood and more extreme galactic environments. 


\subsection{Potential role of additional feedback physics}
\label{sec:otherfeedback}
While we find that protostellar jets dramatically reduce the stellar mass scales to values similar to those observed, these models still have several shortcomings that only additional physics can address. The most significant issues with the current model are that (1) massive stars undergo runaway accretion, creating a top-heavy IMF (see Figures \ref{fig:mass_scale_evol} and \ref{fig:imf_compare}); (2) star formation continues potentially up to $\SFE$ of unity for massive GMCs; and (3) the stellar mass scale set by jets is sensitive to the temperature of the parent cloud, which may potentially violate the observed near-universality of the IMF. 

As discussed in \S\ref{sec:discussion_jets}, one obvious step is the inclusion of radiative heating, which has been argued to be crucial in setting the mass scale of low-mass stars \citep{Offner_2009_radiative_sim,krumholz11a, krumholz_stellar_mass_origin, bate12a, Myers_2013_ORION_radiation_IMF,guszejnov_gmc_imf,guszejnov_feedback_necessity,Cunningham_2018_feedback}. Ionizing radiation of main-sequence stars as well as the stellar winds they emit could also potentially solve the runaway accretion of massive stars \citep{krumholz_2012_orion_sims, li_2018_sf_mhd_jets,Cunningham_2018_feedback}. Furthermore, these feedback processes (along with supernovae) could allow massive stars to disrupt their natal cloud and quench star formation at the observed $\SFE$ levels (\citealt{grudic_mond, krumholz_2019_cluster_review,Li_Vogelsberger_2019_GMC_disrupt}).

\section{Conclusions}\label{sec:conclusions}

In this paper we presented simulations from the STARFORGE project, which are high-resolution MHD simulations of the collapse of a giant molecular cloud that also follow the evolution of individual stars. The runs include progressively more complex physics, starting from isothermal MHD, then adding cooling physics then feedback in the form of protostellar jets. We found that the inclusion of jets dramatically alters the mass spectrum of sink particles (the simulation analogue of the observed stellar IMF). The resulting  mass distribution is broadly similar to the observed IMF in both shape and scale, but additional physics is needed for a complete IMF theory. 

We carried out a large suite of tests to determine the sensitivity of our results to variations in both initial conditions and input physics parameters. We found that the mean sink particle mass set by jets is insensitive to many parameters, but sensitive to the momentum loading of jets and the cold, dense gas and dust temperatures, and potentially the surface density. Based on observed variations in cloud properties these would lead to larger variations in the IMF than observed in the Solar neighborhood and much larger variations in extreme environments (e.g., Galactic Center, starburst galaxies). 

While protostellar jets allowed our simulations to produce a realistic IMF at masses between $0.1-10\,\msun$, massive stars ($>10\,\msun$) undergo runaway accretion, leading to an increasingly top-heavy IMF with time, in increasing conflict with the observed IMF slope. Even though jets can ultimately quench star formation it requires >10\% of the cloud mass to be turned into stars for even low-mass GMCs ($\sim 10^4\,\msun$) so for massive GMCs ($>10^5\,\msun$) star formation would likely continue until an order unity fraction of the gas turns into stars. Meanwhile, observed nearby clouds, whose properties motivate our initial conditions, achieve terminal SFE values of only a few percent. We conclude that additional physics is required to stabilize the IMF and regulate star formation. Candidates for these processes will be explored in future work.

 

\section{Data availability}
The data supporting the plots within this article are available on reasonable request to the corresponding authors. A public version of the {\small GIZMO} code is available at \url{http://www.tapir.caltech.edu/~phopkins/Site/GIZMO.html}.

\section*{Acknowledgements}
DG is supported by the Harlan J. Smith McDonald Observatory Postdoctoral Fellowship. MYG is supported by a CIERA Postdoctoral Fellowship.
Support for PFH was provided by NSF Collaborative Research Grants 1715847 \&\ 1911233, NSF CAREER grant 1455342, and NASA grants 80NSSC18K0562 \&\ JPL 1589742.
SSRO is supported by NSF Career Award AST-1650486 and by a Cottrell Scholar Award from the Research Corporation for Science Advancement. CAFG is supported by NSF through grant AST-1715216 and CAREER award AST-1652522; by NASA through grant 17-ATP17-0067; and by a Cottrell Scholar Award from the Research Corporation for Science Advancement. 
This work used computational resources provided by XSEDE allocation AST-190018, the Frontera allocation AST-20019, and additional resources provided by the University of Texas at Austin and the Texas Advanced Computing Center (TACC; http://www.tacc.utexas.edu).

 


 \bibliographystyle{mnras}
 \bibliography{bibliography} 



\appendix

\section{Dependence of the IMF on initial conditions}\label{sec:scaling_dependencies}
\begin{figure*}
\begin {center}
\includegraphics[width=0.33\linewidth]{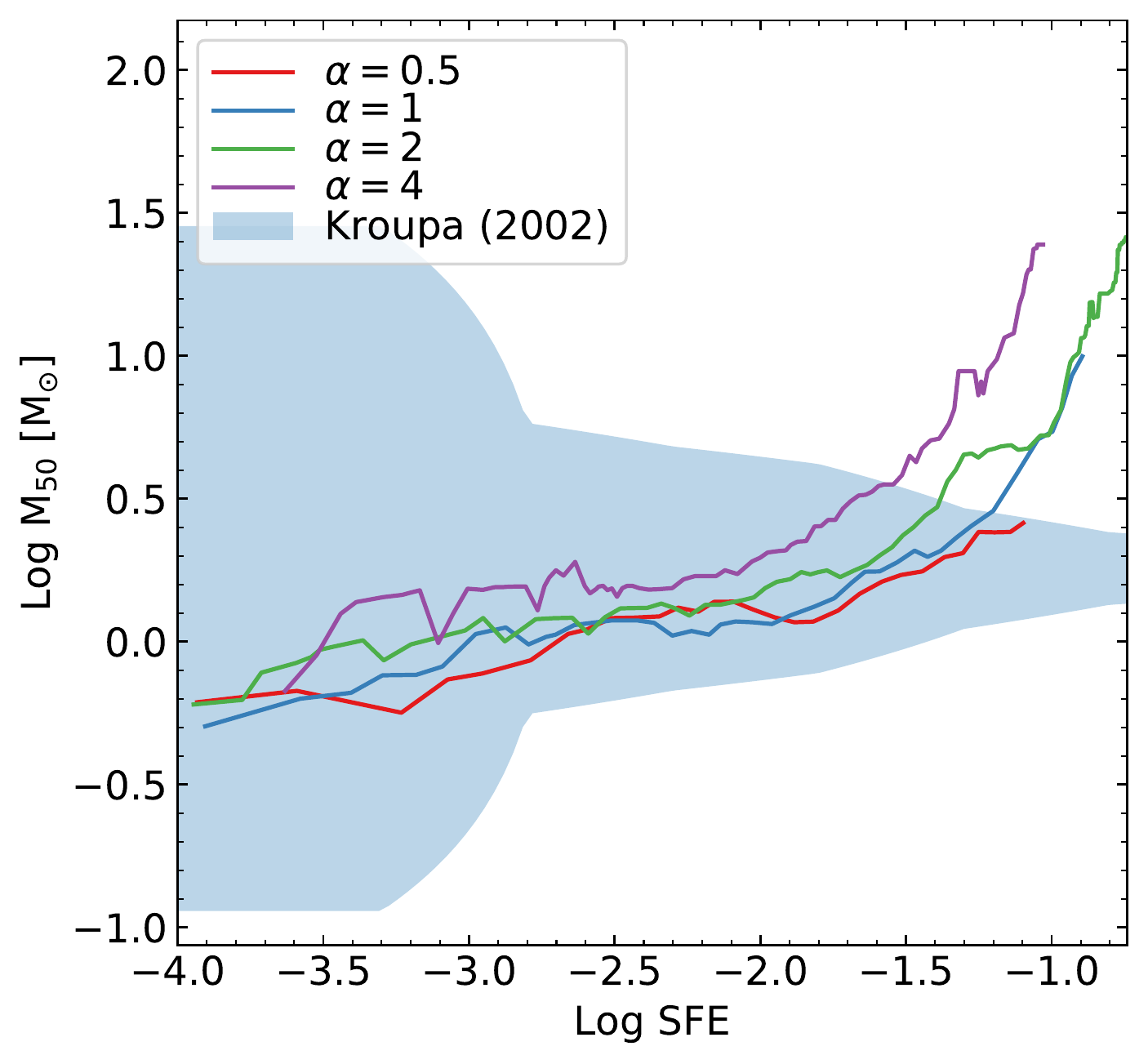}
\includegraphics[width=0.33\linewidth]{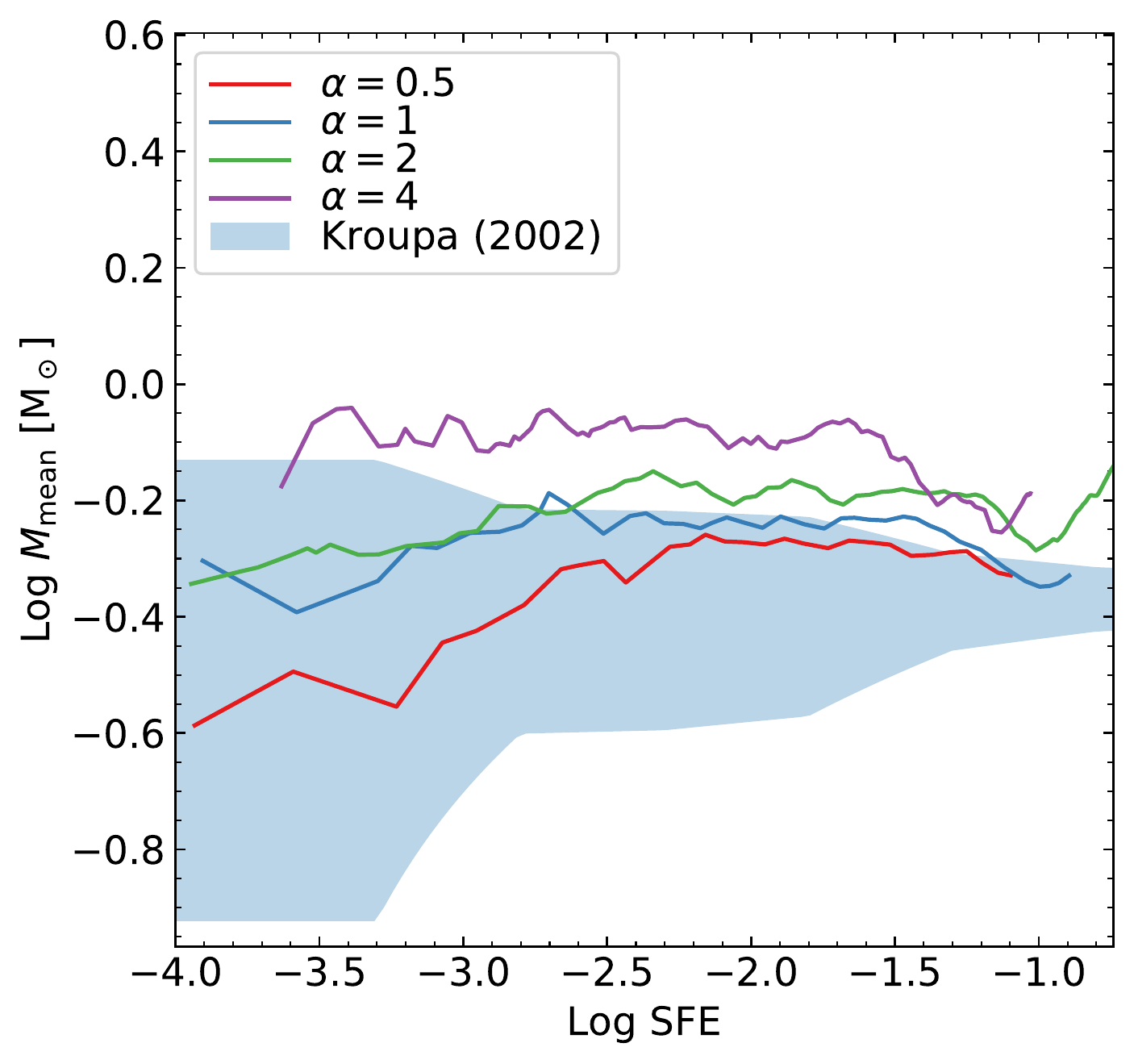}
\includegraphics[width=0.33\linewidth]{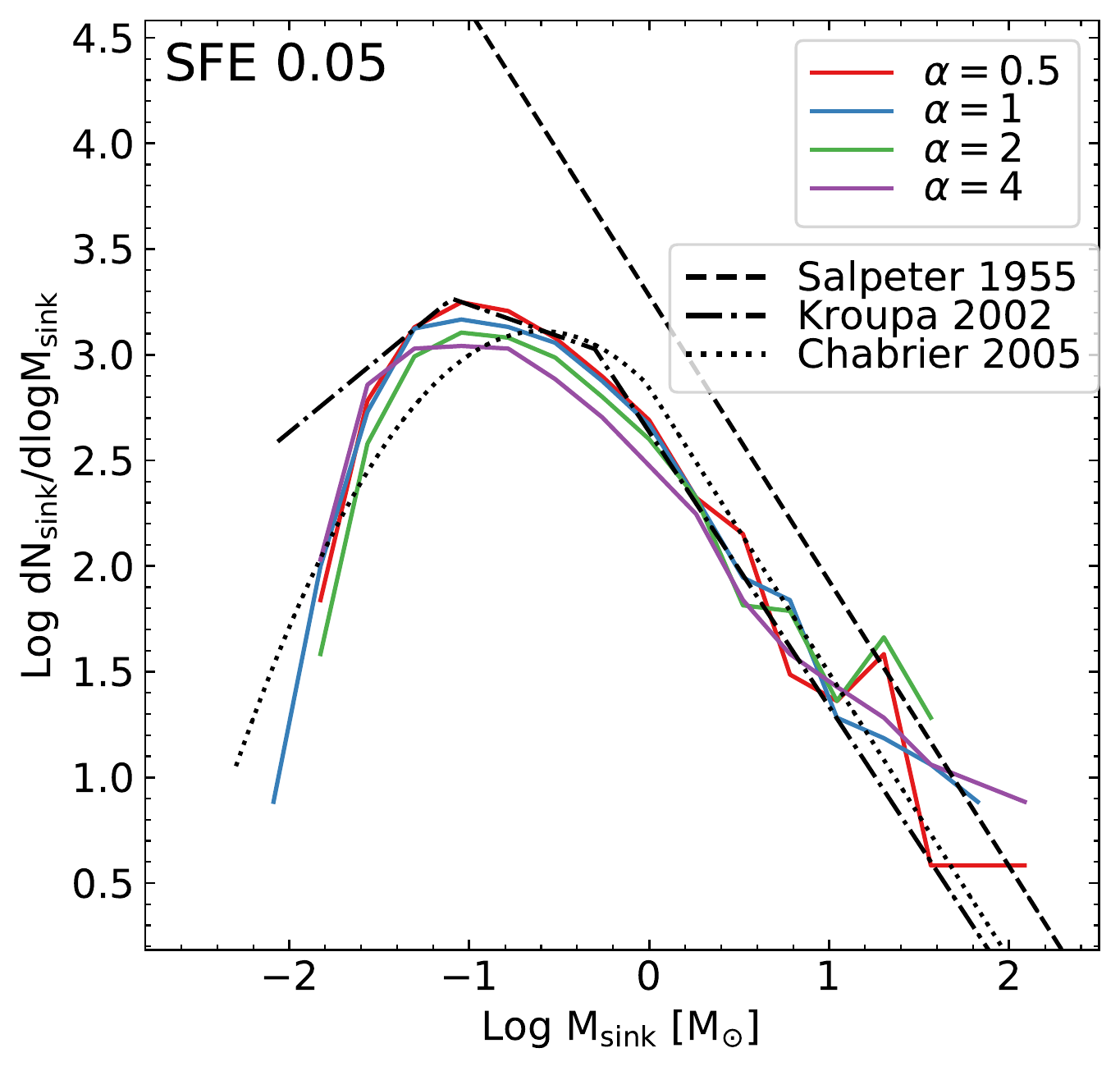}
\\
\includegraphics[width=0.33\linewidth]{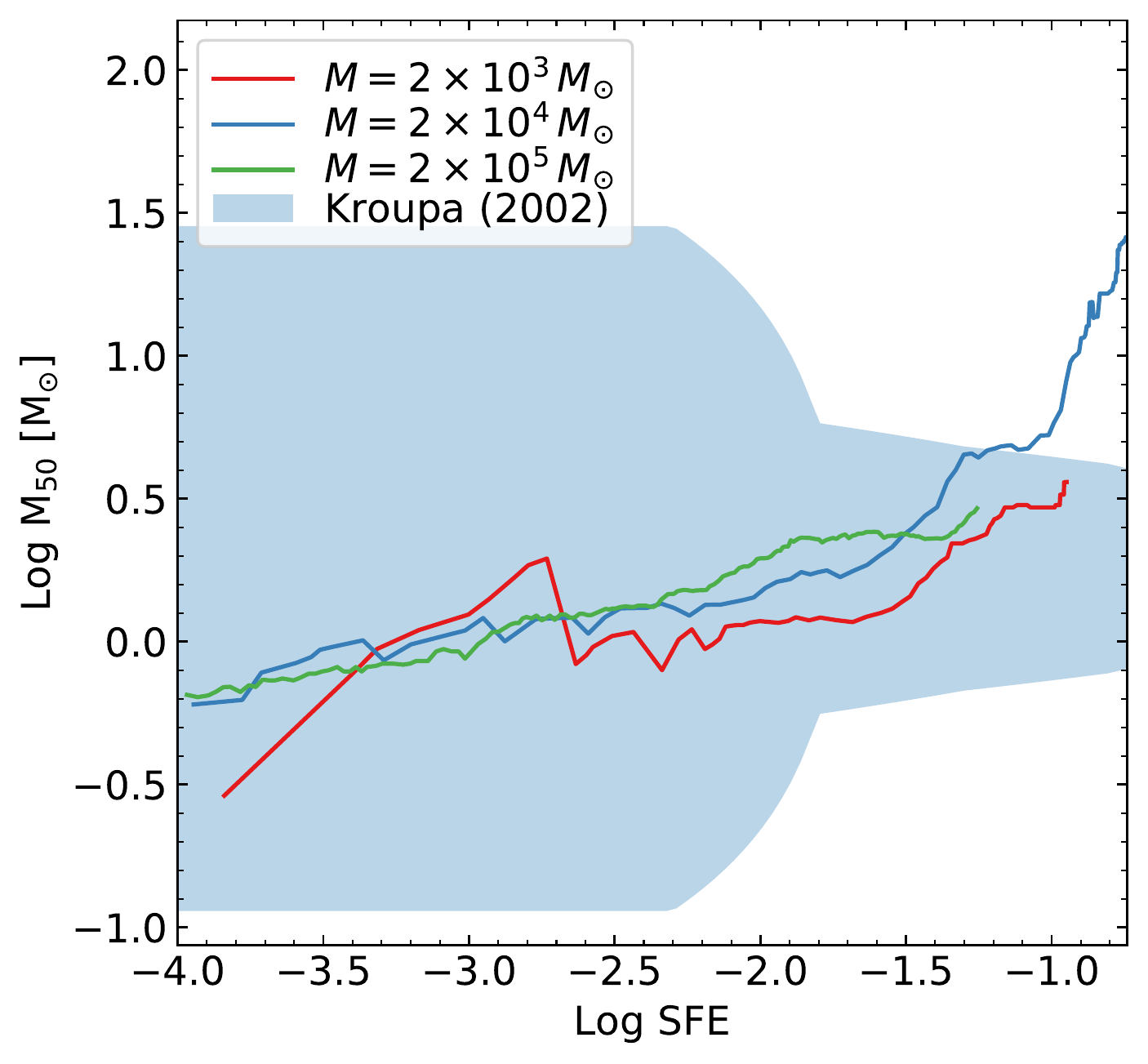}
\includegraphics[width=0.33\linewidth]{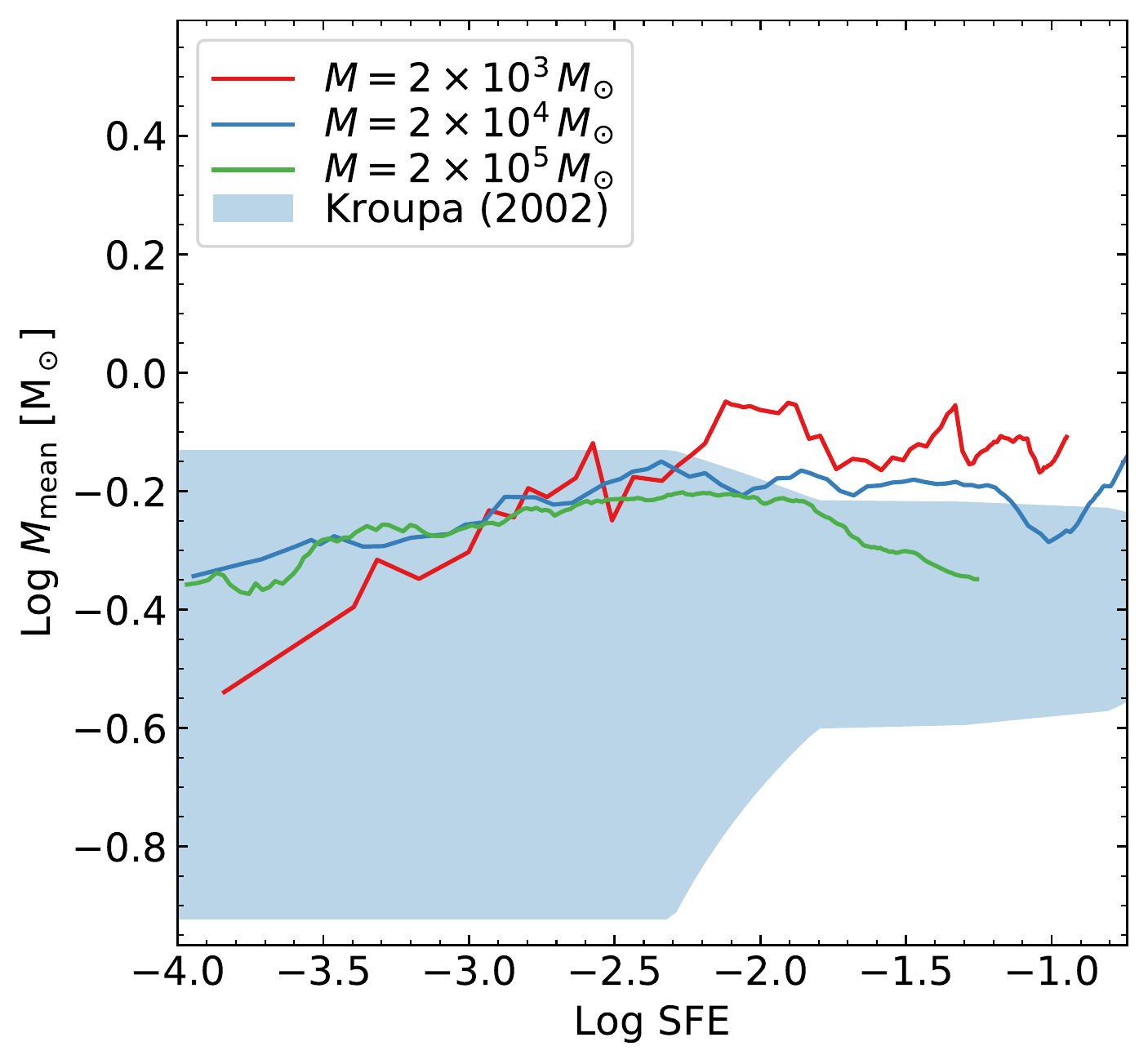}
\includegraphics[width=0.33\linewidth]{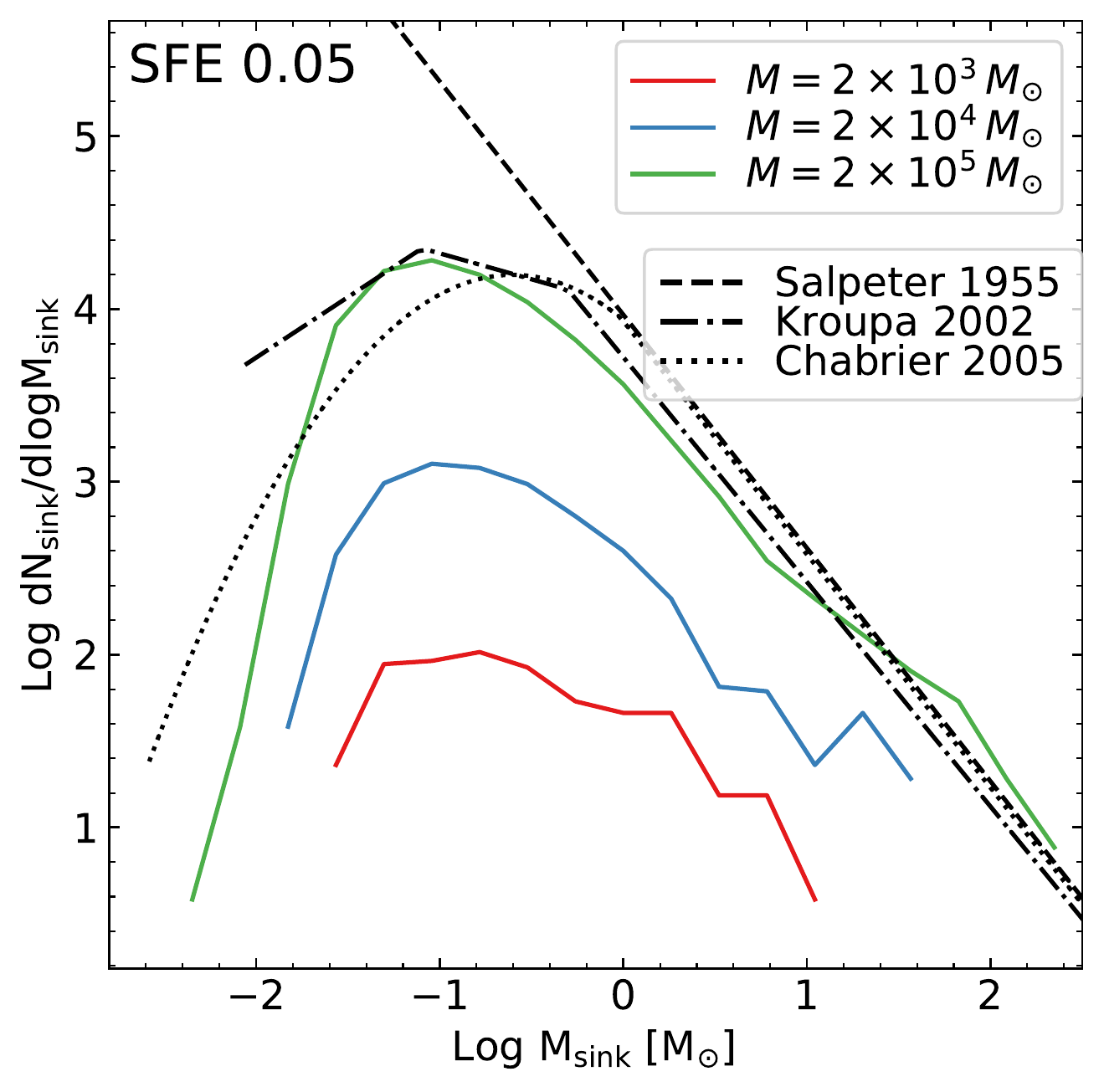}
\\
\includegraphics[width=0.33\linewidth]{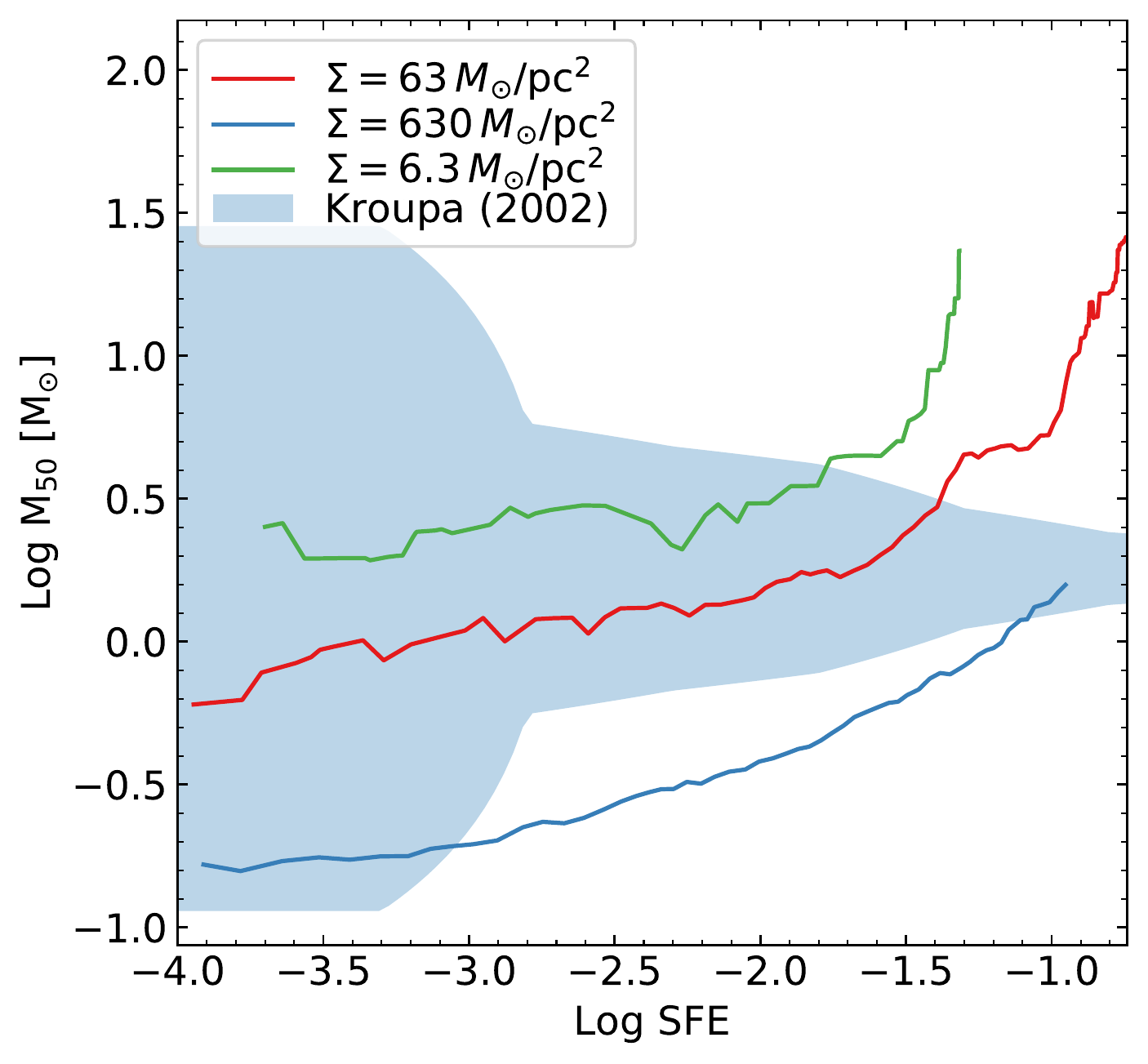}
\includegraphics[width=0.33\linewidth]{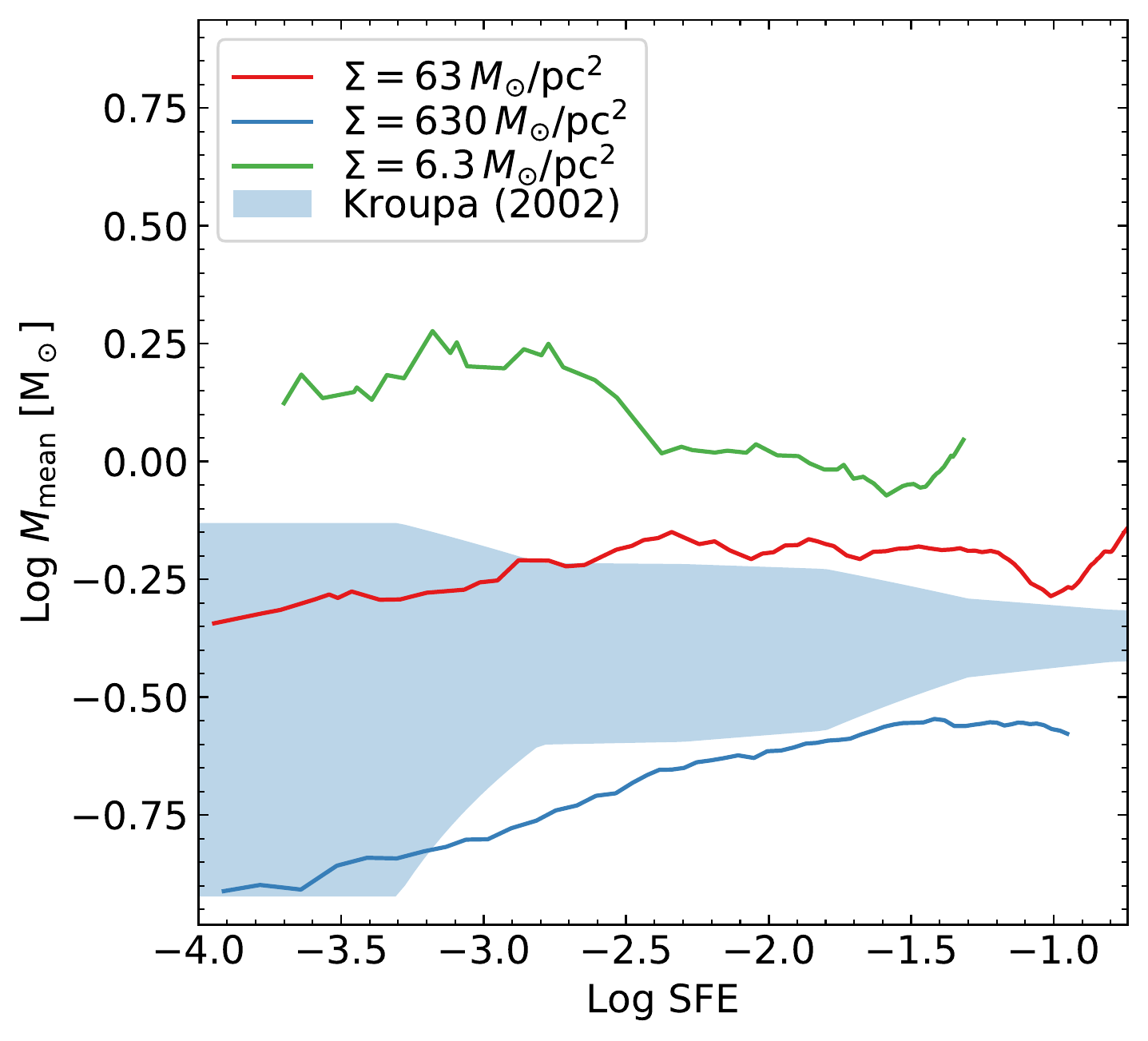}
\includegraphics[width=0.33\linewidth]{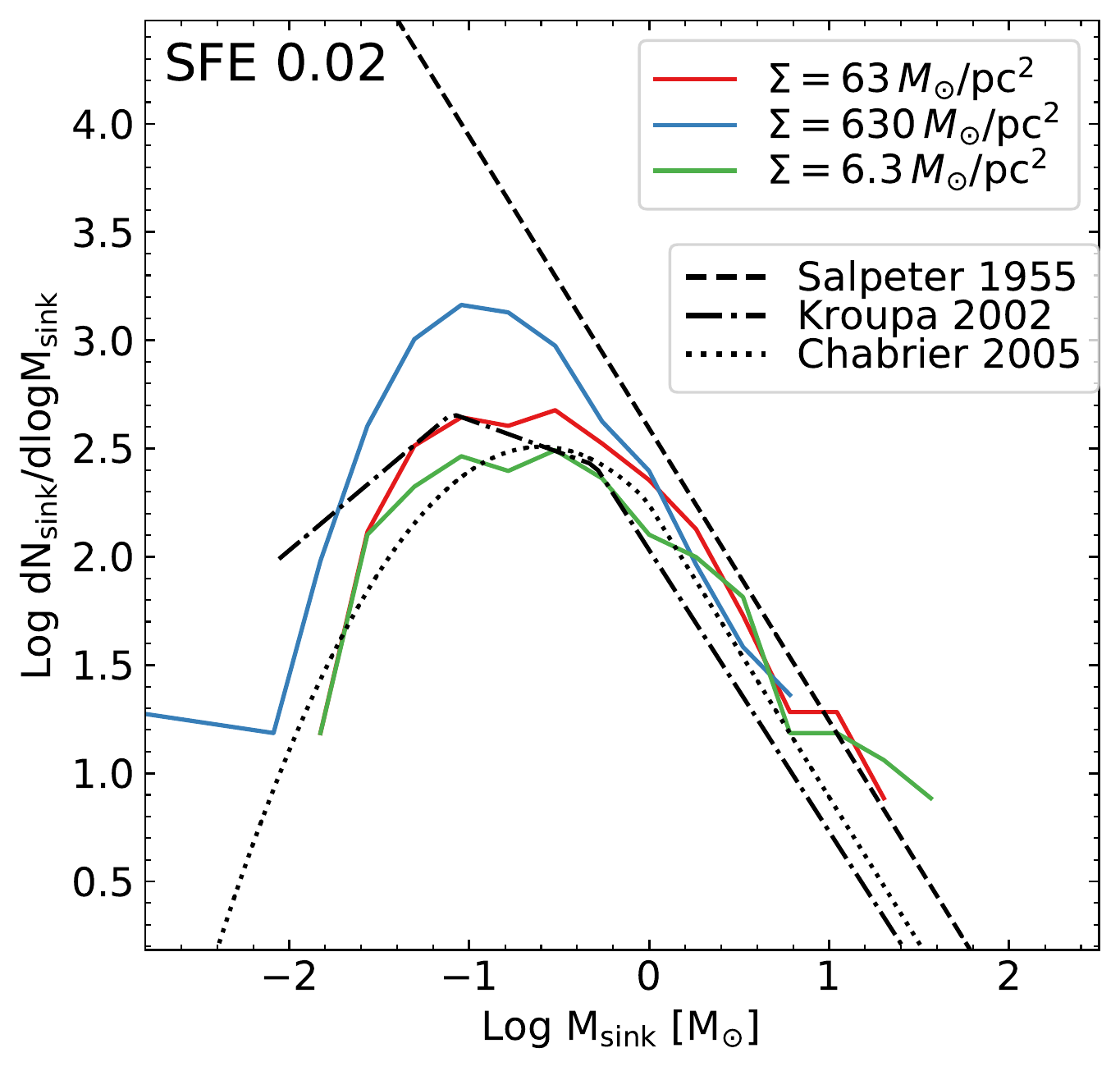}
\\
\includegraphics[width=0.33\linewidth]{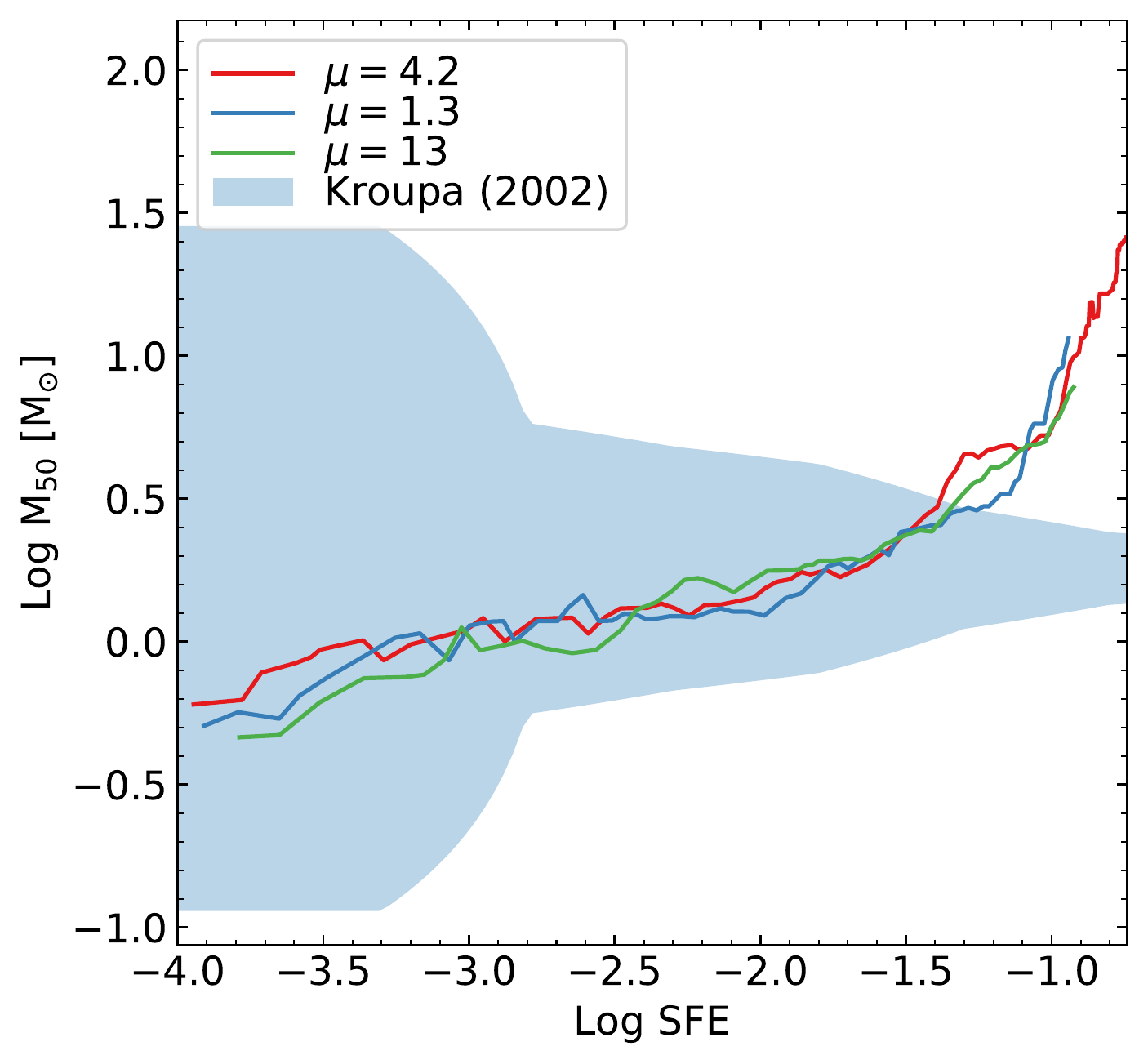}
\includegraphics[width=0.33\linewidth]{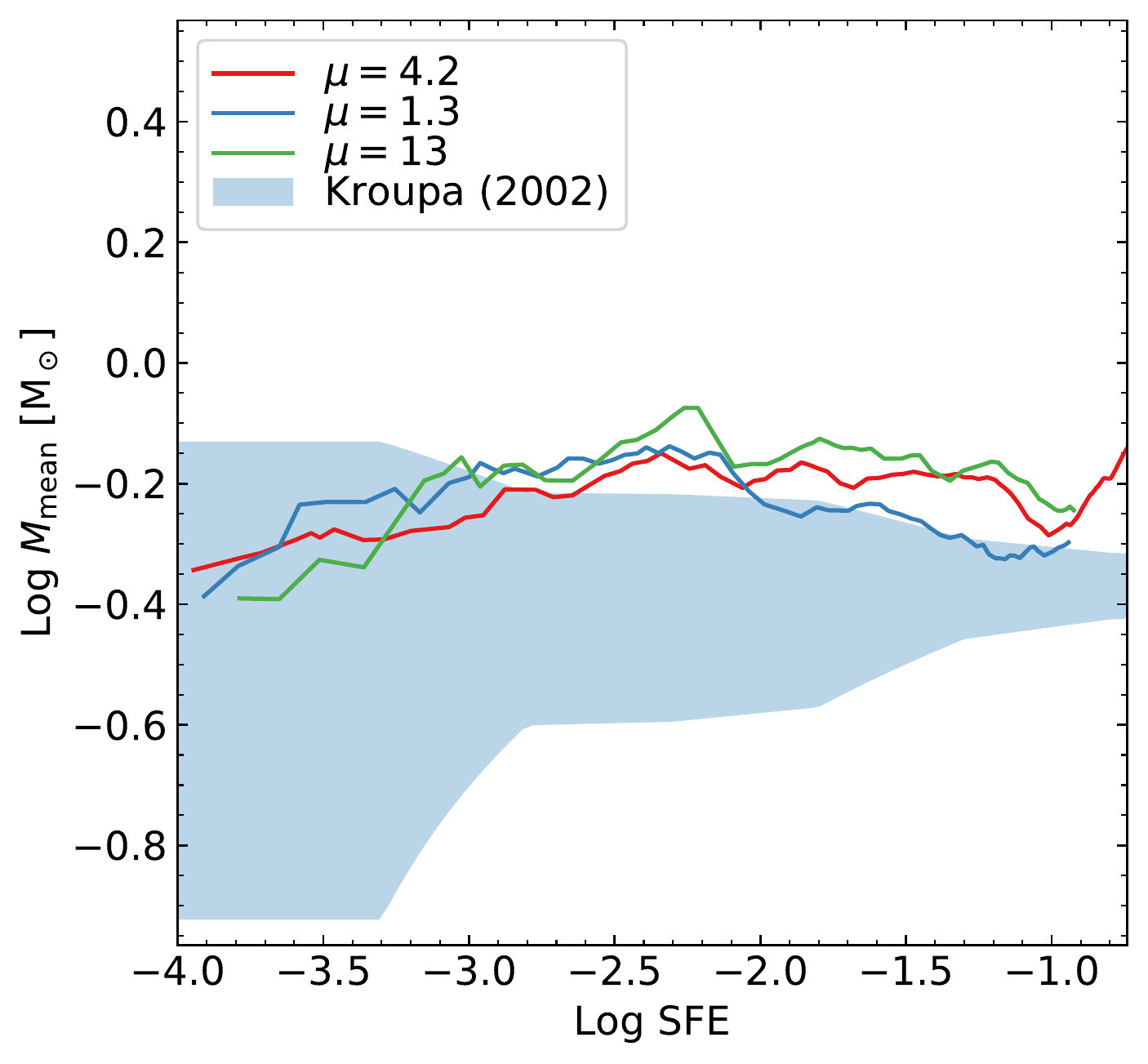}
\includegraphics[width=0.33\linewidth]{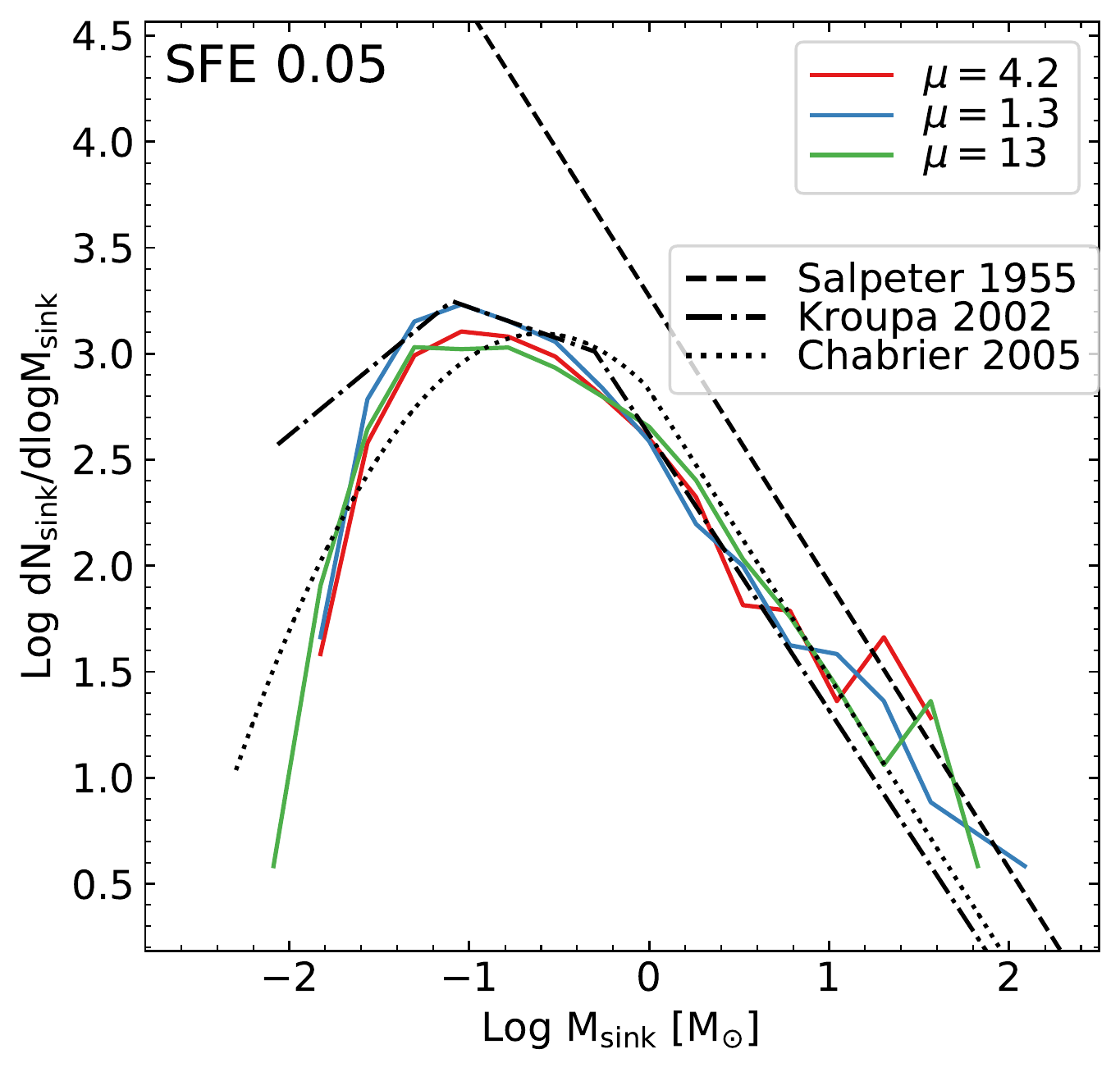}
\\
\vspace{-0.4cm}
\caption{Evolution of $\Mmassmedian$ and the mean sink masses (left and center columns, similar to Figure \ref{fig:mass_scale_evol}) as well as the distribution of sink particle masses at 5\% star formation efficiency (right column) for \textbf{M2e4\_C\_M\_J} (see Tables \ref{tab:physics_ladder}-\ref{tab:IC}) with variations in the initial turbulent virial parameter $\alphaturb$, initial cloud mass $M_0$, cloud surface density $\Sigma$ and normalized magnetic mass-to-flux ratio $\mu$. Note that we plot the IMF at a lower SFE for the surface density test, as the lowest surface density cloud becomes unbound before reaching 5\% SFE.}
\label{fig:imf_sensitivity_1}
\vspace{-0.5cm}
\end {center}
\end{figure*}

In this appendix we present in detail the results of various test runs (see Table \ref{tab:IC}) with protostellar jets enabled. In Figure \ref{fig:imf_sensitivity_1} we find that both the mean and mass-weighted median sink masses are sensitive to the initial properties of the cloud (mass, virial parameter, surface density). The one exception is the initial level of magnetization, which appears to have negligible effects, similar to the isothermal case in \citetalias{Guszejnov_isoT_MHD}.

\begin{figure*}
\begin {center}
\includegraphics[width=0.33\linewidth]{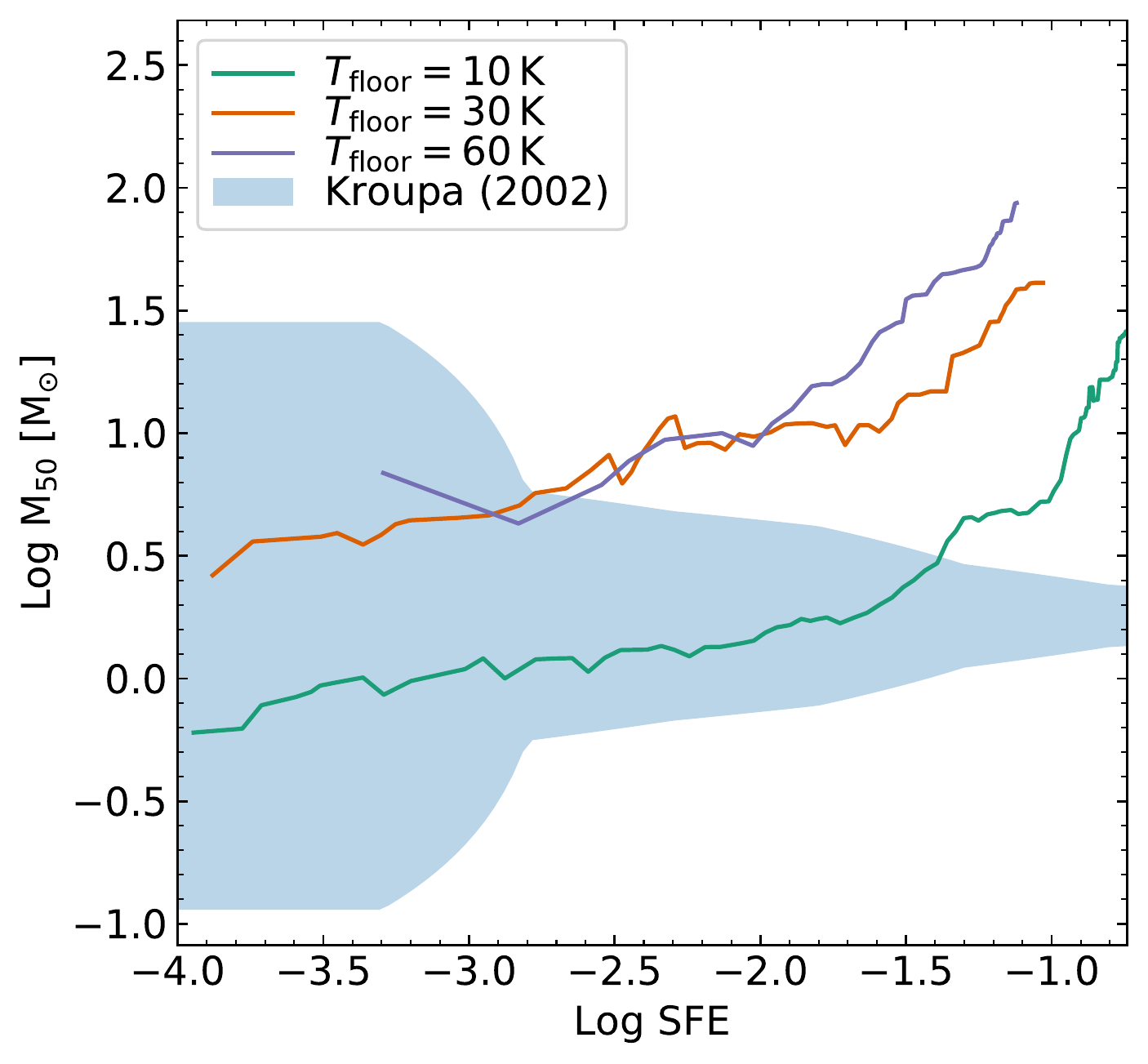}
\includegraphics[width=0.33\linewidth]{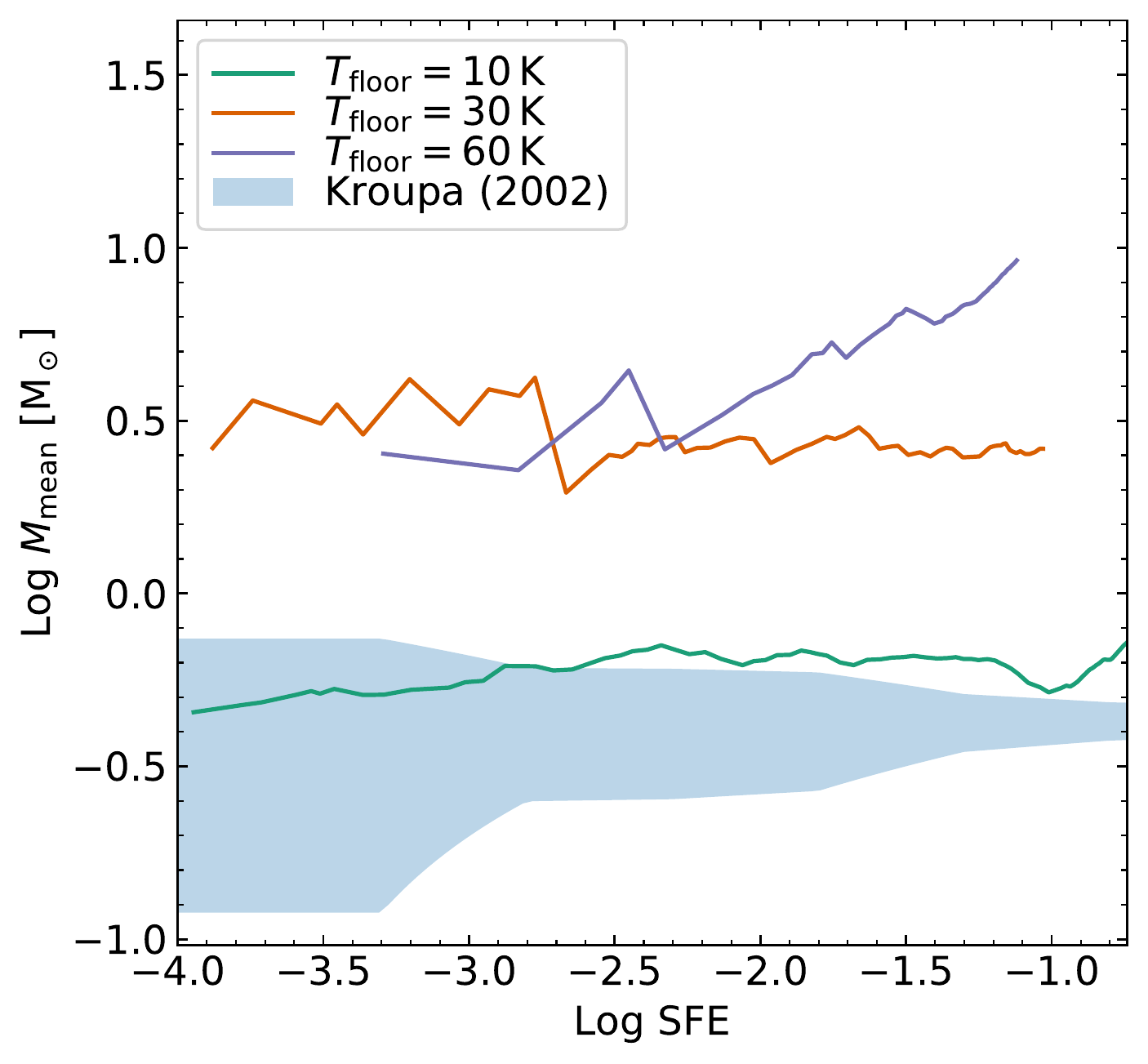}
\includegraphics[width=0.33\linewidth]{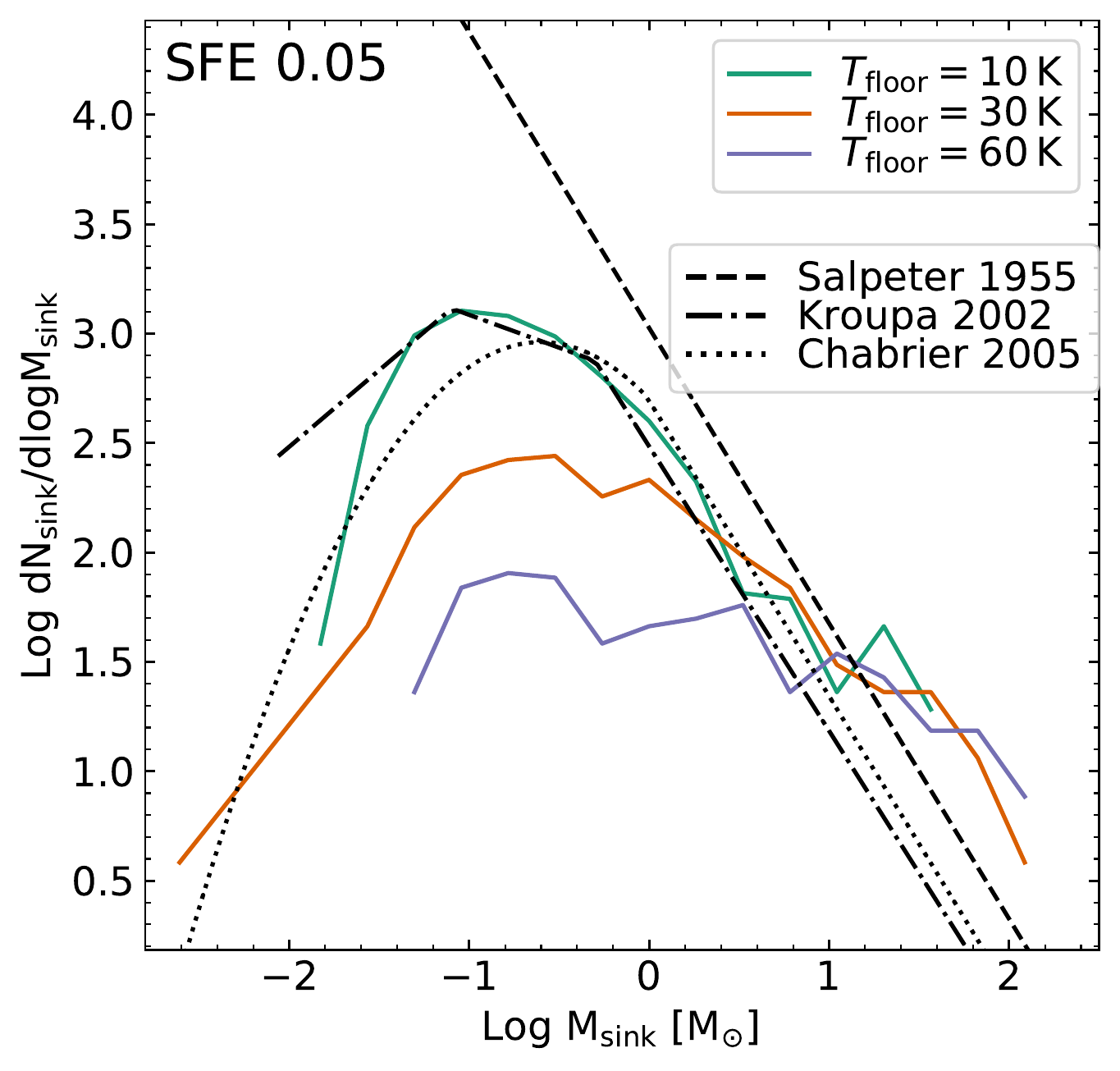}
\\
\includegraphics[width=0.33\linewidth]{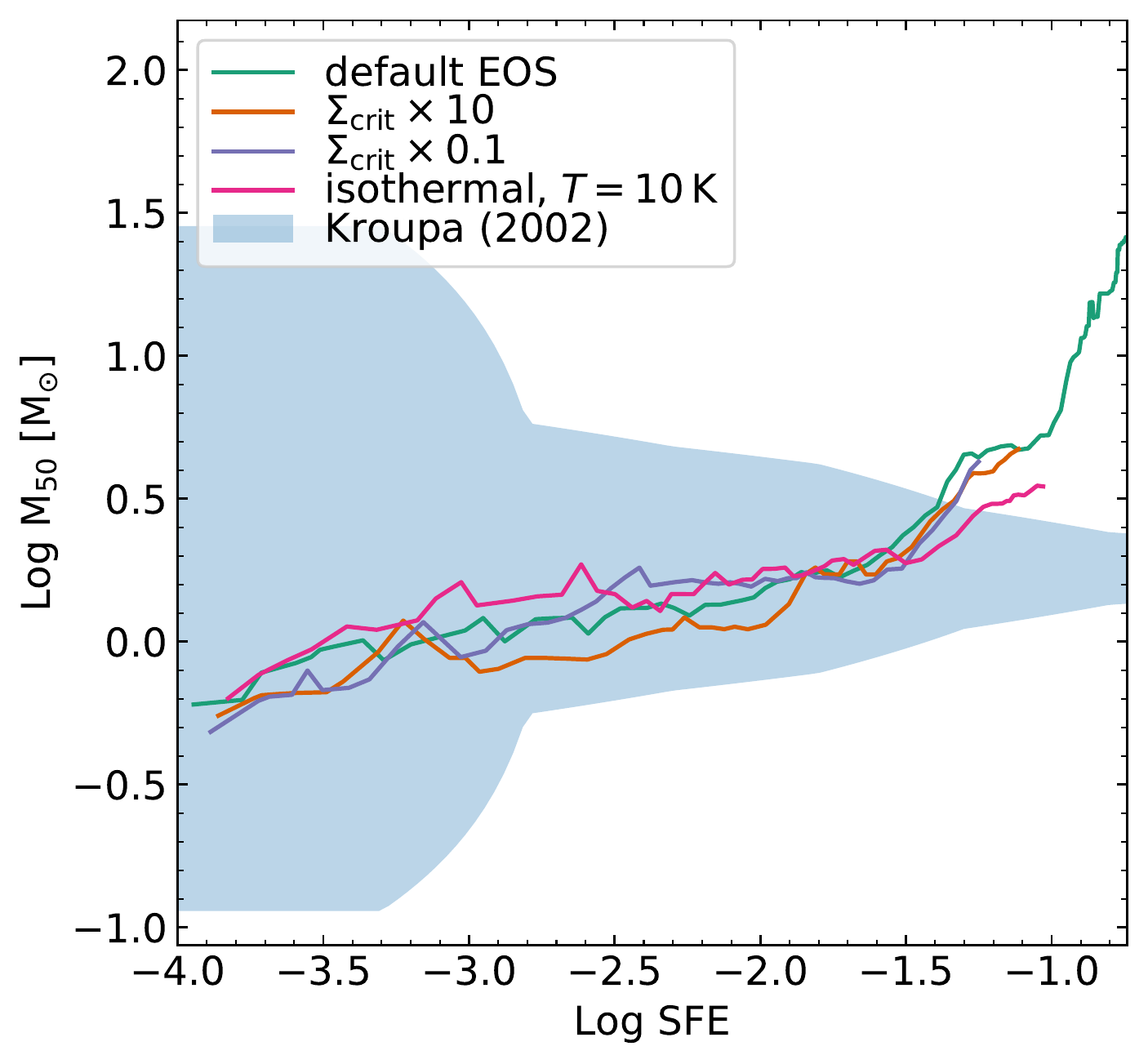}
\includegraphics[width=0.33\linewidth]{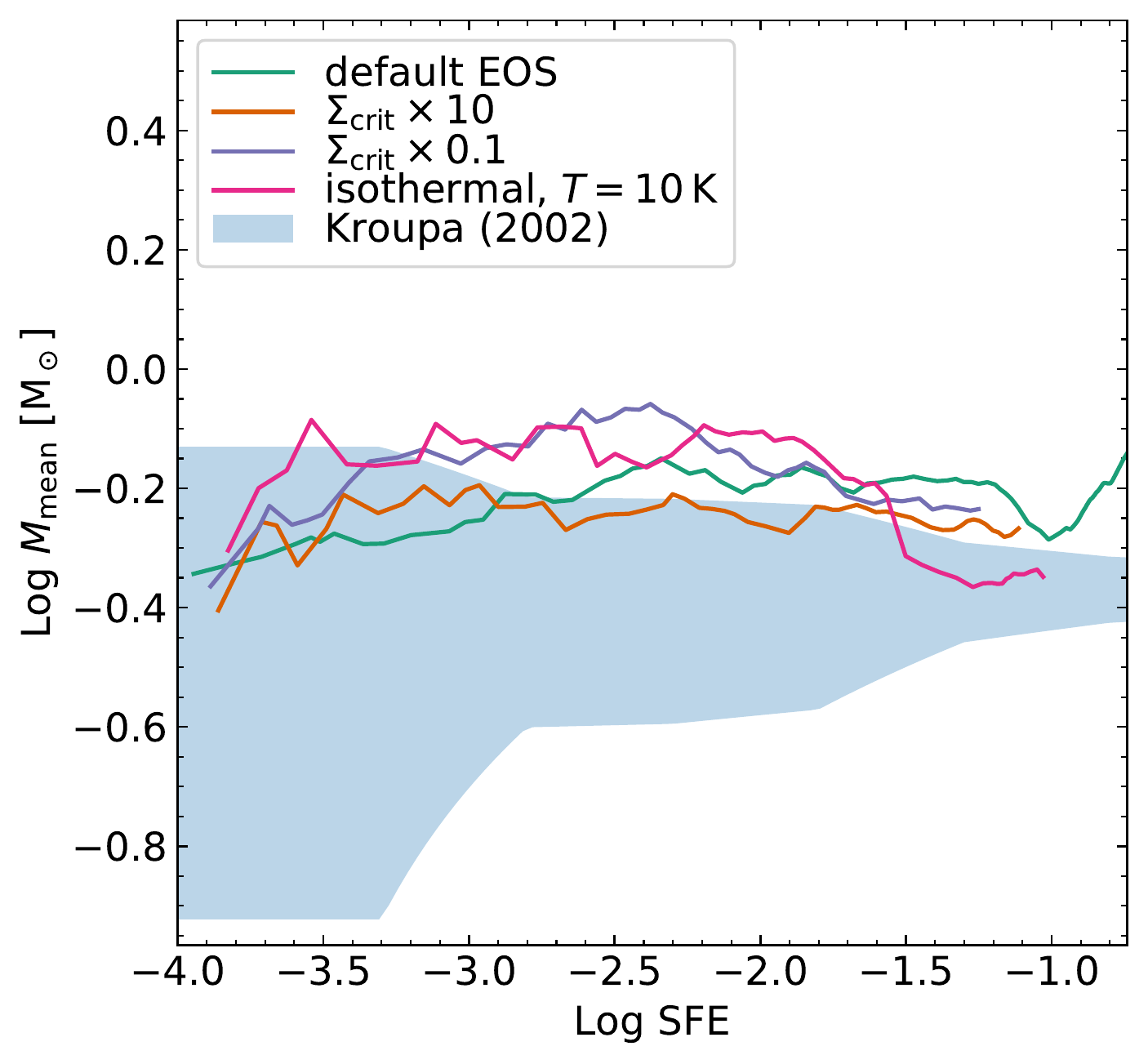}
\includegraphics[width=0.33\linewidth]{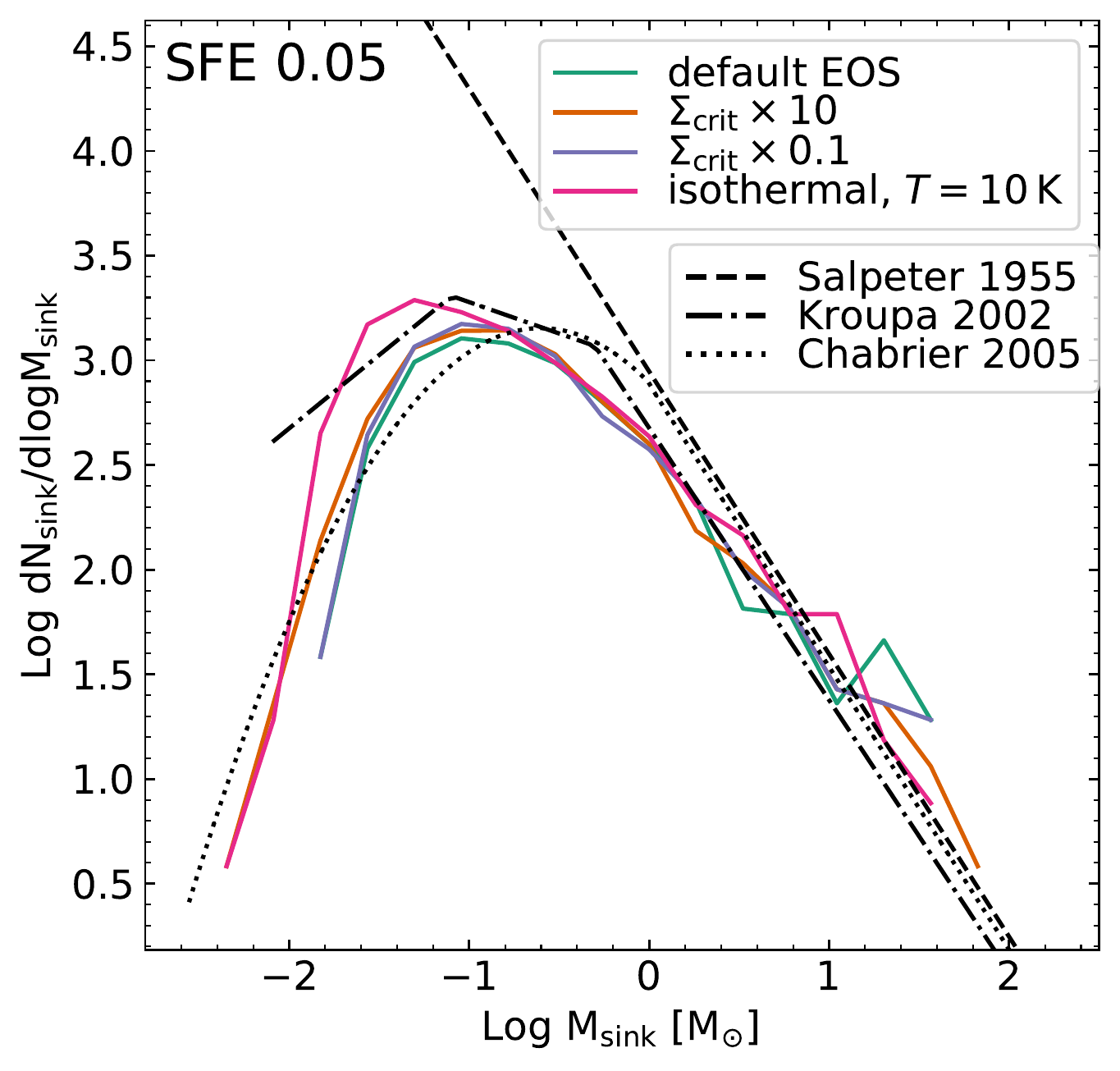}
\\
\includegraphics[width=0.33\linewidth]{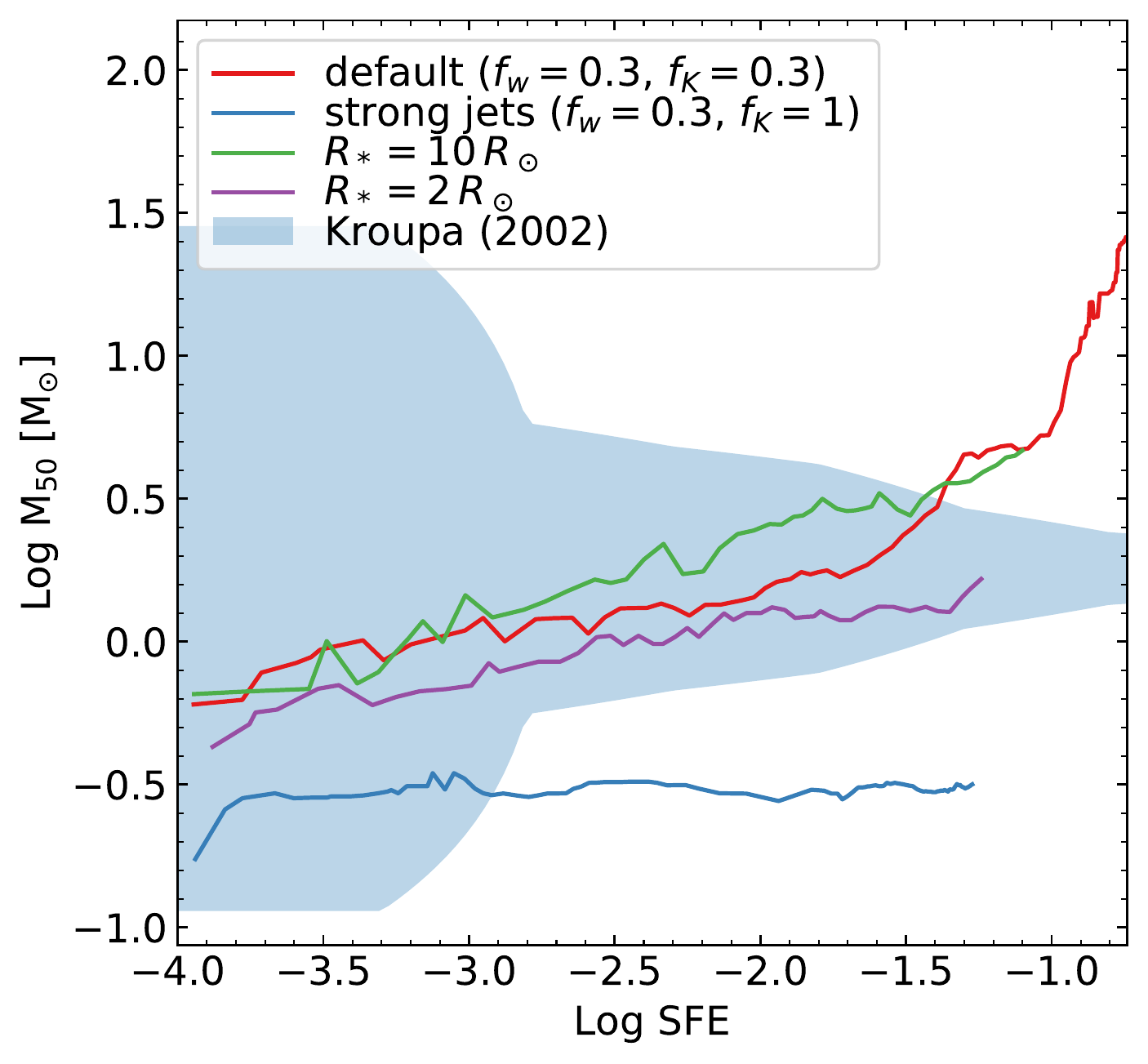}
\includegraphics[width=0.33\linewidth]{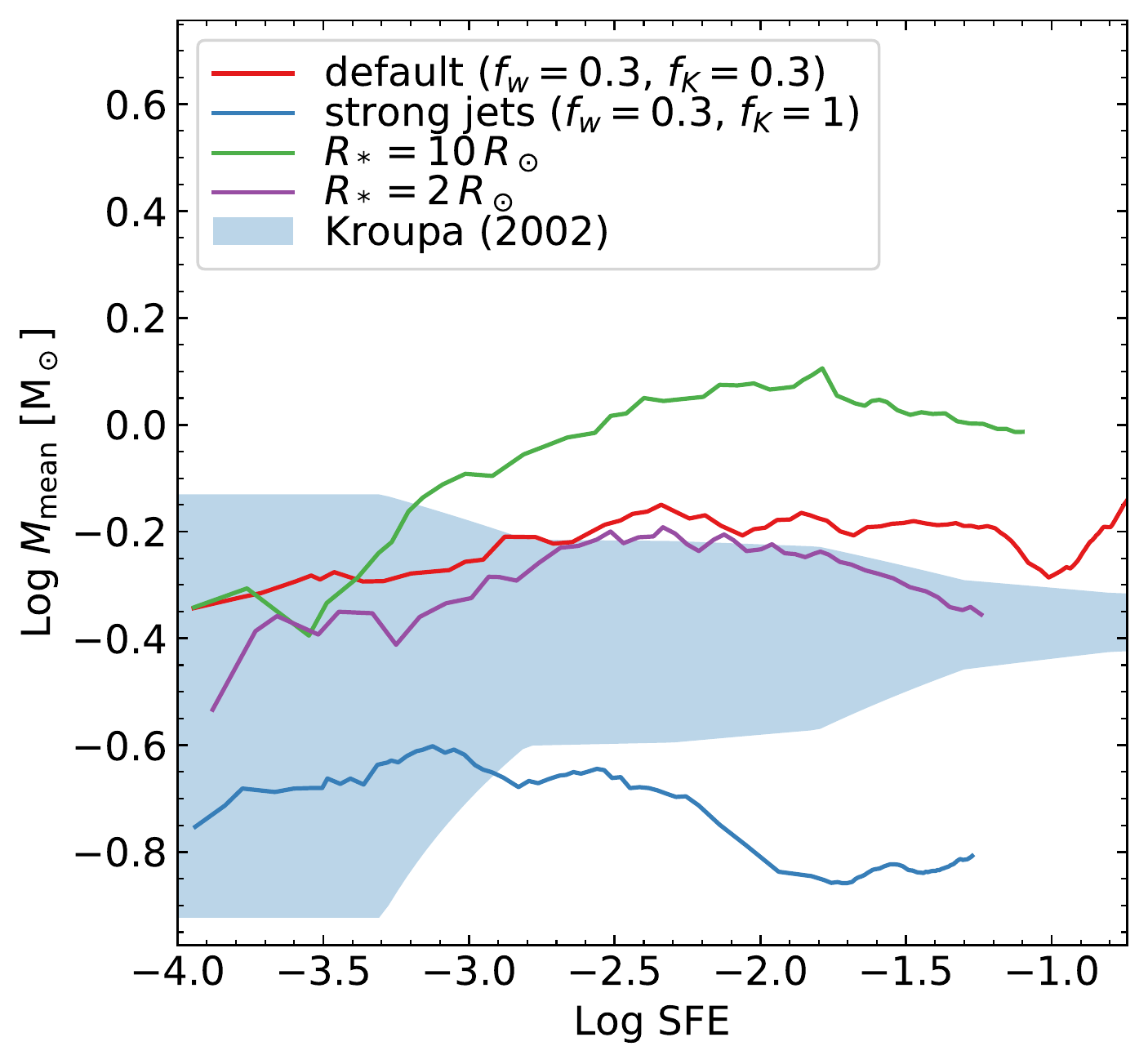}
\includegraphics[width=0.33\linewidth]{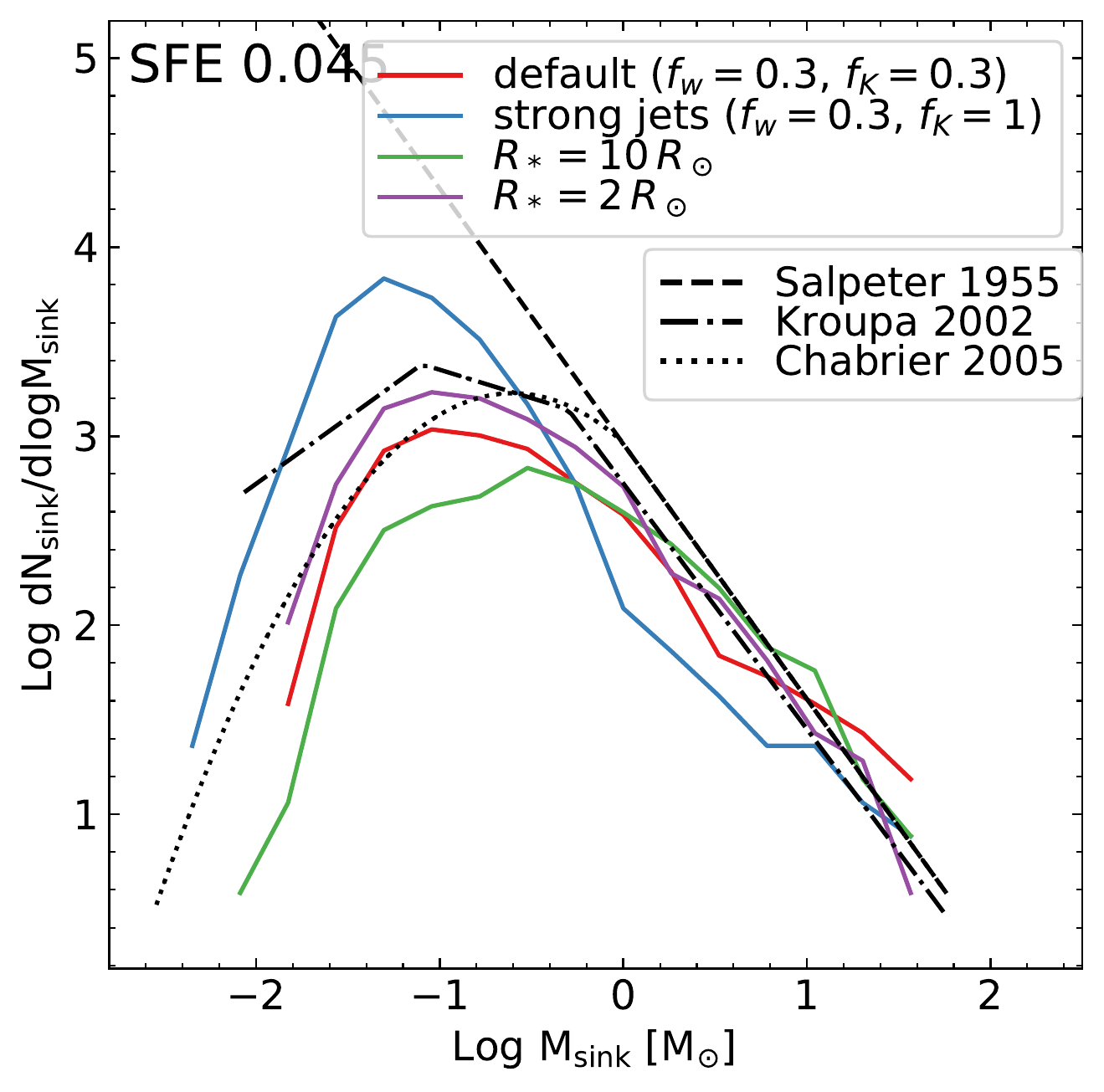}
\\
\includegraphics[width=0.33\linewidth]{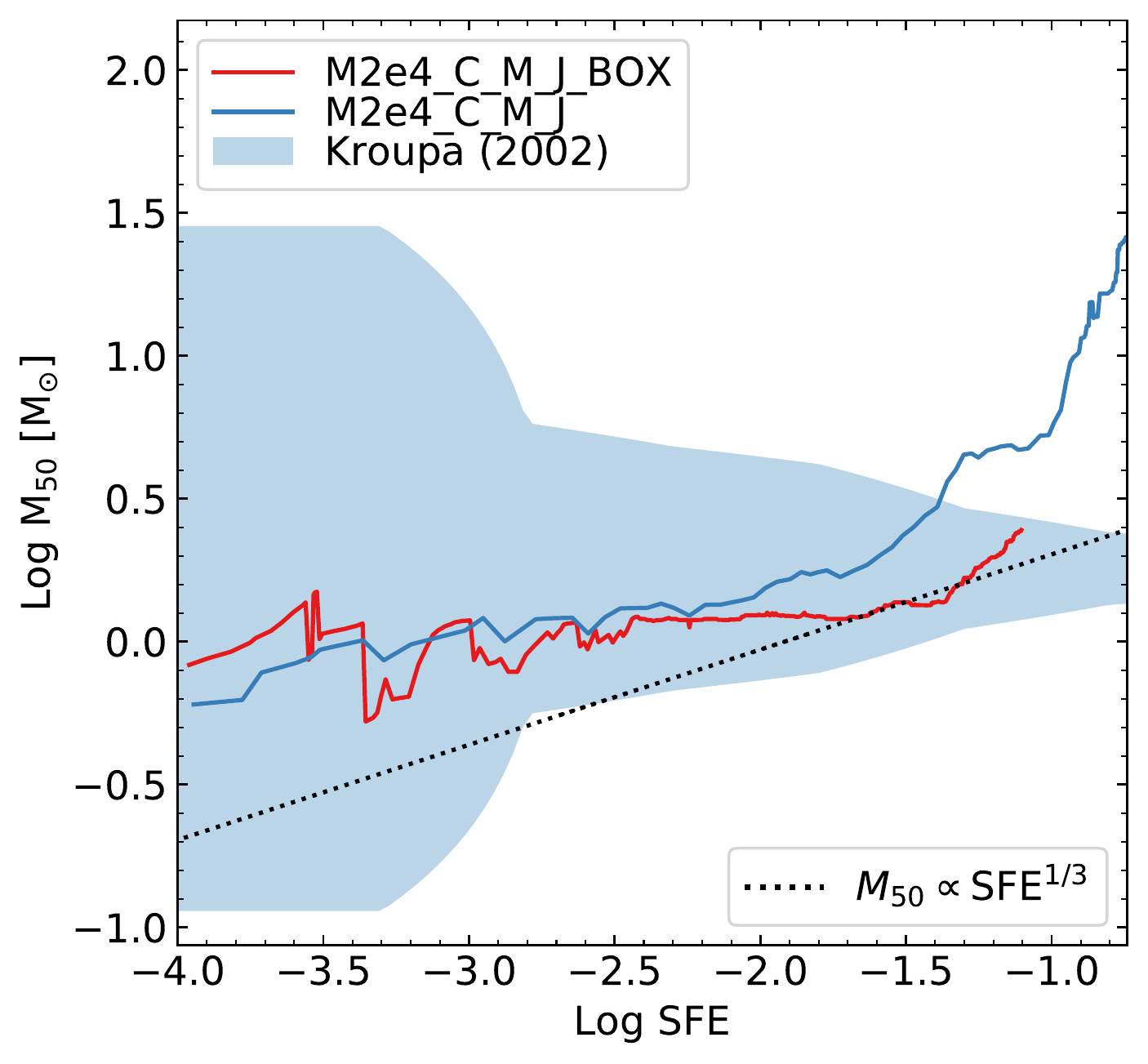}
\includegraphics[width=0.33\linewidth]{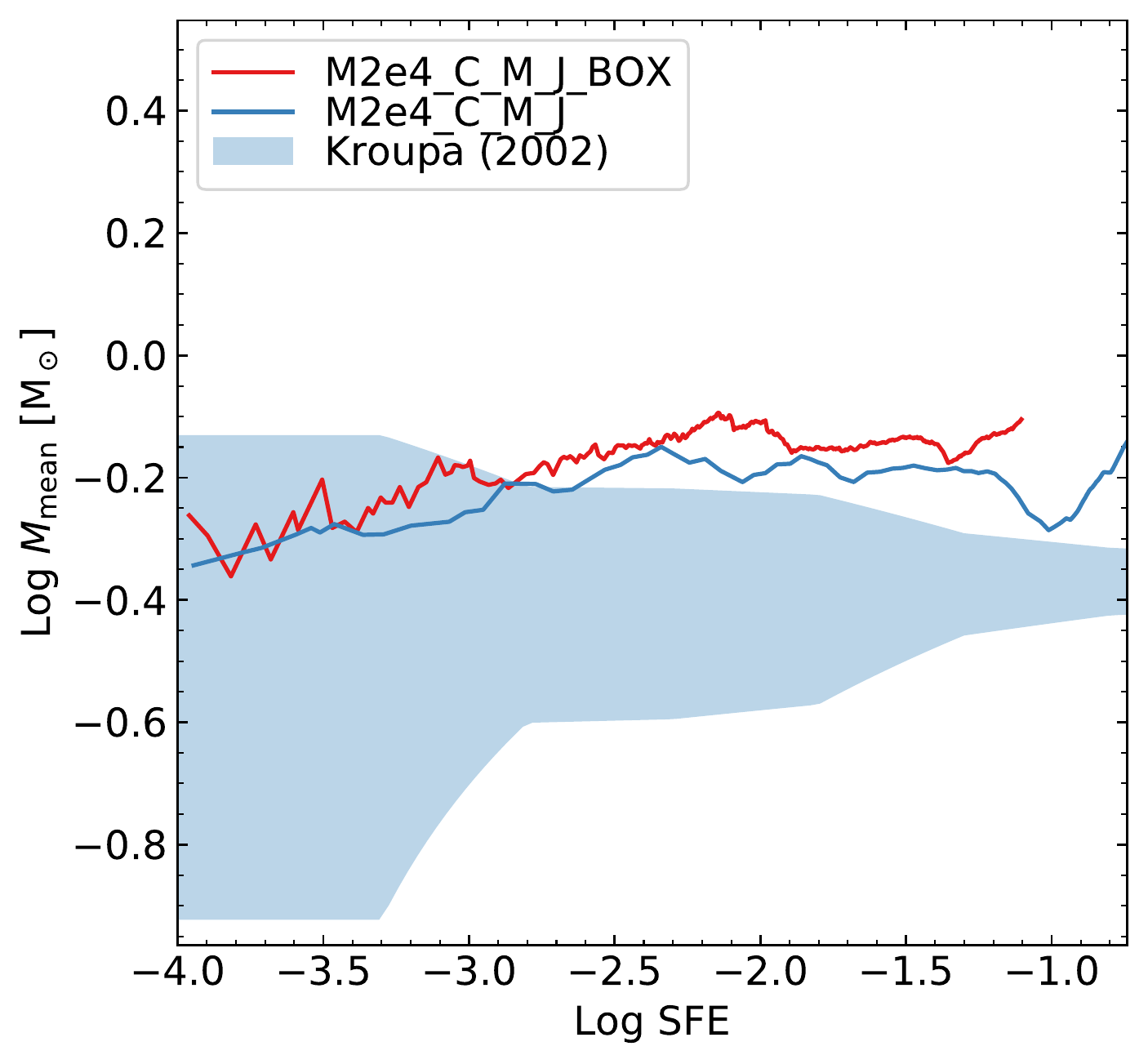}
\includegraphics[width=0.33\linewidth]{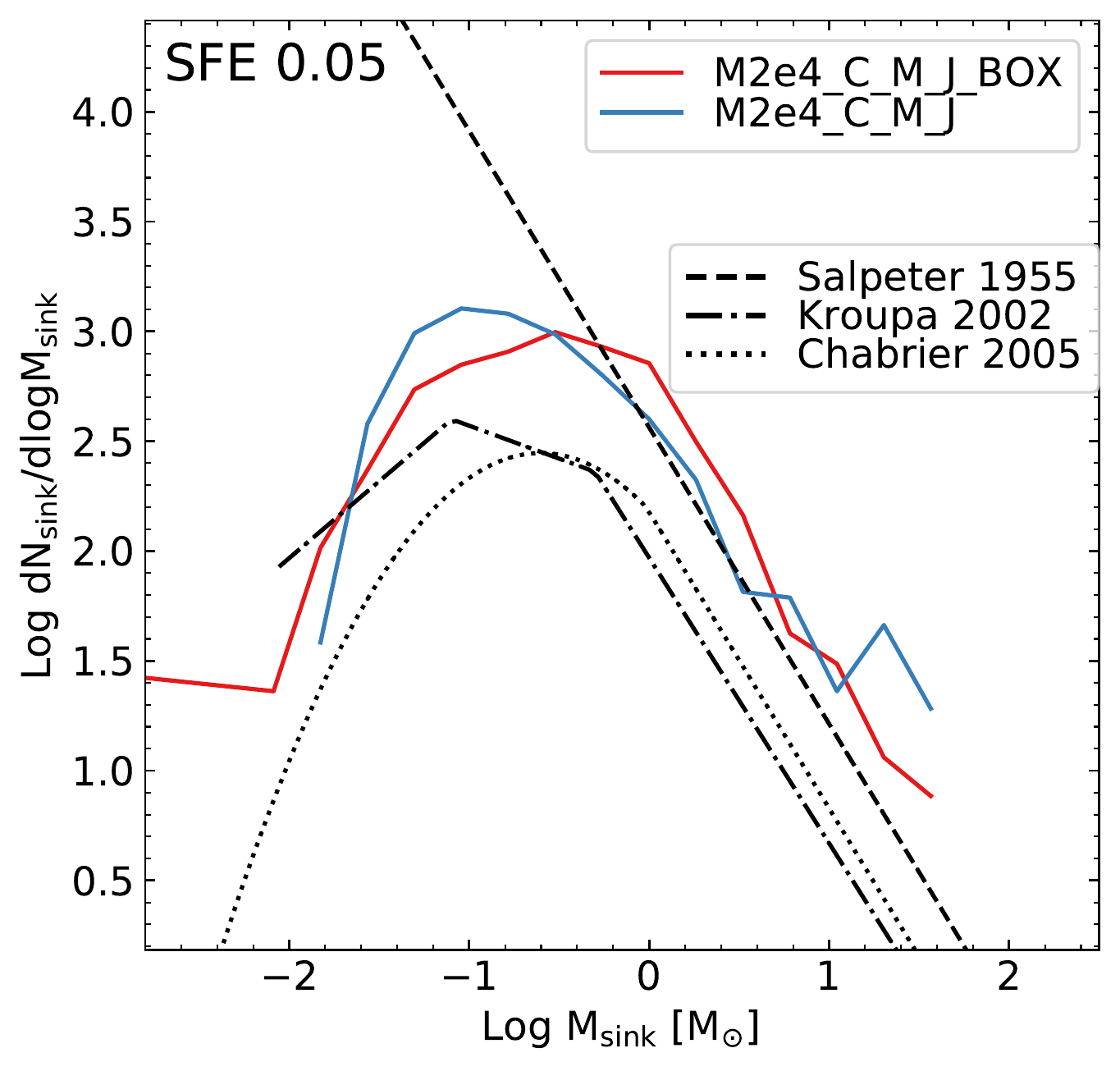}
\\
\vspace{-0.4cm}
\caption{Evolution of the mass-weighted median sink mass $\Mmassmedian$ and the number-weighted mean sink mass $\Mmean$ (left and center columns, similar to Figure \ref{fig:mass_scale_evol}) as well as the distribution of sink particle masses at 5\% star formation efficiency (right column) for \textbf{M2e4\_C\_M\_J} (see Tables \ref{tab:physics_ladder}-\ref{tab:IC}) with variations in the floor temperature $T_\mathrm{floor}$, the thermodynamics of the simulation (varying $\Sigma_\mathrm{crit}$, the transition surface density between optically thin and thick cooling regimes), the parameters of the jet module (different $f_K$ values as well as using a fixed $R_*$ stellar radius, see \S\ref{sec:jets}) and the type of initial condition (Sphere vs Box, see \S\ref{sec:IC}).}
\label{fig:imf_sensitivity_2}
\vspace{-0.5cm}
\end {center}
\end{figure*}

Figure \ref{fig:imf_sensitivity_2} shows the results of further tests where the parameters of the underlying physical models were varied. Figure \ref{fig:imf_sensitivity_2} shows that the mass spectrum is especially sensitive to the floor temperature of the simulation. We carried out an additional test where we varied the critical surface density $\Sigma_\mathrm{crit}$ where the cooling module transitions between optically thin and thick regimes. We found that varying $\Sigma_\mathrm{crit}$ (i.e., the opacity limit) by a factor of 10 in either direction has little effect on the sink mass spectrum. Transitioning to an isothermal equation of state also has only minor effects that arise from the formation of very low mass sinks, which were previously suppressed by the EOS.

As expected, changing the parameters of the jet module has significant effects, we find that the results are sensitive to the momentum loading of the jets, which is set by $\Gamma$, see \S \ref{sec:jet_mom_loading} for details. Note that we also find that launching jets from a constant stellar radius, instead of the one set by the protostellar evolution model of \S\ref{sec:shared_physics}, produces qualitatively similar results (see Figure \ref{fig:imf_sensitivity_2}).



\section{Scaling relations}\label{sec:scaling_derivation}

\begin{figure*}
\begin {center}
\includegraphics[width=0.33\linewidth]{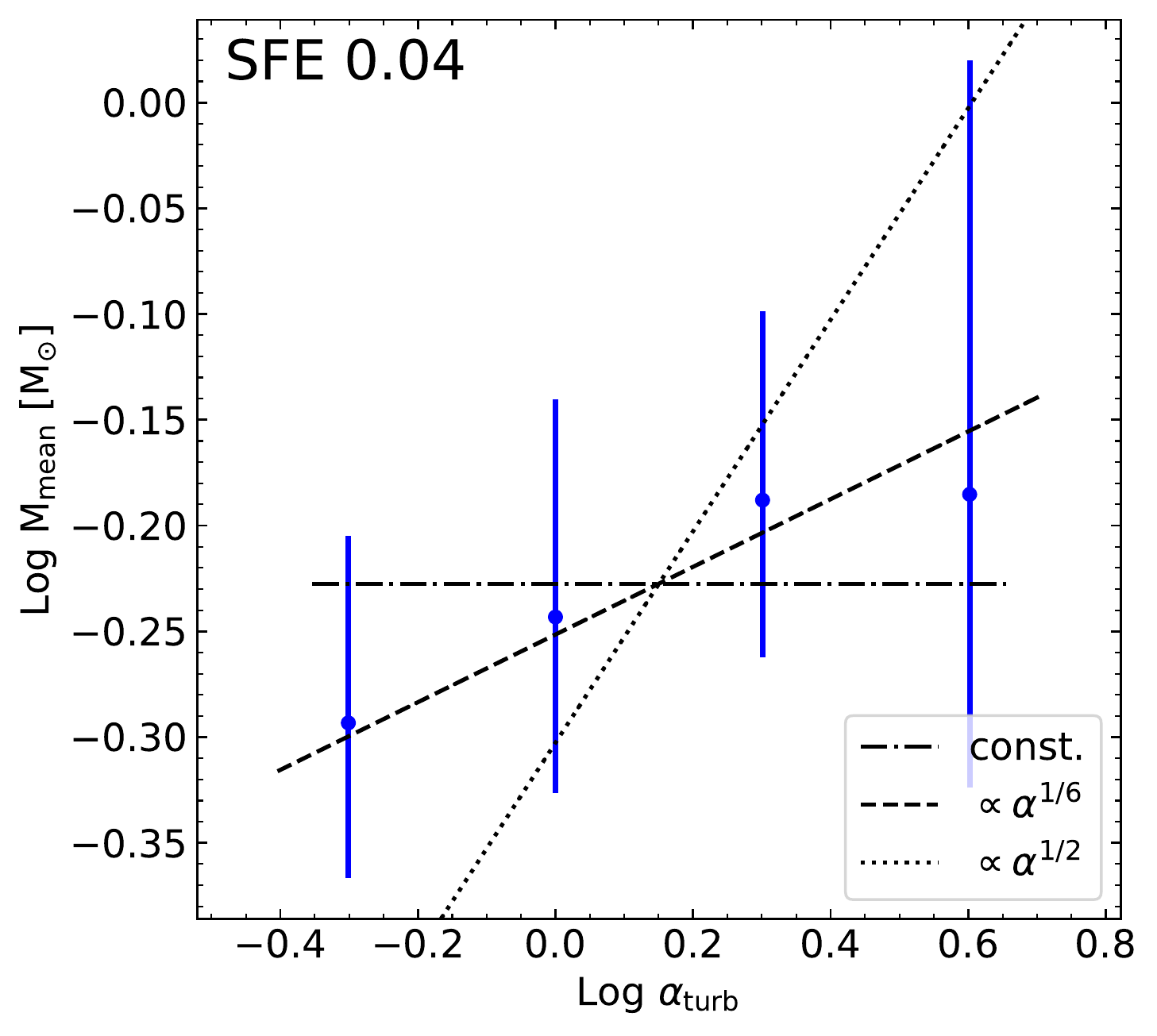}
\includegraphics[width=0.33\linewidth]{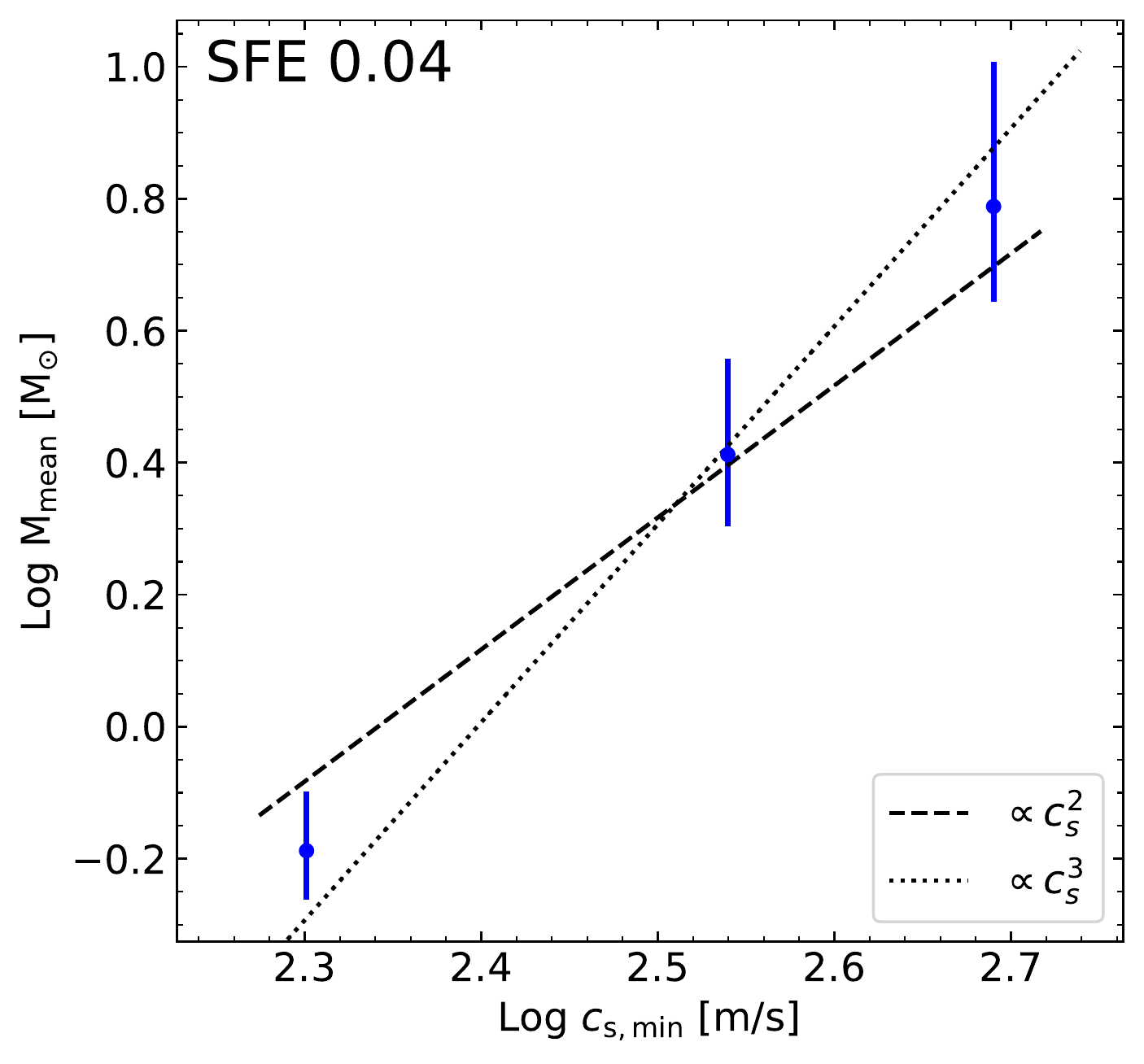}
\includegraphics[width=0.33\linewidth]{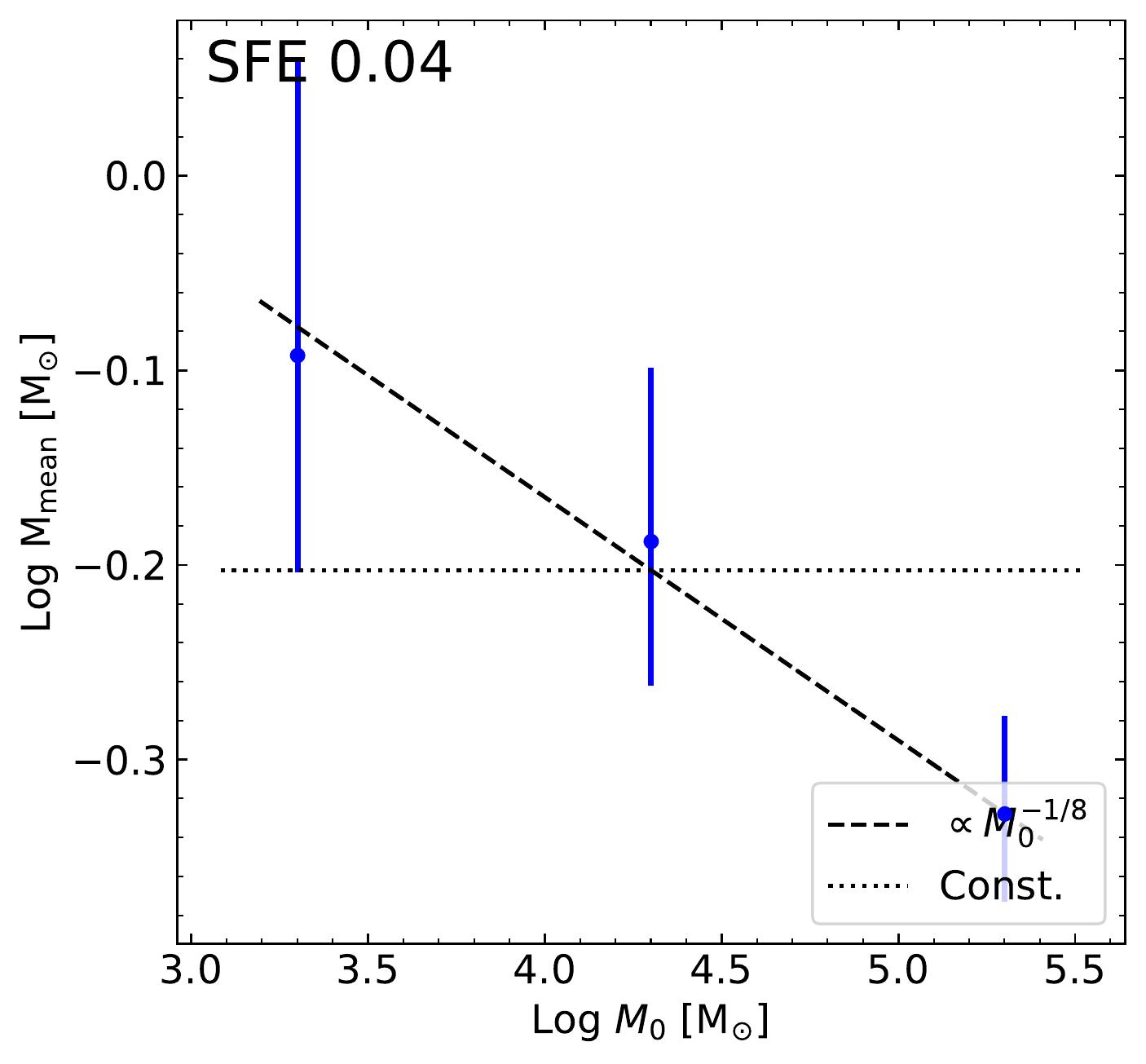}\\
\includegraphics[width=0.33\linewidth]{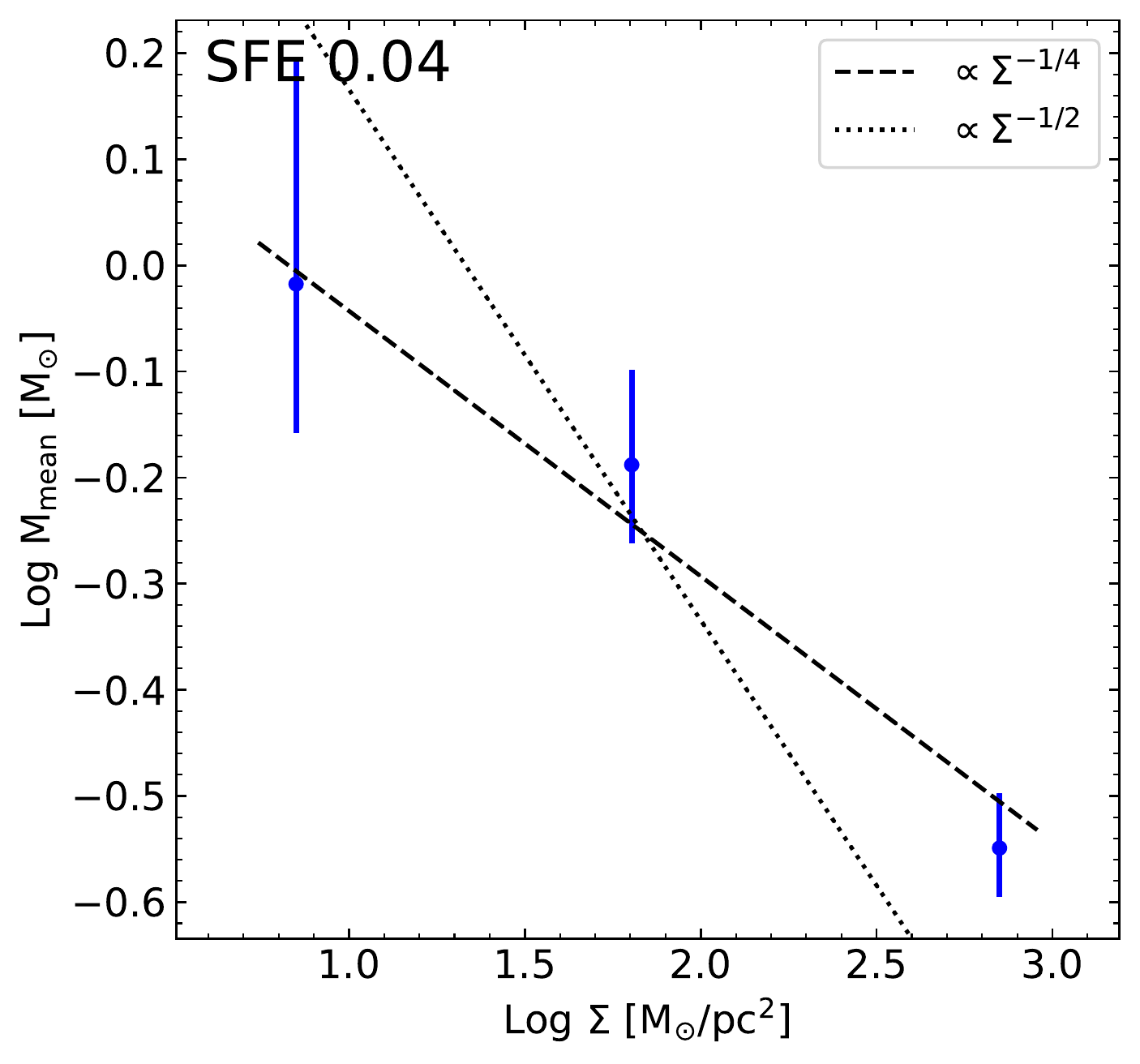}
\includegraphics[width=0.33\linewidth]{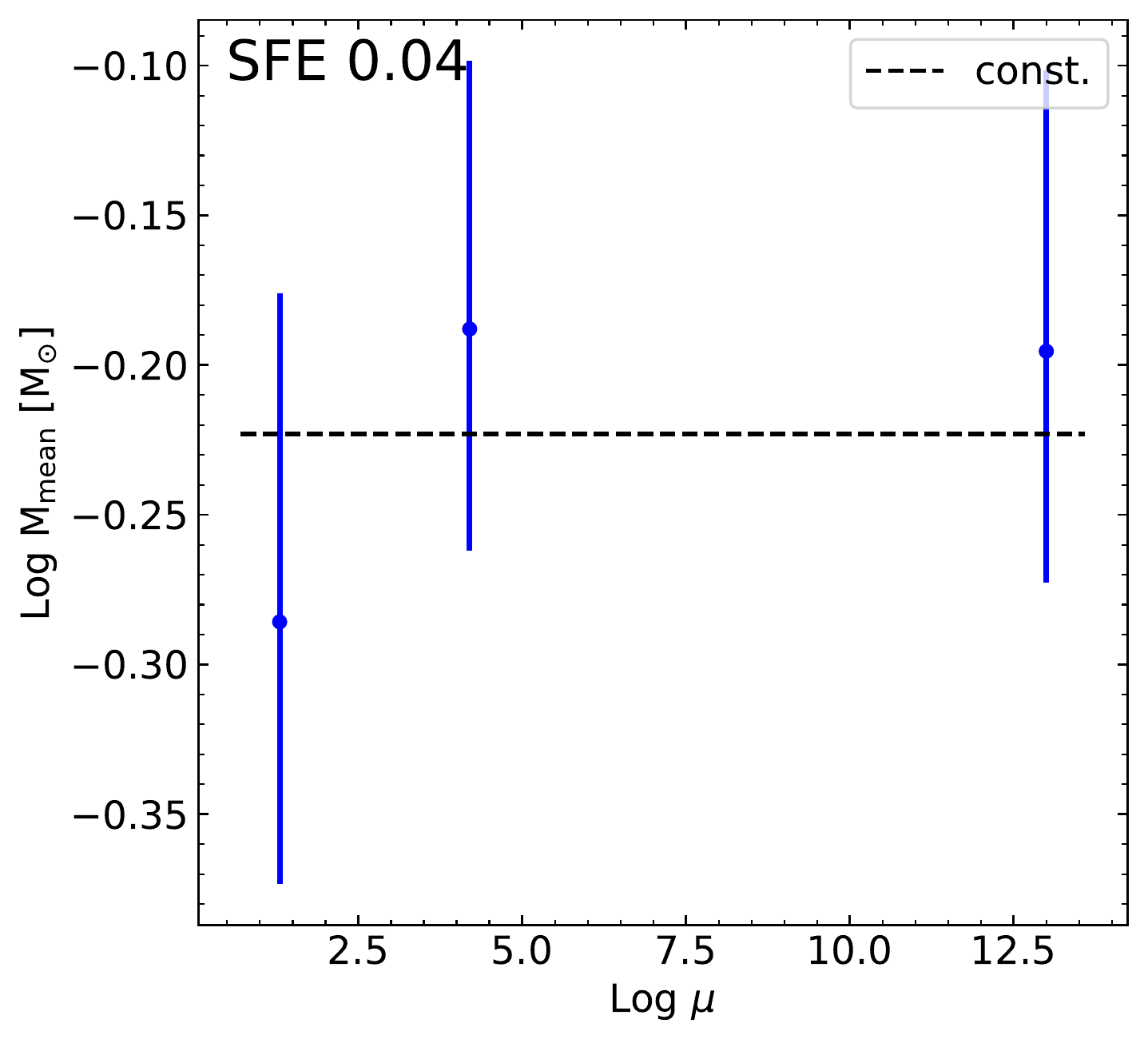}
\includegraphics[width=0.33\linewidth]{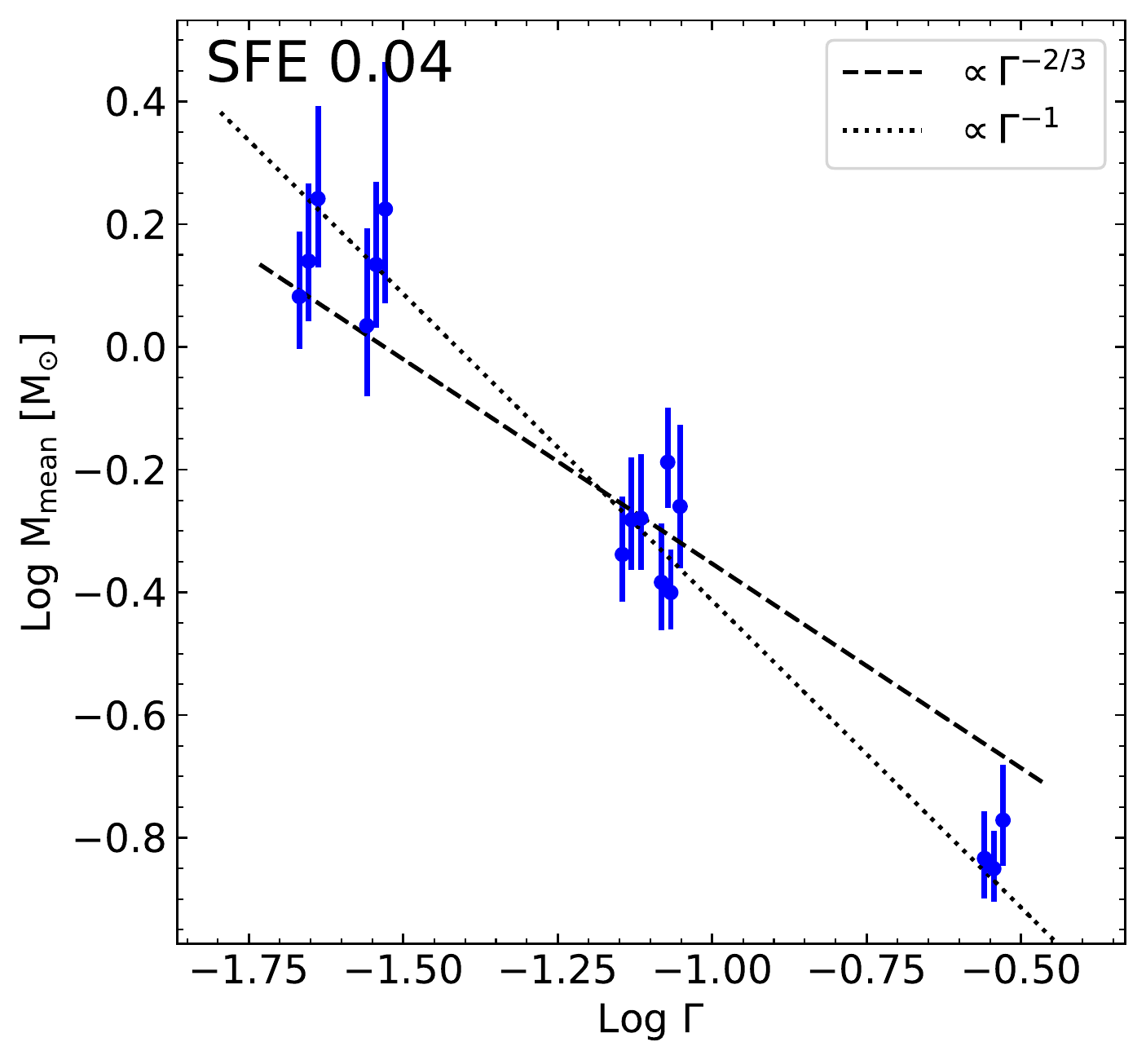}
\vspace{-0.4cm}
\caption{Dependence of the mean sink mass at 4\% SFE on the initial turbulent virial parameter $\alphaturb$ (\textit{top, left}), minimum sound speed $\csmin$ (set by the floor temperature $T_\mathrm{floor}$, \textit{top, middle}), cloud mass $M_0$ (\textit{top, right}), cloud surface density $\Sigma$ (\textit{bottom, left}), normalized magnetic mass-to-flux ratio $\mu$ (\textit{bottom, middle}) and the $\Gamma$ momentum loading of the jets (\textit{bottom, right}). We chose 4\% as the reference SFE because some low mass and surface density runs are disrupted by jets before reaching 5\% SFE. Note that in the case of the jet momentum loading we used runs with different initial turbulent realizations, which are shown slightly offset to make the plot easier to parse.  The errors are estimated by bootstrapping: we resample the sink mass distribution at fixed total stellar mass and calculate the 95\% confidence interval of $\Mmassmedian$ over these realizations, which we denote with errorbars. These scalings are discussed in the main text in \S\ref{sec:sensitivity}}
\label{fig:Mmean_dependence_fit}
\vspace{-0.5cm}
\end {center}
\end{figure*}

In this appendix we examine in detail how the mean sink mass depends on the initial conditions of the cloud (turbulent virial parameter $\alphaturb$, minimum sound speed $\csmin$, normalized magnetic flux ratio $\mu$, surface density $\Sigma$ and initial cloud mass $M_0$), by examining how the characteristic sink mass depends on each of them independently. 
We assume that the relation between the mean sink mass and the initial conditions is described by a multivariate power-law. Using subsets of our runs from Table \ref{tab:IC} where only one of these parameters is varied we carry out least-squares fits to the individual exponents in turn, each at a fixed fiducial SFE value (4\%). To estimate the errors of the fitted exponents we first estimate the errors in the mean sink mass using bootstrapping, which means resampling the sink mass distribution at fixed SFE and calculating the 95\% confidence interval of the mean mass over these new samples (see Figure \ref{fig:Mmean_dependence_fit}). We find the following fitting parameters and errors
\begin{equation}
\Mmean \propto  \Gamma^{-0.65\pm 0.15}\, \cs^{2.5\pm 0.5}\, \Sigma^{-0.3\pm 0.1}\, \alphaturb^{0.15\pm 0.19}\, M_0^{-0.12\pm 0.07},
\label{eq:1Dfitting_func}
\end{equation}
which can be also expressed as
\begin{equation}
\Mmean \propto  \Gamma^{-0.65\pm 0.05}\, \csmin^{2.5\pm 0.5}\, \rho^{-0.2\pm 0.07}\, \alphaturb^{0.15\pm 0.19}\, M_0^{-0.22\pm 0.10}.
\label{eq:1Dfitting_func_rho_app}
\end{equation}
See Figure \ref{fig:Mmean_mass_scale_compare} for a visual representation of the goodness of the fit.

\bsp	
\label{lastpage}
\end{document}